%% file: paper.tex
\documentclass[pdftex,twocolumn,apjfonts,10pt,apj,numberedappendix]{emulateapj}
\usepackage[letterpaper,top=1.875in, bottom=0.14in, left=1.07in, right=0.58in]{geometry}
\usepackage{graphicx}
\usepackage{tikz}
\usepackage{multirow}
\newcommand{\diffblock}[1]{#1}

\begin{document}
\title{Deep HST Imaging in NGC~6397: Stellar Dynamics}
\author{J. S. Heyl$^1$}
\author{H. Richer$^1$}
\author{J. Anderson$^2$}
\author{G. Fahlman$^3$}
\author{A. Dotter$^2$}
\author{J. Hurley$^4$}
\author{J. Kalirai$^2$}
\author{R. M. Rich$^5$}
\author{M. Shara$^6$}
\author{P. Stetson$^3$}
\author{K. A. Woodley$^1$}
\author{D. Zurek$^6$}
\affil{$^1$ Department of Physics and Astronomy, University of British
  Columbia, Vancouver BC V6T 1Z1 Canada}
\affil{$^2$ Space Telescope Science Institute,Baltimore MD 21218}
\affil{$^3$ Herzberg Institute for Astrophysics, National Research
Council, Victoria BC Canada}
\affil{$^4$Centre for Astrophysics \& Supercomputing,
Swinburne University of Technology,
Hawthorn VIC 3122, Australia}
\affil{$^5$Division of Astronomy,
University of California, Los Angeles CA 90095-1562}
\affil{$^6$American Museum of Natural History, New York NY, 10024-5192}
\begin{abstract}
  Multi-epoch observations with ACS on HST provide a unique and
  comprehensive probe of stellar dynamics within NGC~6397. We are able
  to confront analytic models of the globular cluster with the
  observed stellar proper motions. The measured proper motions probe
  well along the main sequence from 0.8 to below 0.1~M$_\odot$ as well
  as white dwarfs younger than one gigayear. The observed field lies
  just beyond the half-light radius where standard models of globular
  cluster dynamics (e.g. based on a lowered Maxwellian phase-space
  distribution) make very robust predictions for the stellar proper
  motions as a function of mass.  The observed proper motions show no
  evidence for anisotropy in the velocity distribution; furthermore,
  the observations agree in detail with a straightforward model of the
  stellar distribution function.  We do not find any evidence that the
  young white dwarfs have received a natal kick in contradiction with
  earlier results.  Using the observed proper motions of the
  main-sequence stars, we obtain a kinematic estimate of the distance
  to NGC~6397 of $2.2^{+0.5}_{-0.7}$~kpc and a mass of the cluster of
  $1.1 \pm 0.1 \times 10^5 \mathrm{M}_\odot$ at the photometric
  distance of 2.53~kpc.  One of the main-sequence stars appears to
  travel on a trajectory that will escape the cluster, yielding an
  estimate of the evaporation timescale, over which the number of
  stars in the cluster decreases by a factor of e, of about 3~Gyr.
  The proper motions of the youngest white dwarfs appear to resemble
  those of the most massive main-sequence stars, providing the first
  direct constraint on the relaxation time of the stars in a globular
  cluster of greater than or about 0.7~Gyr.
\end{abstract}
\maketitle

\section{Introduction}
\label{sec:introduction}

Globular clusters provide unique tests of star formation, stellar
evolution, Galactic structure and celestial mechanics.  As inhabitants
of the Galactic halo they probe the structure of the outer galaxy.  As
the oldest groups of stars in the Galaxy, they give a window into the
formation of the Galaxy and the early Universe in general --- they are
the foremost dig for Galactic archaeologists.  NGC~6397 is the second
closest globular cluster (after Messier 4) to the Earth, so many
studies have focused on observations and models of this cluster.
Globular clusters are bound groups of up to several million stars that
all form more or less at the same time from the same material.  A few
clusters show evidence for multiple generations of stars. In NGC~6397
\citet{2011A&A...527A.148L} and \cite{2011arXiv1110.1077M} find that
the stars exhibit a bimodal abundance distribution.  In about 75\% of
the stars sodium, nitrogen, aluminum and helium are enhanced while
carbon, oxygen and magnesium are depleted relative to the remainder
whose abundances are similar to field stars.  However,
\citet{2010A&A...511A..70D} find that the width of the main sequence
allows only a small variation in helium or CNO abundances.  The
hallmarks of multiple generations of stellar populations in NGC~6397
are much less dramatic than in other clusters such as NGC~2808
\citep{2005ApJ...631..868D}, NGC~1851 \citep{2008ApJ...673..241M}
among others.  The particular abundance differences found by
\citet{2011A&A...527A.148L} and \citet{2011arXiv1110.1077M} do not
significantly affect the stellar fluxes in the bands examined in this
paper (F814W and F606W), so we cannot search for dynamical
differences between the two populations.  In any case given the short
relaxation time of NGC~6397 any such differences should have been
washed out long ago.  The width of main-sequence track of NGC~6397 is
not significantly larger than the observational uncertainties,
supporting the approximate coevality of the stars and revealing a
small binary fraction in the cluster \citep{2008AJ....135.2155D}.  For
these reasons NGC~6397 is an excellent laboratory to understand
stellar evolution
\citep[e.g.][]{2008ApJ...676..594K,2008AJ....135.2141R} and derive
limits on its time of formation \citep[e.g.][]{2007ApJ...671..380H}.

The focus of this paper is the dynamics of NGC~6397.  NGC~6397 is the
nearest cluster with a central cusp in its surface brightness profile,
a theoretical signature of a core collapsed cluster
\citep{1995AJ....109..218T}; on the other hand,
\citet{2008MNRAS.389.1858H} argue that in spite of its prominent core
the possibly closer cluster Messier 4 has experienced core collapse
and is now sustained by binary burning.  Core collapse is a dynamical
process nearly unique to globular clusters in which the timescale to
approach thermodynamic equilibrium (the relaxation time) is much
smaller than the age of the cluster and also much smaller than the
evaporation time (so that the cluster still exists for us to observe).
In a core-collapsed cluster, the stars have achieved equipartition of
their kinetic energies; massive stars typically move more slowly than
less massive ones.  When these stars with different velocity
dispersions ($\sigma$) occupy the same potential, a signature known as mass
segregation (see \S~\ref{sec:mass-segregation}) develops where more
massive stars lie typically closer to the center of the
cluster. Furthermore, core collapse is predicted to proceed after
the bulk of binary burning in the central regions is complete, so 
as the central regions of the cluster continue to lose energy
to the outer regions, the energy transfer drives the system
further from equilibrium, causing the core to collapse in principle to
a singularity, but the formation and tightening of a few hard binaries
in the center is expected to halt the collapse \citep[e.g][]{Spit87}.

NGC~6397 has been the focus of many recent theoretical investigations
including $N-$body simulations such as \citet{2008AJ....135.2129H} and
\citet{2009MNRAS.397L..46H}, Monte Carlo approaches such as
\citet{2009MNRAS.395.1173G} and less recently the Fokker-Planck
treatment of \cite{1995ApJS..100..347D}.  The focus of this paper is
the dynamical observations of the cluster, so the modelling performed
is less detailed than these recent investigations and is more in the
spirit of that outlined in \citet[][MM91]{1991A&A...250..113M} for
NGC~6397 and most recently in \citet[][MAM06]{2006ApJS..166..249M} for
47 Tucanae (47 Tuc).  We did compare our data to the recent
\citet{2008AJ....135.2129H} $N-$body model but found that the number
of the stars in the outskirts of the model were insufficient for a
detailed comparison with the data.  The total number of stars at the
end of this model was about 38,000 of which about 10,000 are useful
for comparison with stars in our field --- this resulted in the
awkward situation where the statistical errors on the model are
comparable to that of the data and made detailed comparision
difficult.  We estimate that a factor of three more stars would be
sufficient.
Our strategy is to fit a particular mass component of a
multi-mass lowered isothermal distribution function
\citep{1966AJ.....71...64K} to the observational data.
\citet{1995ApJS..100..347D} and
\citet{1986AJ.....91..546P,1986AJ.....92.1358P} have pointed out the
limitations of fitting globular clusters with single component models.
We have tried to mitigate these issues to an extent. In particular the
background potential in our case is supplied by an ensemble of lowered
isothermal distribution functions of different stellar masses in
equilibrium such that the total number of stars in each mass bin is
proportional to the observed mass function
\citep{2008AJ....135.2141R}.  As we shall argue, because our field is
centered outside the half-light radius of the cluster, the effects of
mass segregation are more modest than if we had observed the entire
cluster or a field near the core, so a particular mass component (or
value of $\sigma$) is sufficient to characterize the data in our
field.  We use the best-fitting model to assist in deconvolving the
projected distributions and providing an estimate for the mass of the
cluster and the escape velocity for stars within our field.

\subsection{Observational Overview}
\label{sec:observ-overv}

Two sets of observations with the Advanced Camera for Surveys
\citep[ACS,][]{1998SPIE.3356..234F} on the Hubble Space Telescope
(HST) of the globular cluster NGC~6397 spaced over five years provide
sensitive probes of the dynamics of the globular cluster. The
astrometry is sufficiently sensitive to resolve the proper motion of
individual stars in NGC~6397, probing the theoretical model of the
cluster in detail as well as providing constraints on its mass,
distance and relaxation time. This study is closest in spirit to that
of \citet{2006ApJS..166..249M}, so it is quite valuable to introduce
our study by comparing and contrasting it with this work on stars in
the core of 47~Tuc.  Most importantly for this work, 47~Tuc has perhaps ten
times more stars than NGC~6397 and the MAM06 sample focuses on the
core of 47~Tuc, so MAM06 have about 13,000 stars in their proper
motion sample while we have about 3,000 stars.  The larger number of
stars in 47 Tuc and its larger physical size have the dynamical effect
of giving the larger cluster a long half-mass relaxation time of 4~Gyr
\citep{1997ApJ...474..223G} compared with 0.3~Gyr for NGC~6397
(\S~\ref{sec:prop-moti-disp}). The dynamical evolution of 47~Tuc as a
whole has not proceeded to the same extent as that of NGC~6397, so we
expect to see different processes at work.  Second, our sample lies
beyond the half-light radius of the cluster, so our study is more
sensitive to the global properties of the cluster and allows us to
make an estimate of its mass without large extrapolation.

As we have argued, the work of MAM06 provides a natural benchmark and
also a treasure map as well.  Many of the figures, tables and
arguments presented here shall be reminiscent of MAM06; however,
NGC~6397 and the pure ACS astrometry present many new challenges and
new opportunities in the study of globular clusters.

\subsection{Outline}
\label{sec:outline}

Table~\ref{tab:data} presents some of the fundamental properties of
NGC~6397 as well as auxiliary information that affects our
observations of this particular cluster.  We have attempted to give
the original references for the various values; however, as this is
not intended to be a thorough review, this list may not be exhaustive.
The table also contains results from this paper with the corresponding
section and acts as an index of sorts to delve directly into the results
of interest.  When a result relies on data presented here combined
with other work (typically MM91), the second paper is also listed.
\begin{table*}
\begin{center}
\caption{Basic data on NGC~6397}
\label{tab:data}
\begin{tabular}{lll}
\hline \hline
\multicolumn{1}{c}{Property} & 
\multicolumn{1}{c}{Value} & 
\multicolumn{1}{c}{Reference} \\
\hline
Cluster center (J2000) & 
$\alpha=17^\mathrm{h}40^\mathrm{m}42^\mathrm{s}\!\!.09, \delta=-53^\circ40'27''\!\!.6$
& \citealt{2010AJ....140.1830G}\\
Galactic coordinates & $l=338^\circ\!\!.1650, b=-11^\circ\!\!.9595$ &
\citealt{2011AJ....142...66G} \\
Apparent Magnitude & $V_\mathrm{tot} =   5.73$ &
\citealt{1995AJ....109..218T} \\
Luminosity & $L_V = 4.6 \times 10^4 d_{2.53}^2 \mathrm{L}_\odot$ &
\citealt{1995AJ....109..218T} \\
Integrated colors & $B-V=0.73, U-V=0.85$ & \citealt{1996AJ....112.1487H}  \\
Metallicity & $[\mathrm{Fe/H}] =-2.03 \pm 0.05$ & \protect{\citealt{2003A&A...408..529G}} \\
Foreground reddening & $E(B-V) = 0.18 \pm 0.01$ & \protect{\citealt{1998AJ....116.2929R,2003A&A...408..529G}} \\
Foreground absorption & $A_V = 0.56$ & \citealt{1989ApJ...345..245C} \\
                      & $A_\mathrm{F814W} = 0.33$ & \citealt{2005PASP..117.1049S} \\
Field contamination ($V \leq 21$) & $\Sigma_\mathrm{fore} \approx 23$
stars arcmin$^{-2}$ & \citealt{1985ApJS...59...63R}\\
\phantom{Field contamination }($V \leq 29$) & $\phantom{\Sigma_\mathrm{fore}} \approx 240$
stars arcmin$^{-2}$ & \\
Structural Parameters: \\
~~Total Mass$^*$ & $1.1 \pm 0.1  \times 10^5 d_{2.53}^3 \mathrm{M}_\odot$ & This
paper, \S~\ref{sec:mass-ngc-6397} \\
~~Core Radius & 3~arcseconds & \citealt{1995AJ....109..218T} \\
~~Half-Light Radius & $R_h=2.9$~arcminutes & \citealt{1995AJ....109..218T} \\
Multi-Mass King Model: & & \\
~~Central Escape Velocity & 2.31 mas/yr & This paper,
\S~\ref{sec:modelling-ngc-6397} \\
~~Central Escape Velocity & 2.8 $d_{2.53}^{-1}$ mas/yr & \citealt{1999ApJ...522..935G} \\
~~$\sigma-$parameter $(19.5<\mathrm{F814W}<24.5)$  & 1.01 mas/yr & This
paper, \S~\ref{sec:modelling-ngc-6397} \\
~~$\sigma-$parameter ($\mathrm{F814W}<16$) & 0.54 mas/yr &  This
paper, \S~\ref{sec:modelling-ngc-6397} \\
Heliocentric distance: & & \\
~~Subdwarf fit & $2.53 \pm 0.05$~kpc &
\protect{\citealt{2003A&A...408..529G}} \\
& $2.67 \pm 0.25$~kpc & \protect{\citealt{1998AJ....116.2929R}} \\
~~White-dwarf fit             & $2.55 \pm 0.11$~kpc & \protect{\citealt{2007ApJ...671..380H}} \\
~~Kinematic & $2.0\pm 0.2$~kpc & \protect{\citealt{1991A&A...250..113M}};
this paper, \S~\ref{sec:kinem-dist-ngc} \\
Timescales at 5': & & \\
~~Crossing time from proper motions & $\tau_c \approx r/\sigma \approx
0.6$~Myr & This paper,
\S~\ref{sec:prop-moti-disp} \\
~~Evaporation time & $\tau_e = \left(d\ln N/dt\right)^{-1} \approx 3$~Gyr & This paper,
\S~\ref{sec:stellar-escapers} \\
~~Relaxation time from white-dwarfs & $\tau_r \gtrsim 0.7$~Gyr & This
paper, \S~\ref{sec:prop-moti-distr} \\
~~Dynamical relaxation time (in our field) & $\tau_r \approx 1$~Gyr & This paper, \S~\ref{sec:prop-moti-disp} \\
~~Dynamical relaxation time (at $R_h$) & $\tau_{rh} \approx 0.3$~Gyr & This paper, \S~\ref{sec:modelling-ngc-6397} \\

\hline
\end{tabular}
\end{center}
All of the error intervals are ninety-percent confidence regions.
$^*$The value $d_{2.53}$ is the distance of NGC~6397 from Earth divided by 2.53~kpc.
\end{table*}

The following section (\S~\ref{sec:observations}) outlines the
observations.  Next we describe the sample selection
(\S~\ref{sec:star-selection}), the field geometry
(\S~\ref{sec:acs-field}) and the analysis techniques.  In particular
\S~\ref{sec:modelling} outlines the theoretical model and the
statistical estimators to probe it (an appendix applies this model to
proper motions), and \S~\ref{sec:error-estimation} describes the
estimation of errors.  A presentation of the results follows in
\S~\ref{sec:results}: 
\S~\ref{sec:modelling-ngc-6397} presents a theoretical model for the
current state of the cluster, 
\S~\ref{sec:mass-segregation} outlines the
evidence for mass segregation in the cluster,
\S~\ref{sec:prop-moti-isotr} looks for anisotropy in the velocity
distribution, \S~\ref{sec:prop-moti-disp} describes the dispersion in
proper motion as a function of stellar type and position in the
cluster, and \S~\ref{sec:prop-moti-distr} examines the proper motion
distribution in further detail for various subsamples.  In each of
these sections the results of the model are presented with the
observations.   We also obtain an estimate of
the distance of the cluster (\S~\ref{sec:kinem-dist-ngc}), its mass
(\S~\ref{sec:mass-ngc-6397}) and the candidate stellar escapers
(\S~\ref{sec:stellar-escapers}); all of these areas build directly
upon the model of the cluster.  Next, \S~\ref{sec:discussion}
discusses the consequences of these results for our understanding of
the dynamics (\S~\ref{sec:dynamics}), the global properties of
NGC~6397 (\S~\ref{sec:properties}) and future directions
(\S~\ref{sec:future-directions}).

\section{Observations}
\label{sec:observations}

The first epoch photometry used in this paper is the same as that in
the various papers on this cluster we developed from our HST Cycle 13
project
\citep{2006Sci...313..936R,2008AJ....135.2114A,2008AJ....135.2141R}.
There are two ACS epochs that went into determining the stellar proper
motions.  Nine new orbits were added in 2010 to re-image the cluster
in F814W only, in order to obtain the best proper motions possible.
Because all the astrometry was performed on F814W images and used the
distortion corrections for the F814W filter
(\citep{2006acs..rept....1A}), there should not be any color-dependent
effects on the astrometry.  These observations are summarized in
Table~\ref{tab:obs}.  We constructed a sample of likely stars using
the techniques outlined in \cite{2008AJ....135.2114A} to remove
artifacts and galaxies from the sample.

\subsection{Second-Epoch Astrometry}
\label{sec:second-epoch-astr}

We reduced the second epoch images using the publicly available
routine {\tt img2xym\_WFC.09x10}, described in
\citet{2006acs..rept....1A}.  This is a one-pass photometry program
that goes through each exposure pixel-by-pixel and identifies local maxima
that are sufficiently bright and isolated.  The routine then measures
positions and fluxes for these putative stars by means of a spatially
variable library PSF.  We identified every single local maximum as a
potential star and measured it with the PSF.  This gave us a list of
1.5M ``sources'' for each exposure.  The vast majority of these are
noise flucutations, but some are real stars.

In order to determine proper motions, we must compare the positions of
the stars in the first epoch with those in the second epoch.  To do
this, we must transform the positions of the stars in each exposure
into the reference frame.  We thus took a list of the main-sequence
stars (i.e.,\ those that lie along the main-sequence ridgeline, MSRL) and
identified the same stars in each of the second epoch images.  We took
these pairs of positions to construct a general six-parameter linear
transformation from the second-epoch exposure into the reference frame.
We examined the residuals of this transformation and noted that the
distortion solution had changed somewhat over the time from the first
epoch (in 2005 pre-SM4) to the second epoch (in 2010, post-SM4).  This
visible residual distortion could be characterized by clear linear
trends (with an amplitude of $\sim$0.03 pixel) within each chip, so we
decided that instead of transforming the entire detecor with a global
transformation, we would transform each chip indepndently with a
global transformation.  The residuals of this approach showed no
significant remaining trends.  We validate this with local
transformations below (\S~\ref{sec:global-vs-local}).

To compute the displacement for each star between the first and second
epochs, we used the above transformations to map each ``peak''
detection in each image found above into the reference frame.  We did
this separately for the 252 F814W deep first-epoch exposures and for
the 18 deep second-epoch exposures.  To identify the most likely
position of each of the 46,785 sources in each epoch, we collected all
the peaks found within five pixels of the target location and identified
the place where we found the largest number of detections within a
radius of 0.75 pixel.  We then performed an iterative sigma clipping
(with a $\sigma$-threshold of 3.5) to determine an average position
and a error in the average, based on the RMS about the average and the
number of peaks remaining.  We did this for the first and second
epochs.

Although we started out with ``good'' positions for the first-epoch
stars, the method we used to find the positions for these
displacements was different enough from the original finding algorithm
that we needed to be robust against the impact of artifacts and
neighbors.  For instance, a faint star within five pixels of a brigher
star (or a PSF artifact) might have the position of the brighter star 
identified as its position, rather than its own position.  To avoid this,
we used the detection-concentration method to re-find the first-epoch
positions.  If the new first-epoch position is not consistent with the
original first-epoch position, then it is likely that the second-epoch
position will also be bogus, so we flag these stars as not having
valid proper motions.  If, on the other hand, first-epoch data shows
that the most significant concentraion of detections within five pixels
is in fact the source itself, then we can be confident that the
second-epoch images will not mis-identify it as a neighbor or
artifact, and we can trust the proper motion.

To determine a displacement, we simply subtract the first-epoch
position obtained above from the second-epoch position.  The random
error in the displacement is simply the sum in quadrature of the first- and
second-epoch errors (dominated of course by the second epoch).  The
displacement errors for the bright stars are typically 0.002 pixel and
those for the faintest detectable stars are 0.20 pixel.  We convert
this into a proper motion by dividing by the time baseline of 4.936
years.  We neglect the statistical error related to the global linear
transformations, since they were based on well over 1,000 stars in
each chip.  We show below (\S~\ref{sec:global-vs-local}) that if we
apply a local correction to the transformations, we get residuals that
are at about the level of the quoted measurement errors.  Much of this
is due to the small number of stars used to do the local
transformations and the internal dispersion.  However, it is clear that
transformation error is no larger than the quoted random measurement
error.

To measure the proper motions for the bright stars, we used the 40s
short exposures from the two epochs.  The first epoch has two 40s
exposures and the second epoch has three.  We reduced the
pixel-based-CTE-corrected images as above for the deep exposures, then
as before we found the transformations into the reference frame using
the common stars along the MSRL and global transformations for each
chip.  We then computed displacements from first to second epoch (and
the corresponding errors) and report the proper motion in pixels per
year.  This adds 999 stars to the proper motion list.  About 85
sub-giant-branch, red-giant-branch and horizontal-branch stars are
saturated in the 40s exposures.  Because of CTE concerns in the
extremely low-background 5s and 10s exposures, we decided not to
attempt to compute proper motions for these stars.

We thus have an estimate of the proper motion for the faint stars from
the set of long exposures and an estimate of the proper motions for
most of the bright stars from the short exposures.  This gives us
motions for essentially all stars at or below the turnoff.
Fig.~\ref{fig:allpm} gives the proper motion relative to the mean
cluster motion of all the stellar objects measured in the field with
proper-motion errors less than 0.4 mas/yr. The arrow gives the
direction of the cluster center from the ACS field center.  It is
coincidental that this arrow happens to point toward the field stars
in the proper-motion diagram.

\subsection{Global vs. Local Transformations}
\label{sec:global-vs-local}

When we performed the single-chip global transformations above, we did
not notice any remaining trends of motion with position, indicating
significant residual distortion, but this is worth verifying.  To do
this, we followed the example of MAM06 and computed for each star a
local correction for its proper motion based the average motion of the
nearest 75 MSRL cluster stars.  The typical cluster star has a random
motion of 0.095 pixel per year with respect to the systemic cluster
motion, with an error of 0.0045 pixel per year (about 5\%).  We found
that the median systematic proper motion residual was 0.008 pixel per
year.  Since this was based on the average of 75 stars, we would
expect to see 0.007 pixel per year without any true systematic error
being present at all.  The fact that we observe slightly more than
this could be indicative of non-random errors of 0.0045 pixel per
year, similar to the random measurment error.  At any rate, our choice
of using global transformations to transform each chip into the
reference frame will not have a significant impact on our motions.

\subsection{Charge-Transfer Efficiency Correction}
\label{sec:charge-transf-effic}

Before any positions were measured for the second epoch, the images
were corrected for imperfect charge-transfer efficiency (CTE) using the
pixel-based correction described in \citet{2010PASP..122.1035A}.  
The CTE correction has been applied to the calibrated ({\tt \_flt})
images, as is the practice in the ACS pipeline.  The background in the
deep exposures is about 125 electrons per pixel.  This large
background shields faint sources from the majority of traps they would
experience.  The result of CTE losses would be to cause the stars to
be shifted towards the gap, the faint stars more than the brighter
stars.  The single-chip-based transformations we used compensated
somewhat for the average CTE, since it naturally allowed for an
arbitrary rescaling of the $y$ axis from first to second epoch.  So
all we are sensitive to is differential CTE shifts: a shift of the
faint stars relative to the bright stars.

To get a sense of the overall amplitude of the astrometric CTE effect, 
we reduced the CTE-corrected and the CTE-uncorrected images in an 
identical way and compared the output positions as a function of the 
star's flux.  We found that the brightest stars had a $+0.02$-pixel 
relative shift at the gap, and the faintest stars had a $-0.04$-pixel 
shift.  The CTE model is not perfect, but it can be counted on to remove 
roughly 75\% of the signal \citep{2010PASP..122.1035A}.  As a result, we are 
confident that the residual CTE-related astrometric error will be less 
than 0.005 pixel in total displacement, which corresponds to a proper 
motion of 0.001 pixel per year.  Such a small signature is not possible 
to see in the data, since the random motions of the stars are about 
10 times this.

\subsection{Incompleteness Corrections}
\label{sec:incompl-corr}

In this paper we make no use of incompleteness corrections. We are not
considering very faint stars in this work. From our first epoch
incompleteness corrections
\citep{2007ApJ...671..380H,2008AJ....135.2141R,2008AJ....135.2114A}
the faintest stars in our current sample are almost 85\% complete, and
we recovered all the first epoch stars in our subsequent epochs;
consequently, incompleteness should not be a serious issue.  In fact
in \S~\ref{sec:mass-segregation} (Fig.~\ref{fig:radmio}) we present
the completeness fraction as a function of magnitude to estimate how
differential incompleteness could bias the radial distributions and
find it to be much smaller than the statistical uncertainties.

\begin{table}
\begin{center}
\caption{ACS observations of NGC~6397}
\label{tab:obs}
\begin{tabular}{lcrcr}
\hline
\hline
\multicolumn{1}{c}{Data Set} &
Program ID & 
\multicolumn{1}{c}{$N_\mathrm{obs}$} &
Filter &
\multicolumn{1}{c}{Total Exposure} \\
\hline
Richer 2005 &                10424    &                      126     &
F606W        &  93.442 ksec \\
Richer 2005  &              10424    &                      252
&             F814W        & 179.704 ksec \\
Rich 2010    &               11633  &                          18
&              F814W     &      24.705 ksec \\
\hline
\end{tabular}
\end{center}
\end{table}




\begin{figure}
\includegraphics[width=\columnwidth,clip,trim=0 0.8in 0 0.8in]{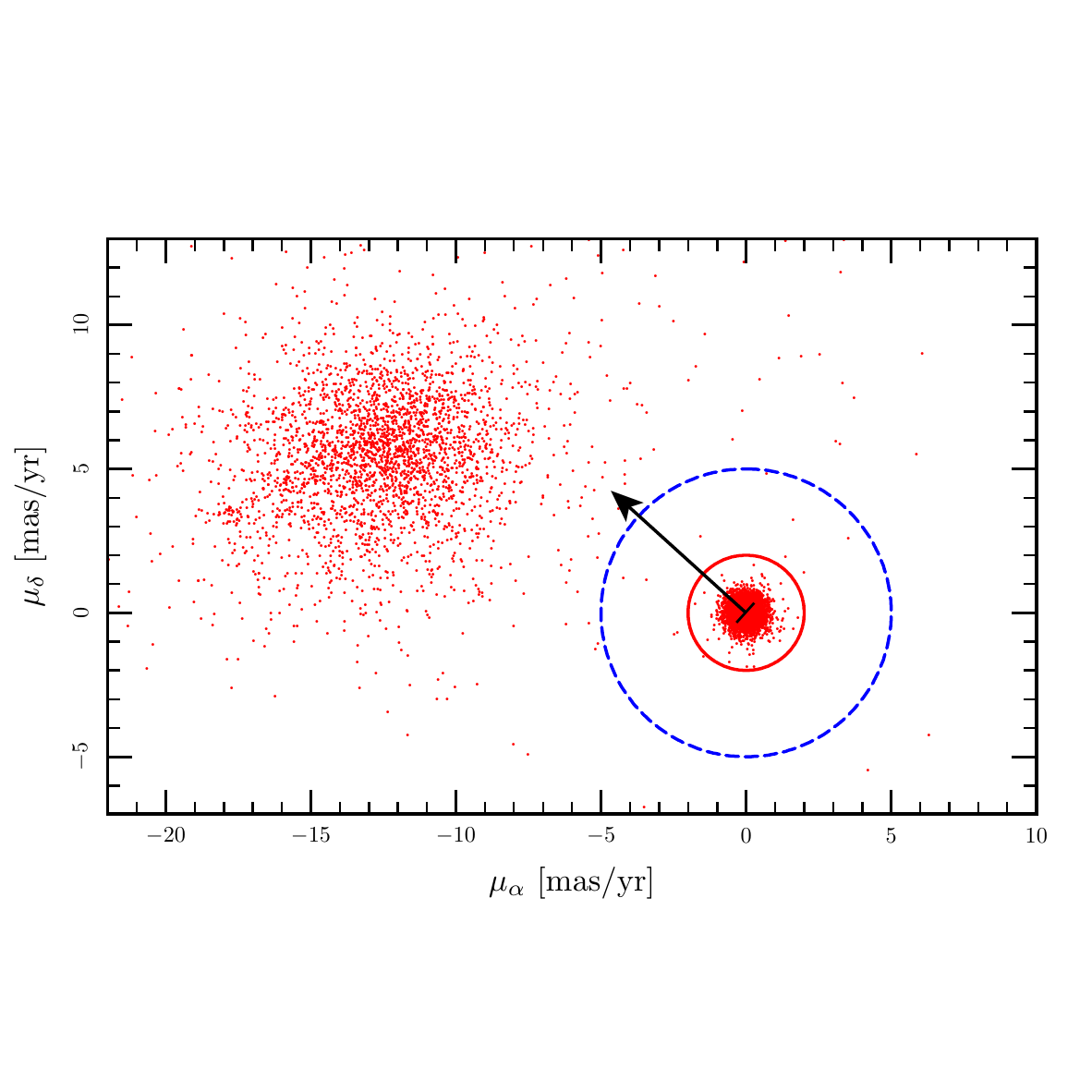}
\caption{Proper motions of the objects identified as stars in the ACS
  field.  Since we used cluster members to define the reference frame,
  the zero point of the vector-point diagram corresponds to the bulk
  motion of the cluster.  Those stars within the inner red circle
  comprise the sample to determine the color-magnitude diagram of the
  cluster.  The stars within the outer blue dashed circle comprise the
  sample used to probe the proper-motion distribution. The arrow
  indicates a proper motion in the direction of the cluster center
  from the center of the ACS field.  The swarm of stars centered
  around $\mu_\alpha\approx -12$ are field stars.  }
\label{fig:allpm}
\end{figure}

\begin{figure}
\includegraphics[width=\columnwidth]{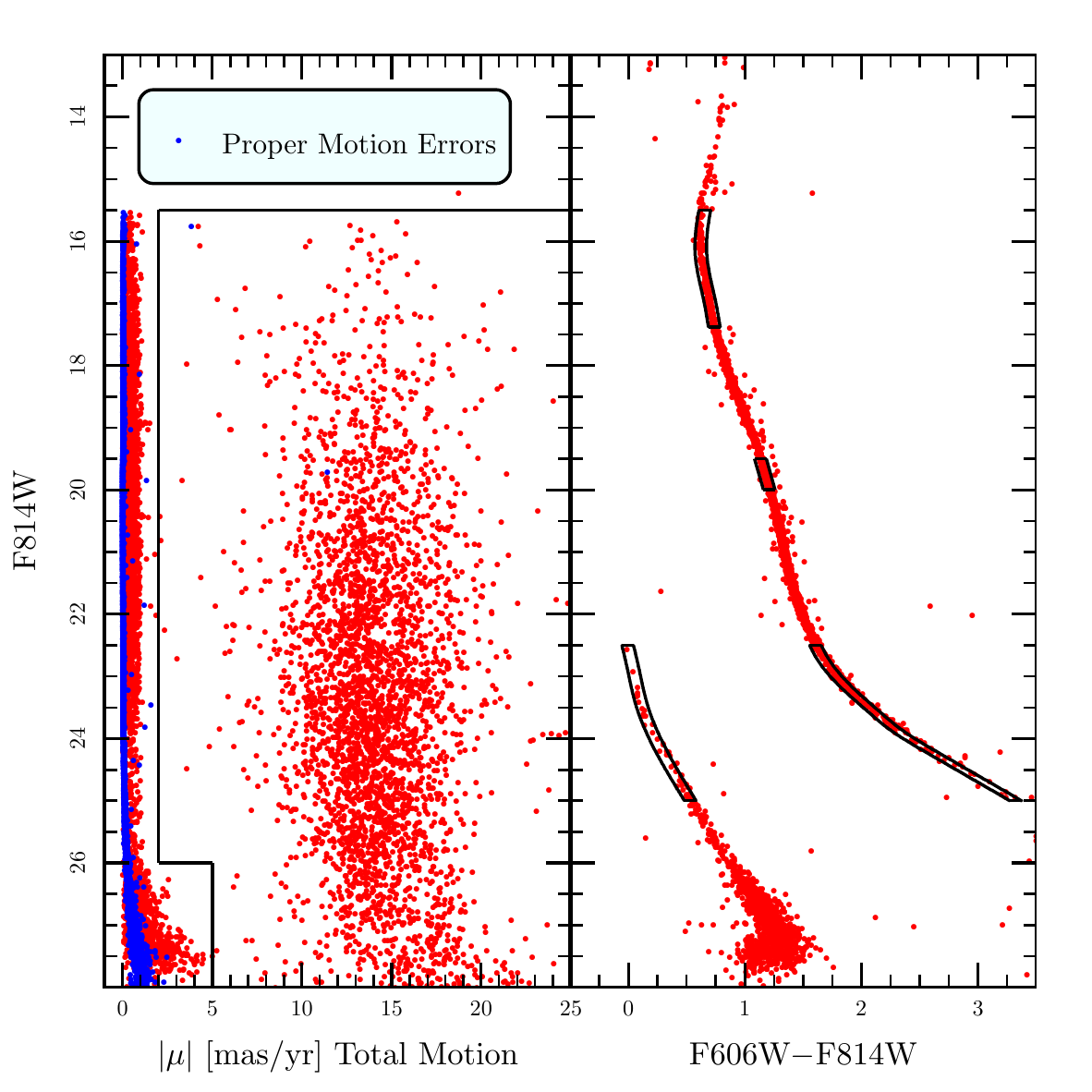}
\caption{Proper motions, proper-motion errors and colors as a
  function of magnitude.  The left panel give the observed total
  proper motions as a function of apparent magnitude in red, and the
  proper motion errors in blue. The right panel plots the apparent
  magnitude against the observed color.  Furthermore, the boxes depict
  the four subsamples outlined in \S\ref{sec:star-selection}.}
\label{fig:pmcmd}
\end{figure}

\section{Methods}
\label{sec:methods}

\subsection{Star Selection}
\label{sec:star-selection}
To construct a member-only color-magnitude diagram, we select only
those stars with proper motions (Fig.~\ref{fig:allpm}
and~\ref{fig:pmcmd}) relative to the mean motion of the cluster stars
of less than 2 mas/yr for $15.5 < \mathrm{F814W} < 26$ (red circle)
and proper-motion errors (blue points in Fig.~\ref{fig:pmcmd} and all
subsequent figures) of less 0.4 mas/yr. All stars brighter than
$\mathrm{F814W}=15.5$ are used to construct the color-magnitude
diagram, but because proper motions were not determined for these
stars, they do not end up in the proper-motion sample. For stars
fainter than $\mathrm{F814W}=26$, the proper-motion cutoff is 5 mas/yr
and there is no error cutoff. field stars outside of the cluster
comprise the broader distribution centered around $|\mu|\approx
15$~mas/yr on the left panel of Fig.~\ref{fig:pmcmd}.

Selecting the stars moving with the cluster yields the color-magnitude
diagram (right panel of Fig.~\ref{fig:pmcmd}) from which the
proper-motion sample is constructed. To prevent biasing the selected
proper-motion sample, the cutoff in proper motion for this sample is 5
mas/yr and a proper-motion error cutoff of 0.4 mas/yr for all
magnitudes. The sample is further winnowed by including only stars
with colors that are within 0.05 magnitudes of the main-sequence and
the white-dwarf cooling track.  Table~\ref{tab:rmserr} gives some
statistical properties of the objects that we identify as cluster and
field stars.  We focus on two large subsamples to probe the mass
segregation of the cluster in radius.  In particular, sample R1
consists of main-sequence stars with $17<\mathrm{F814W}<19.5$ that
have a typical mass of about $0.5\mathrm{M}_\odot$.  The second sample
R2 probes fainter MS stars ($21<\mathrm{F814W}<22.5$) with a typical
mass of $0.2\mathrm{M}_\odot$.  Here we want to maximize the size of
the samples to be sensitive even to subtle differences in the radial
distributions without allowing the two samples to overlap.  The
samples R1 and R2 represent a compromise.

We also need to select some samples to investigate the dynamics of the
cluster through the proper motions.  Our results concerning the proper
motions of the population as a whole focus on those stars with the
most precisely measured proper motions, the ``Best PMs'' sample, the
main-sequence stars with $19.5 < \mathrm{F814W} < 24.5$ --- all of
these PMs were measured using the long exposures, making a uniform
sample.  To probe the dynamical relaxation of the cluster we will
define four smaller samples.  Two of these are subsamples of
the ``Best PMs'' sample, and two come from other regions of the
color-magnitude diagram:
\begin{enumerate}
\item White dwarfs with $22.5 < \mathrm{F814W} < 25$ (``Bright White Dwarfs''),
\item Main-sequence stars with $22.5 < \mathrm{F814W} < 25$ (``Faint MS''), 
\item Main-sequence stars with $19.5 < \mathrm{F814W} < 20$ (``Middle MS''),
\item Main-sequence stars with $15.5 < \mathrm{F814W} < 17.38$ (``Bright MS'').  
\end{enumerate}
Sample 1 contains 46 white dwarfs with the best measured proper
motions.  Sample 2 contains 254 main-sequence stars with well-measured
proper motions over the same range of apparent magnitude as the white
dwarfs.  The size of sample 2 sets the goal size for the remaining
subsamples.  Sample 3 consists of 255 main-sequence stars, the
brightest stars with proper motions measured from the long exposures.
Furthermore, these main-sequence stars are expected to have a mass
similar to the white dwarfs.  Finally, sample 4 consists of the
brightest 255 main-sequence stars with proper motions (here measured
in the short exposures).  This final sample provides our best estimate
of the dynamical properties of the progenitors of the white dwarfs.
The three main-sequence samples have the nearly the same size,
allowing a fair comparision each sample with the white-dwarf sample.
All four subsamples are depicted in the
color-magnitude diagram.  Table~\ref{tab:rmserr} gives the astrometric
errors of the entire sample as well as the smaller subsamples.

\begin{table*}                                                                                 
\begin{center}                                                                                 
  \caption{Proper-motion dispersions, root-mean-squared errors, mean errors                          
    for the various subsamples}                                                                
 \label{tab:rmserr}                                                                            
  \begin{tabular}{lrccccccr}
\hline \hline                                                                                  
          &        & \multicolumn{2}{c}{Median}   & $\sigma_\mu$ & ${\hat \sigma}(\mu)$ & Mean                                  
    Error & RMS Error & $f$  \\                                                                
     Sample & Number & F814W & Mass & [mas yr$^{-1}$] & [mas yr$^{-1}$] & [mas                 
           yr$^{-1}$] & [mas yr$^{-1}$] & [\%]                                                 
    \\ \hline                                                                                  
  ~~All stars                                          &   6345~ &  21.6 & --- &  5.42 &  1.31 &  0.07 &  0.10  &  0.30  \\   
  ~~All MS stars                                       &   2880~ &  20.3 & 0.34 &  0.36 &  0.37 &  0.03 &  0.04  &  0.58  \\   
  ~~All WD stars                                       &    186~ &  26.1 & 0.53 &  0.50 &  0.47 &  0.23 &  0.25  & 15.67  \\   
  ~~All field stars                                 &   3279~ &  23.3 & --- &  4.40 &  2.79 &  0.09 &  0.12  &  0.10  \\   
Radal Samples: \\
  ~~R1 (MS: $17 < \mathrm{F814W} < 19.5$)              &    788~ &  18.3 & 0.58 &  0.37 &  0.37 &  0.05 &  0.05  &  1.09  \\   
  ~~R2 (MS: $21 < \mathrm{F814W} < 22.5$)              &    736~ &  21.6 & 0.18 &  0.38 &  0.37 &  0.03 &  0.03  &  0.28  \\   
Proper-Motion Samples: \\
  ~~Best PMs (MS: $19.5-24.5$)      &   1899~ &  21.0 & 0.24  &  0.37 &  0.37 &  0.03 &  0.03  &  0.30  \\   
  ~~Bright WD stars (1, $22.5-25$)                   &     46~ &  24.3 & 0.53 &  0.34 &  0.32 &  0.08 &  0.09  &  4.15  \\   
  ~~Faint MS stars (2, $22.5-25$)                    &    254~ &  23.1 & 0.11 &  0.37 &  0.38 &  0.05 &  0.05  &  0.83  \\   

  ~~Middle MS stars (3,  $19.5-20$)                  &    255~ &  19.8 & 0.42 &  0.34 &  0.35 &  0.02 &  0.02  &  0.23  \\   
  ~~Bright MS stars (4, $15.5-17.38$)                &    255~ &  16.7
  & 0.74  &  0.34 &  0.34 &  0.04 &  0.05  &  1.04  \\   
\hline                                                                                         
  \end{tabular}                                                                                
\end{center}                                                                                   
\smallskip

The quantity $f=1 - {\hat \sigma}_{n,\mathrm{true}}/{\hat \sigma}_{n,\mathrm{obs}}$ gives
the relative decrease in the value of the velocity dispersion after
correcting for the uncertainty in the observed proper motions.
The ``All Stars'' sample has no proper-motion cutoff; whereas the non-fieldx
other samples have a proper-motion cutoff of 5 mas/yr.  All samples
have an proper-motion-error cutoff of 0.4 mas/yr.   The masses of
main-sequence stars are given in solar masses using the models of
\citet{2008AJ....135.2129H} and assuming a distance of 2.53~kpc.  The
masses for the white dwarfs come from the spectroscopic measurements
of \citet{2008ApJ...676..594K}.
\end{table*}

\subsection{The ACS Field}
\label{sec:acs-field}

\begin{figure}
 \includegraphics[width=\columnwidth]{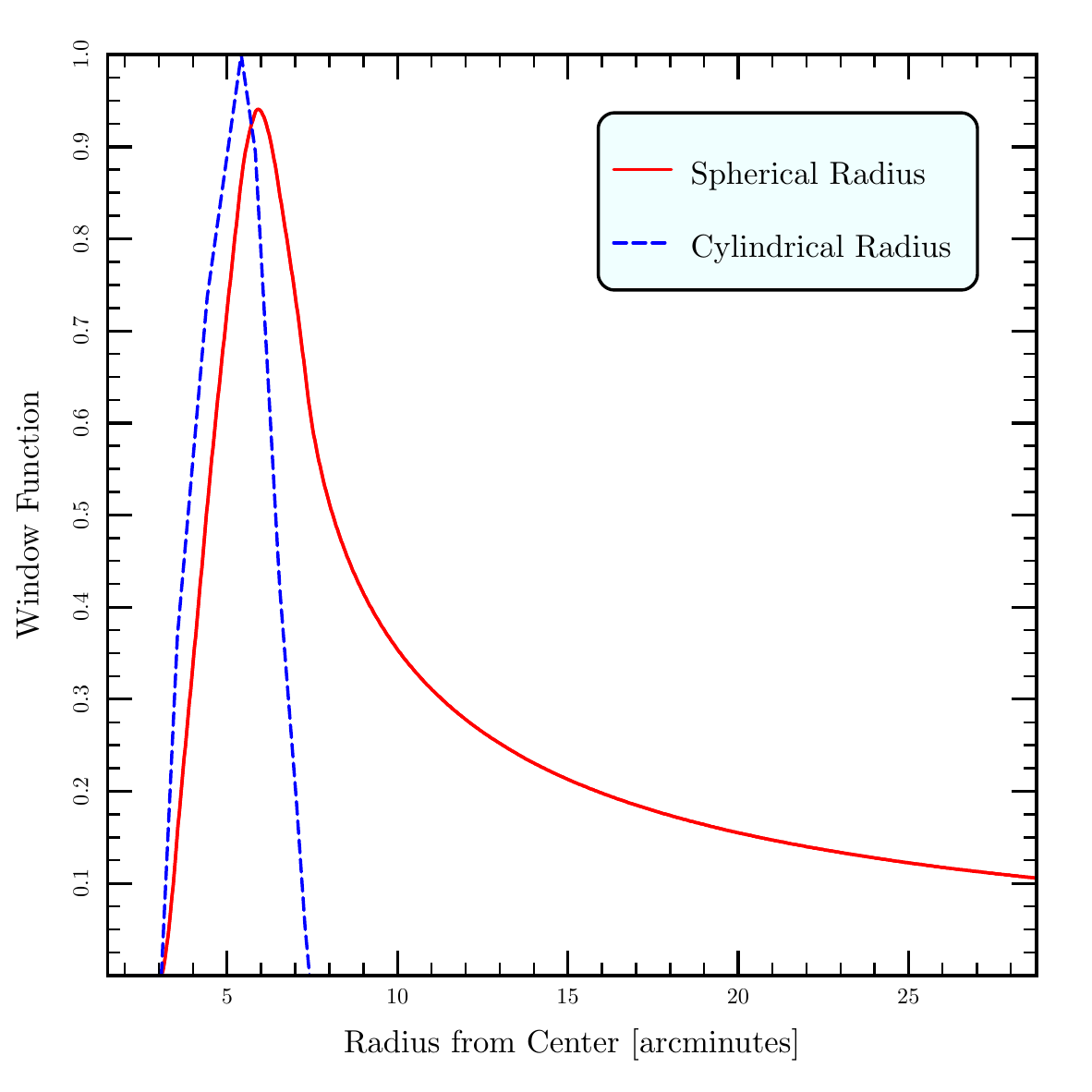}
\caption{The projected (blue dashed) and spherical radii
  (red solid) probed by the ACS field.}
\label{fig:windowfunk}
\end{figure}

To efficiently average over the ACS field (depicted in
Fig~\ref{fig:xy_escape} with the potential stellar escapers), we
define a window function that quantifies the regions of the cluster
probed by the field.  In the plane of the sky, the window function is
simply defined to be proportional to the ratio of the area of the ACS
field at a given projected radius from the center to the total area at
that radius.  The geometry of the field provides the initial guess of
the window function.  The estimate of the window function is optimized
using the assumption that those stars in the field that are not cluster
members are uniformly distributed.

For convenience the window function is normalized to unity at its peak.  The
three-dimensional window function is defined similarly: the ratio of
the total volume within the field as extended through the cluster at a
given three-dimensional radius from the center to the total volume at
that particular radius.  Again this function is normalized to unity at
its peak.  These window functions, as depicted in
Fig.~\ref{fig:windowfunk}, provide a useful technique to generate
either a Monte Carlo sample or integrate over the three-dimensional
structure of the cluster.  The window function is simply the
probability that a particular star in the entire Monte Carlo sample
ends up in the sample that represents the field.

\subsection{Theoretical Model}
\label{sec:modelling}
\label{sec:proper-motions}
To obtain estimates of the total mass of the cluster and the escape
velocities of individual stars, we will rely on a theoretical model of
the cluster.  We will use the model as one method of deprojecting our
velocity and surface density measurements to constrain the dynamical
properties of the cluster.  This section outlines the model and the
statistical measures that we will used to characterize both the model
and the data.  An appendix derives several useful results from the
model that further our interpretation of the data.

To model the stars in the globular cluster we will assume that the
phase-space density is given by a lowered Maxwellian distribution
\citep{1963MNRAS.126..499M,1966AJ.....71...64K},
\begin{equation}
f(E) = N \left[ \exp (-E/\sigma^2) - 1 \right ]
\label{eq:1}\end{equation}
where $N$ is the normalization, $E$ is the energy of a star per unit
mass (less than zero because the stars are bound) and $\sigma$ is a
parameter with dimensions of velocity that sets the velocity scale for
a particular group of stars in the cluster.  Let $\Psi(r)$ be the
negative of the potential energy, which vanishes at the edge of the
cluster (the tidal radius, $r_t$) so the number density of stars at a radius $r$ is
\begin{equation}
n(r) = \frac{\rho_1}{\left(2\pi\sigma^2\right)^{3/2}}
\int_0^{v_e} 4\pi v^2 \left [ e^{(v_e^2-v^2)/(2\sigma^2)} - 1 \right ] dv
\label{eq:2}\end{equation}
where $\rho_1$ is the number density normalization, $v$ is the speed of the star and $v_e^2 = 2\Psi(r)$ is the speed
needed to escape the cluster by reaching the tidal radius.  Therefore,
the number density is given by
\begin{equation}
n(r) = \rho_1 \left [ e^{v_e^2/(2\sigma^2)} \mathrm{erf} \left
    (\frac{v_e}{\sqrt{2}\sigma}\right ) -
  \sqrt{\frac{2}{\pi}} \frac{v_e}{\sigma} \left ( 1 + \frac{1}{3} \frac{v_e^2}{\sigma^2}
  \right ) \right ].
\label{eq:3}
\end{equation}
Similarly the local three-dimensional velocity dispersion is given by
\begin{equation}
\overline{v^2}(r)=\frac{J_2}{J_0}
\label{eq:4}\end{equation}
where
\begin{equation}
J_n = \int_0^{v_e} \left [ e^{(v_e^2-v^2)/(2\sigma^2)} - 1 \right ] v^{n+2} dv.
\label{eq:5}\end{equation}
To build a model star cluster we take a range of stellar masses each
with in its own value of $\sigma$ in Eq.~\ref{eq:3}; the stars of
different masses are assumed to be in thermal equilibrium, {\em i.e.}
$m_i \sigma_i^2$ is a constant.  This provides the total mass density
of stars as a function of the local potential.  We select a value for
the central potential and integrate the Poisson equation for the
gravitational potential using the total mass density of the ensemble
of stars as a source.  For each value of the central potential, we
vary the normalizations of the mass densities ($\rho_1)$ for each mass
to match the observed mass function of NGC~6397
\citep{2008AJ....135.2141R}.  This generates the value of the
potential as a function of radius, yielding the number density, column
density and the one-dimensional velocity dispersion as a function of
radius for any subpopulation of stars with a particular value of
$\sigma$.  We then choose the values of the central potential and
$\sigma$ that yield the best match to the photometry of
NGC~6397\citep{1995AJ....109..218T}.  \S\ref{sec:modelling-ngc-6397}
compares the observational results from this work and previous work
with the theoretical model of the cluster outlined here, and the
appendix derives several properties of this model that are useful to
interpret the observational results.

We would like to compare these model distributions against the data
and characterize these distributions using standard deviation of the
velocity distribution along a particular direction.  The standard
deviation has several advantages including the ease of calculation
from both the models and the data and a straightforward physical
interpretation, in the Jeans equation, for example.  However, the
standard deviation is extremely sensitive to outliers.  In fact a
single interloping star, say from the Galaxy, in the sample could ruin
our estimate of the standard deviation of the stars within the
cluster.  There are several robust estimators of the scale or width of
a distribution, such as the median absolute deviation or for
two-dimensional distributions like the proper motion, the median
magnitude of the proper motion.  These two estimators are maximally
insensitive to outliers; up to half of the points can be shifted to
arbitrarily large values without affecting the result.  For a normal
distribution these estimators yield values proportional to the
standard deviation; however, as the distributions deviate from
normality such as those present in the lowered isothermal sphere, these
constants of proportionality can change.

Instead of these estimators we use the first quartile of the
differences of the proper motions along a particular direction
to estimate the standard deviation of the distribution:
\citep[the estimator $Q_n$ defined by][]{Rous93} 
\begin{equation} 
  {\hat \sigma} = d_n \times \text{first quartile of}
  \left( \left| \mu_i - \mu_j \right| : i < j \right) .
\label{eq:41}
\end{equation}
where $\mu_i$ and $\mu_j$ are the measured proper motions of stars in
our sample or the model along a particular direction and $d_n$ is a
factor that depends on the size of the sample that ensures the ${\hat \sigma}$
is an unbiased estimator of the standard deviation for normally
distributed data.  As the sample size diverges ($n\rightarrow
\infty$), $d_n$ approaches $1/(2 \mathrm{erf}^{-1}(1/4)) \approx
2.219$.  Like the medians discussed earlier, the estimator ${\hat
  \sigma}$ has a breakdown point of 50\%, meaning that one
could shift up to one half of the values $\mu_i$ to infinity without
affecting the estimator.  

The estimator $\hat \sigma$ has several further advantages.  First, it
can be determined for each direction, unlike the median magnitude of
the proper motion.  Second, the value of $\hat \sigma$ approximates
well the standard deviation for distributions that differ
significantly from Gaussian such as a uniform distribution.  Finally,
the estimator $\hat \sigma$ is statistically efficient, meaning for a
sample of a given size the typical error is smaller than that which
results for the median absolute deviation or median proper-motion
magnitude.



We calculate ${\hat \sigma}$ from the proper-motion data, the radial
velocity sample and from the models.  Additionally we treat ${\hat
  \sigma}$ as a proxy for the standard deviation for much of our
analysis.  In particular, when we correct our observed proper motions
for the estimated errors in the proper motions, we will use
\begin{equation}
{\hat \sigma}^2_{\mu,\mathrm{true}} = {\hat \sigma}^2_{\mu,\mathrm{obs}} -
\frac{1}{n} \sum_{i=1}^n \epsilon_i^2
\label{eq:19}\end{equation}
where $\epsilon_i$ is the proper-motion error of each star.  
Table~\ref{tab:rmserr} shows that these corrections
are generally small.  Finally as we shall see in
\S~\ref{sec:mass-ngc-6397} the Jeans equation uses the velocity
variance to provide an estimate of the mass of the cluster; therefore,
we shall use both observed proper-motion variances and $\hat \sigma$
to estimate the mass of NGC~6397.  We shall find that the values of
$\hat \sigma$ and the standard deviation are approximately equal for
most of the subsamples that we investigate, so this choice makes little
difference to the final mass estimates.

\subsection{Error Estimation}
\label{sec:error-estimation}

The statistical bootstrap, the cousin of the Quenouille-Tukey
jackknife, was introduced by Efron in 1977 \citep{Efron79}.  In this
section we introduce or rather reintroduce the bootstrap and outline
how we use it to estimate the errors in our derived proper-motion
dispersions, column density distributions and cluster mass.  (Refer to
\citealt{1993stp..book.....L} for further details).  Observations of
NGC~6397 draw a sample of $n$ stars with positions and velocities
$\{(\vec x_i, \vec v_i)\}_0$ from the phase--space distribution
function, $f(\vec x,\vec v, t_0)$.  We use various statistics of the
observed stars to make conclusions about the distribution function and
its evolution: for example, the mean position along the $\alpha$-axis
of the sample:
\begin{equation} 
\label{eq:20}
\label{eq:zcmsum}
\bar x_\alpha = {1 \over n} \sum_i x_{\alpha,i}
\end{equation} 
and its associated error
\begin{equation} 
\label{eq:21}
\sigma^2_{\bar x_\alpha} = {1 \over n} {1 \over n-1} \sum_i (\bar x_\alpha - x_{\alpha,i})^2. 
\end{equation} 
By performing these summations, we assumed that the statistics of the
set $\{(\vec x_i, \vec v_i)\}$ are a good approximation to those of
the actual phase--space distribution function, $f(\vec x, \vec v,
t_0)$.  This is also the assumption behind statistical bootstrapping.

However, we could have taken a more complicated route and created
another distribution function (in fact the distribution that maximizes
the likelihood of the observations),
\begin{equation} 
f^* = {1 \over N} \sum_i \delta (\vec x - \vec x_i, \vec v - \vec
v_i), 
\label{eq:22}
\end{equation} 
and then calculated mean position by taking the integral over the
region probed by the ACS observations,
\begin{equation} 
\bar x_\alpha = \int_\mathrm{ACS~Field} x_\alpha f^* d\vec x d\vec v.
\label{eq:23}
\end{equation} 
Now we have two methods to estimate the error in our determination of 
the center of mass: calculating an integral analogous to the second
summation (Equation~\ref{eq:21}), or drawing a variety of samples from
our new distribution function $f^*$ and investigating the distribution
of $\bar x_\alpha^*$.  Our original sample, $\{(\vec x_i, \vec
v_i)\}$, is now only one of many possible realizations of $f^*$, and
the distribution of $\bar x_\alpha^*$ over the various resamplings
approximates the distribution of $\bar x_\alpha^*$ over $f^*$
\citep{1993stp..book.....L} which we in turn assume to approximate
$f(\vec x, \vec v, t_f)$.  Thus, through resampling we can estimate
the error in our determination of $\bar x_\alpha^*$ given by
Equation~\ref{eq:21}.
This Monte--Carlo resampling is less expedient than using
Equation~\ref{eq:21}, but often we do not have the luxury of a
statistic defined as simply as $\bar x_{\alpha}$ is in
Equation~\ref{eq:20}.  Even more rarely do we have the luxury of a
straightforward analytic error estimator like Equation~\ref{eq:21}.

So how is the bootstrap implemented with ACS observations?  Ordinarily
when one calculates a physical value from a sample, one uses all the
stars, counting each one once.  To use the bootstrap, one simply needs
to select from an $n$--star sample, $n$ stars {\em with
  replacement}. To create a bootstrapped sample, we uniquely label
each of the stars in an $n$-star sample from 1 to $n$.  We then select
$n$ integers from 1 through $n$ and use the stars with the appropriate
labels to create a new ensemble.  Some of the stars in the original
sample are included in the new ensemble several times and others not
at all.  We often use the value of ${\hat \sigma}$ for a sample of
proper motions to characterize them. We resample each group of stars
1,000 times, determine the value of ${\hat \sigma}$ of each
resampling, sort the results and quote the fifth and ninety-fifth
percentiles as a ninety percent confidence region.  In
\S~\ref{sec:mass-ngc-6397} we estimate the mass of NGC~6397.  In this
case we calculate a mass estimate for each resampling, sort these
results and calculate the ninety-percent confidence region for the
final result.

\section{Results}
\label{sec:results}

This section begins with a theoretical model for the cluster
(\S~\ref{sec:modelling-ngc-6397}) to provide a context for various
characteristics of the data including the radial distribution
(\S~\ref{sec:mass-segregation}), proper motions
(\S\S~\ref{sec:prop-moti-isotr}-\ref{sec:prop-moti-distr}) and some
derived properties of these data including its distance
(\S~\ref{sec:kinem-dist-ngc}) and mass (\S~\ref{sec:mass-ngc-6397}).

\subsection{A Model for NGC~6397}
\label{sec:modelling-ngc-6397}

\begin{figure}
 \includegraphics[width=\columnwidth]{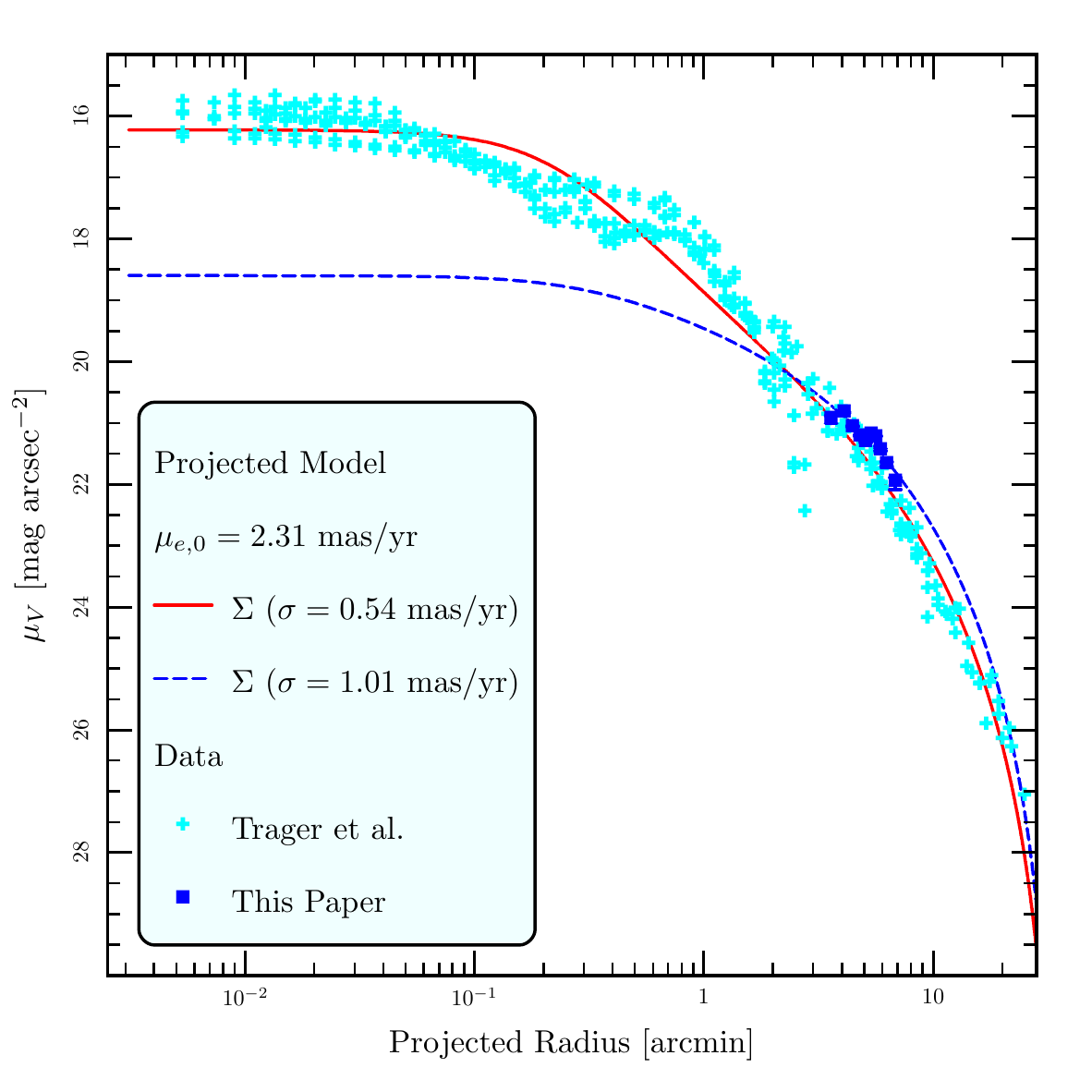}
 \caption{Surface photometry as measured by
   \citet{1995AJ....109..218T} (cyan crosses) and surface star counts
   from this work (blue squares) compared the the model surface density
   for the best fitting stars in the model (red solid curve) and the most
   massive stars in the model (blue dashed curve).}
\label{fig:model_trager}
\end{figure}

The mass function of \citet{2008AJ....135.2141R} and the observed
concentration of the cluster from \citet{1995AJ....109..218T} provide
a useful starting point to construct a preliminary multi-mass King
model for the cluster from which we can draw some global inferences.
Fig.~\ref{fig:model_trager} compares the model surface density for the
best fitting values of the central escape velocity and dispersion
parameter ($\sigma$) with the observed photometry from
\citet{1995AJ....109..218T}.  We have also included the star counts
within our field; \S~\ref{sec:prop-moti-disp} and
Fig.~\ref{fig:qn_npm_r} presents these counts in further detail.  To
convert the star counts to an estimate of the surface brightness we
convert the ACS magnitudes used here to the Johnson system using the
synthetic transformations of \citet{2005PASP..117.1049S}.  This yields
the total flux from all of the stars in the CMD sample defined in
Fig.~\ref{fig:pmcmd} of $V_{\mathrm{total}} = 9.75$ within the ACS
field of 10.9 square arcminutes.  This gives a mean surface brightness
of 21.23~$V$ magnitudes per square arcsecond or equivalently a mean
$V-$magnitude of 17.94 for each main-sequence star in the
proper-motion sample used to calculate the results depicted in
Fig.~\ref{fig:qn_npm_r}.  The star counts as a function of projected
radius are scaled by this mean magnitude per star, yielding the blue
squares in Fig.~\ref{fig:model_trager}.

\begin{figure}
 \includegraphics[width=\columnwidth]{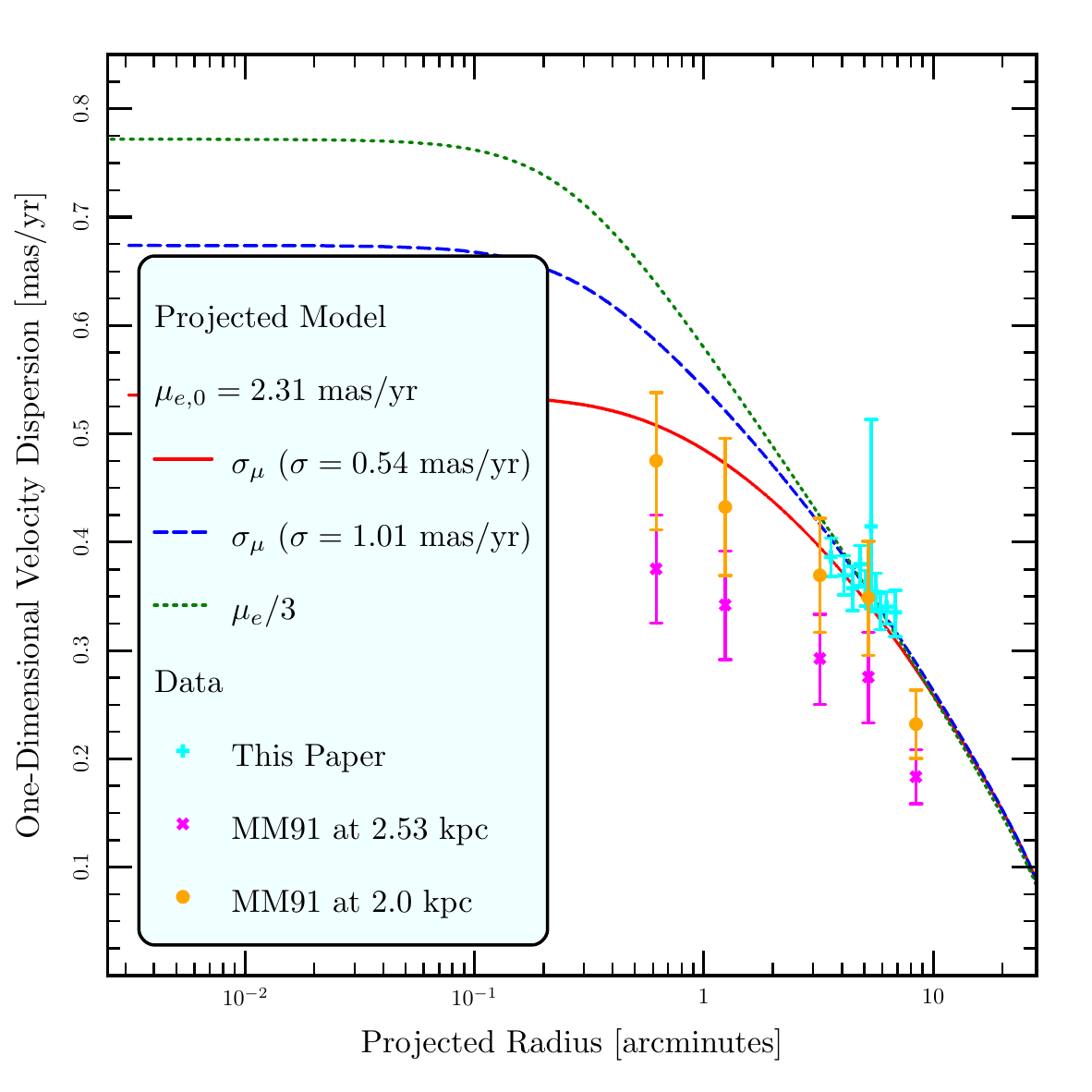}
 \caption{Projected velocity dispersion as a function of projected
   radius: model (best-fitting stars in red and most massive stars in
   green), radial-velocity and proper-motion measurements.  The
   velocity measured of \citet{1991A&A...250..113M} (MM91) have been
   converted to proper motions assuming a distance of 2.55~kpc
   \citep{2008AJ....135.2141R} and 2~kpc to bring agreement between
   the proper-motion and radial-velocity measurements.  The error bars
   from this paper are ninety-percent confidence intervals obtained by
   bootstrapping.  The errorbars on the MM91 data are 1.64-$\sigma$.}
\label{fig:disp_radius}
\end{figure}

The radial, magnitude, velocity normalizations, central velocity
dispersion and $\sigma-$parameter of the models are arbitrary, and a
comparison with the data is used to set the scale of the model.  We
use the models of \citet{2008AJ....135.2129H} to set the ratio of the
$\sigma-$parameter of the turn-off stars, which are assumed to provide
the bulk of the light, and that of the proper-motion sample.  We take
the mass of the turn-off stars to be $0.84~\mathrm{M}_\odot$ and the
proper-motion sample to be $0.24~\mathrm{M}_\odot$.  Furthermore, we
assume that the values of $\sigma$ following the equiparition relation
with same value $m_i \sigma_i^2$ to the various populations.  The
values of $\sigma$ and $\mu_{e,0}$ are determined with an unweighted
least-squares fit between the models and the data.  The bulk of the
points in this fit are provided by the \citet{1995AJ....109..218T}
photometry; however, a measurement of the velocity dispersion is
required to set the velocity scale of the models.  Here, we also use
the data from \S~\ref{sec:prop-moti-disp} and Fig.~\ref{fig:qn_npm_r}.
Fig.~\ref{fig:disp_radius} depicts the observed proper motions, those
found in the model and the observed radial velocities of MM91 for
different assumed distances.

Both the gross properties of the cluster such as its mass and the
profile near the half-light radius depend little on the properties of
the central regions.  Operationally the radial scale and the ratio of
the central escape velocity to the $\sigma$-parameter are determined
mainly by the surface photometry (Fig.~\ref{fig:model_trager}) and the
velocity normalization is determined by the value of ${\hat \sigma}$
as a function of radius (Fig.~\ref{fig:disp_radius}).  The two crucial
parameters for the model are the central escape proper motion of
$\mu_{e,0} = 2.31$~mas/yr and the $\sigma$-parameter of 0.54~mas/yr
for the turn-off stars that dominate the light and 1.01~mas/yr for the
stars in the proper-motion sample.  For the brightest stars, the value
of the $\sigma$-parameter is approximately the central velocity
dispersion along one dimension because $\mu_{e,0} \gg \sigma$.  The
ratio of these two quantities is determined mainly by the surface
brightness profile while the normalization is determined exclusively
by the proper-motion data depicted in Fig.~\ref{fig:disp_radius}.
The observed velocity dispersion, either in the radial direction or in
a particular direction of proper motion, is integrated over the
line-of-sight so we can project the model onto the plane of the sky.
Figure~\ref{fig:disp_radius} depicts both of these results.  Beyond
one half-light radius the observed velocity dispersion depends only 
very weakly on the $\sigma$-parameter, so although mass segregation is
important in generating the radial distribution, the velocity
dispersions are nearly independent of stellar mass.  Furthermore, the
escape velocity at a given radial distance from the center of the
cluster is typically three times the velocity dispersion within
five-percent agreement with the analytic treatment in 
the appendix (Eq.~\ref{eq:13}).

Fig.~\ref{fig:model_trager} shows that the change in the slope of the
surface brightness profile as a function of the $\sigma$-parameter (or
mass) is largest around one arcminute from the center of the cluster.
The difference in slopes is more modest at the projected radius of the
ACS field; therefore, the mass segregation in the model is weaker
here.   The analytic treatment outlined in the appendix
although illustrative, is not strictly applicable here, because the
escape proper motion is approximately 1~mas/yr and the
$\sigma$-parameter is 1.01~mas/yr, so the value of $\sigma$
approximates that of $\mu_e$ and does not greatly exceed it.  The
field is not in the regime where the distribution function is strictly
linear, however, as we can see from Fig.~\ref{fig:disp_radius}.

The model velocity and distance normalizations also yield a model mass
for the cluster of
\begin{equation}
M = 1.1 \times 10^5 d_{2.53}^3 \mathrm{M}_\odot
\label{eq:29}
\end{equation}
where $d_{2.53}$ is the distance to NGC~6397 from Earth divided by 2.53~kpc.
We can combine the mass estimate in Eq.~\ref{eq:29} with the mean
stellar mass of about $m_* = 0.35~\mathrm{M}_\odot$
\citep{2007AJ....134..376D,2008AJ....135.2141R} to obtain an estimate
of the relaxation time at the half-mass radius
\citep{1971ApJ...164..399S}
\begin{equation}
\tau_{rh} =  0.138 \frac{M^{1/2} R_h^{3/2}}{m_* G^{1/2} \ln \left
    (0.4 M/m_*\right)}
\label{eq:38}
\end{equation}
to yield $\tau_{rh} = 0.45 d_{2.53}^3$~Gyr where we have suppressed the
additional logarithmic dependence on distance through the Coulomb
logarithm.

\subsection{Mass Segregation}
\label{sec:mass-segregation}

\begin{figure}
\includegraphics[width=\columnwidth]{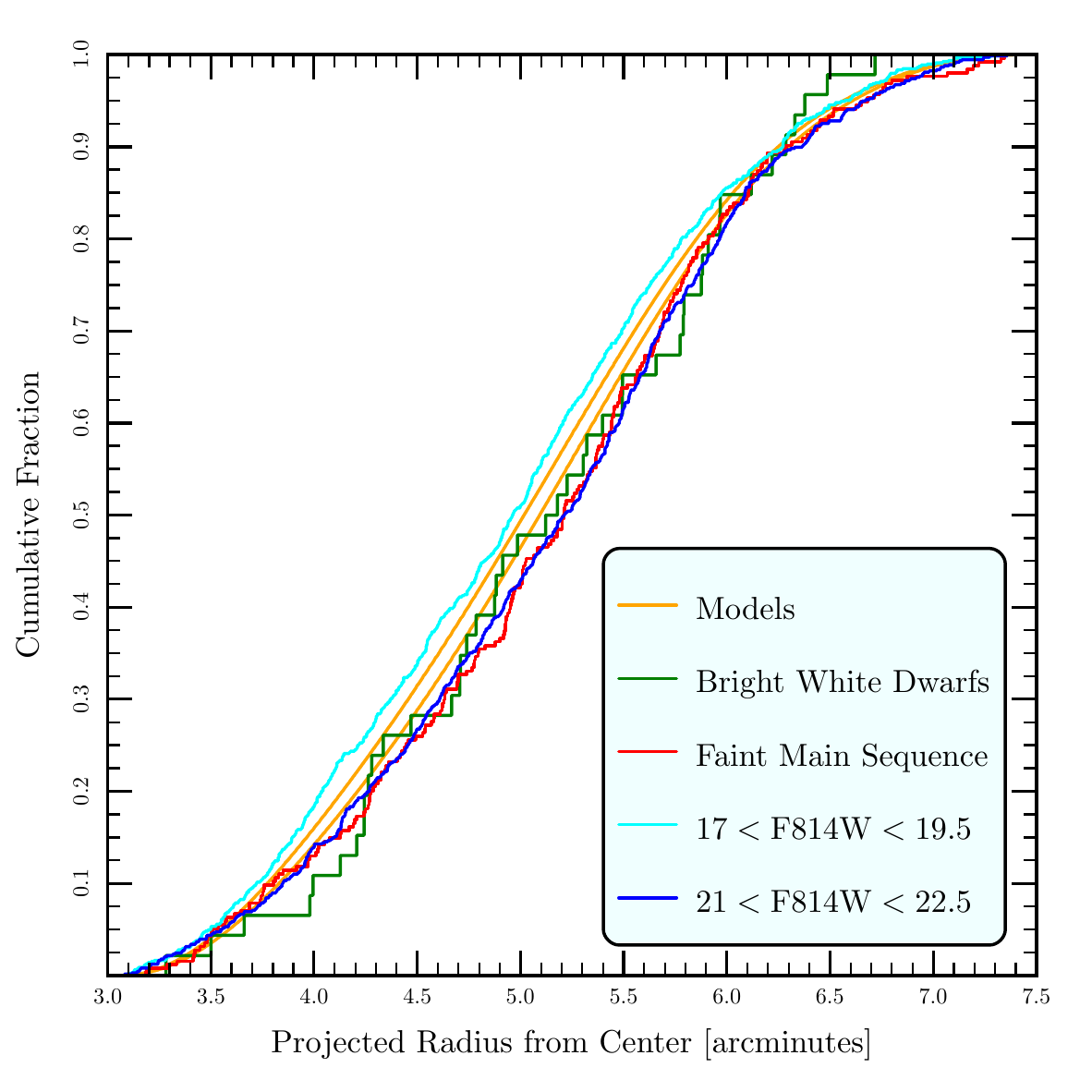}
\caption{Cumulative projected radial distribution for various
  subsamples. The three model curves give the best-fitting model in
  the middle (a mass of 0.35~M$_\odot$ or F814W $\sim 21$) and models
  for two other masses assuming that $m_i \sigma_i^2$ is constant.  The
  models of \citet{2007AJ....134..376D} give the correspondence
  between masses and magnitudes.  The left-hand model traces the
  distribution for a stellar mass of 0.58~M$_\odot$ (F814W $\sim 18$),
  and the right hand for a mass of 0.18~M$_\odot$ (F814W $\sim 22$).}
\label{fig:cum_rdist}
\end{figure}

Fig.~\ref{fig:cum_rdist} depicts the cumulative radial distribution of
several groups of stars in the sample, including two of the subsamples
described in \S~\ref{sec:star-selection} and Table~\ref{tab:rmserr}.
If we compare the main sequence stars with $17 < \mathrm{F814W} <
19.5$ and $21 < \mathrm{F814W} < 22.5$, we obtain a $p-$value of $7
\times 10^{-5}$ by the Wilcoxon rank-sum test and twice this value for
the Kolmogorov-Smirnov (KS) test, indicating that it is highly
unlikely that these two groups of stars follow the same radial
distribution.  The two main sequence samples show significant
segregation, even greater than the model distributions.  The model
distributions are typically more centrally concentrated than the stars
in the ACS field because they are based on the surface photometry as
outlined in \S~\ref{sec:modelling-ngc-6397}.  Because the field lies
outside the half-light radius, following the arguments of
\S~\ref{sec:modelling}, we do not expect mass segregation to be strong
in this region and the model exhibits only weak mass segregation.
Either the cluster has stronger mass segregation than the model, or
perhaps the samples are incomplete especially for fainter stars near
the center.  Because the incompleteness fraction is low (less than 7\%
for the stars in these two samples), we can discount the second
possibility. Although we will present the observed column density of
stars within the field and use this to estimate the radial density
gradient, we use this gradient as input for only two of several
independent mass determinations in \S~\ref{sec:mass-ngc-6397}, the
results of which agree with the mass determination from surface
brightness measurements.

Fig.~\ref{fig:cum_rdist} serves a second purpose.  Although the faint
MS and bright WD samples span the same magnitude range, the stars in
the WD sample are typically fainter than in the MS sample as shown in
Table~\ref{tab:rmserr}.  In \S~\ref{sec:prop-moti-distr} we shall argue
that these two samples have significantly different proper-motion
distributions; in principle, this could result from the two samples
having different radial distributions.  The typically fainter white
dwarfs might be more difficult to find toward the center of the
cluster, biasing this sample outward and biasing the proper-motion
distribution toward smaller values.  The cumulative distributions
depicted in Fig.~\ref{fig:cum_rdist} demonstrate that this is
unlikely.  A comparison of the radial distribution of the two samples 
yields a $p-$value of 89\% by the KS test and 92\% by the Wilcoxon 
rank-sum test.

We also compare the radial distribution of the 46 bright white dwarfs
with the two larger main sequence samples and find 
probabilities of 98\% and 48\% from the KS tests, and
92\% and 32\% Wilcoxon rank-sum probabilities.
Furthermore, when we compare the radial distribution of bright white
dwarfs ($22.5 < \mathrm{F814W} < 25$) with a sample of 97
fainter white dwarfs ($25 < \mathrm{F814W} < 26.5$ -- not shown in
Fig.~\ref{fig:cum_rdist}), we find 54\% KS and 51\%
Wilcoxon probabililties.  This fails to support the conclusions of
\citet{2008MNRAS.383L..20D} which used a smaller sample to conclude
that the distribution of young white dwarfs was radially extended
relative to that of older ones and argue that a possible explanation
for this phenomenon was a white-dwarf natal kick.

\begin{figure}
\includegraphics[width=\columnwidth]{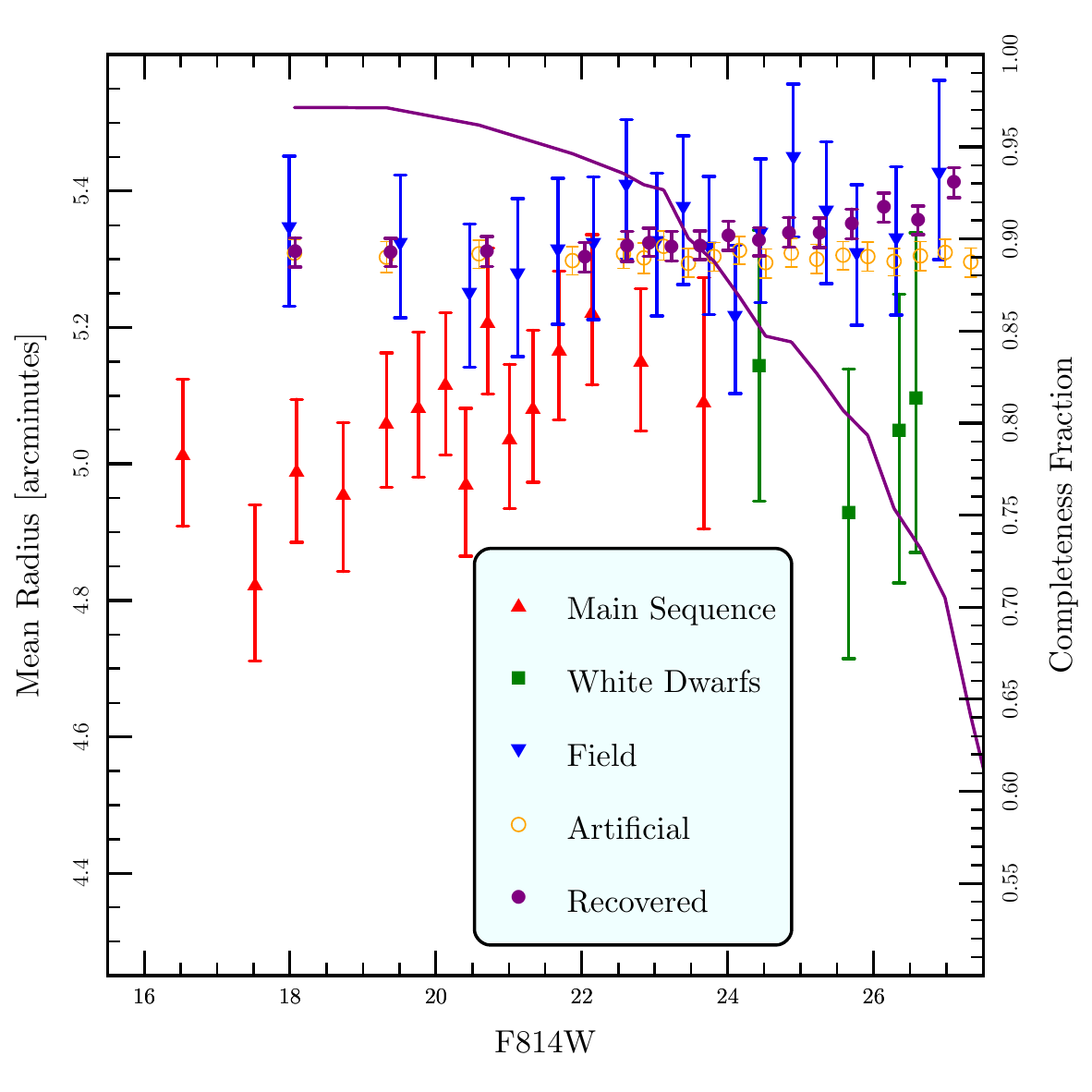}
\caption{Mean projected radius of stars as a function of F814W.  The
  open circles give the mean radius for the input artificial stars,
  and the closed circles give the mean radius of those recovered. The
  errorbars here and in subsequent figures represent ninety-percent
  confidence limits and are obtained by bootstrapping. The solid curve
  gives the completeness fraction as a function of apparent magnitude
  as calculated from the artificial star tests.  This curve is read
  using the right-hand axis.}
\label{fig:radmio}
\end{figure}

A possible explanation for the differing distributions as a function
of apparent magnitude is differential incompleteness.  It is more
difficult to find faint stars in more crowded areas of the sky near
the centre of the cluster.  We have addressed this question in several
ways in Fig.~\ref{fig:radmio}.  First we have looked at the stars in
the image and sorted them by apparent magnitude into bins of 50
white-dwarf stars or 200 main-sequence or field stars.  A priori we do
not expect the field stars (at any magnitude) to be centrally
concentrated with respect to the cluster, so a change in the mean
radius of the field stars would have to result from differential
incompleteness.  Because the number of field stars is limited to about
3,000, to get a better statistical handle on this effect, we inserted
nearly 200,000 artificial stars throughout the image of various
magnitudes and tried to recover them.  The recovery fraction depicted
by the solid curve in Fig.~\ref{fig:radmio} gives an estimate of the
completeness fraction.  The mean radius of the input stars as a
function of magnitude is given by the open circles.  The mean radius
of the recovered stars is given by the closed circles.  We have sorted
the artificial stars into bins of 5,000 objects. We see that fainter
than $\mathrm{F814W} \approx 23$ the recovered stars are on average
further from the cluster centre than the input stars, the hallmark of
differential incompleteness by about 0.05 arcminutes; however, the
effect is much more subtle than the change in the mean radius along
the main sequence and white-dwarf tracks.  It is unlikely to be strong
enough to account to the differences in the radial distirbutions
depicted in Fig.~\ref{fig:cum_rdist} where the median radius of the
models differs from the data by more than a tenth of an arcminute..

\subsection{Proper-Motion Isotropy}
\label{sec:prop-moti-isotr}

\begin{figure}
\includegraphics[width=\columnwidth,clip,trim=0 0.8in 0 0.8in]{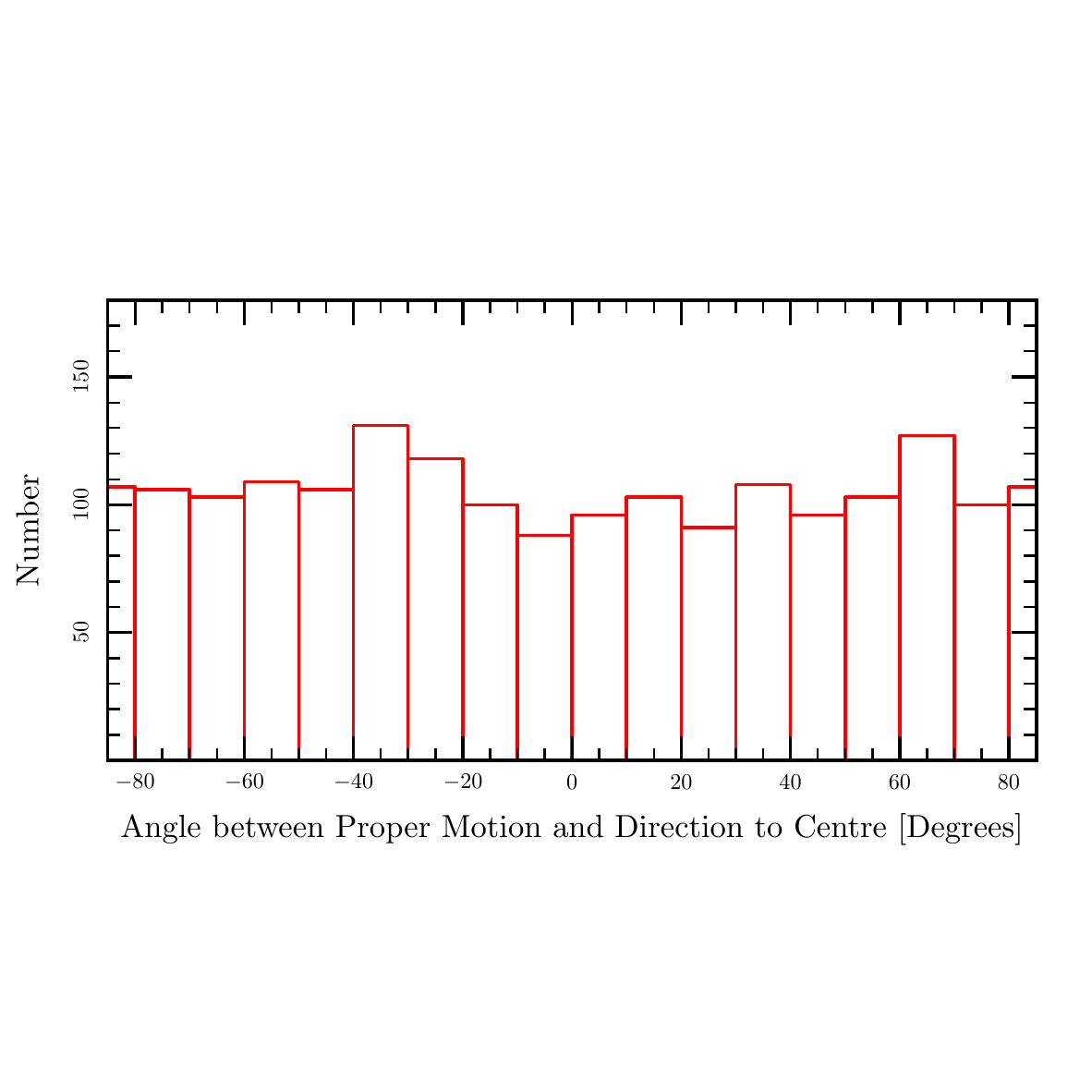}
\caption{Histogram of proper-motion angle relative to the center of
  NGC~6397.}
\label{fig:ang_hist}
\end{figure}

We measure the angle of the proper motion relative to the cluster
center of the main-sequence stars with $19.5 < \mathrm{F814W} < 24.5$,
proper-motion errors less than 0.4 mas/yr and total proper motions
less than 5 mas/yr. The distribution of this angle is depicted in
Fig.~\ref{fig:ang_hist}. There is no evidence for anisotropy from this
distribution --- a KS test indicates that the cumulative distribution
of 35\% of samples drawn from a uniform distribution will differ from
uniform at least as much as the distribution in
Fig.~\ref{fig:ang_hist} does.  The $p-$value for the Kuiper test
(\citeyear{Kuip60}, a generalization of the KS test for a distribution
on a circle) is smaller at 13\%.  If we look at the direction
of the proper motions relative to right ascension and declination, the
$p-$value for the KS test increases to 39\% and for the Kuiper test to
32\%.  The directions alone do not reveal any anisotropy.

\begin{figure}
  \includegraphics[width=\columnwidth]{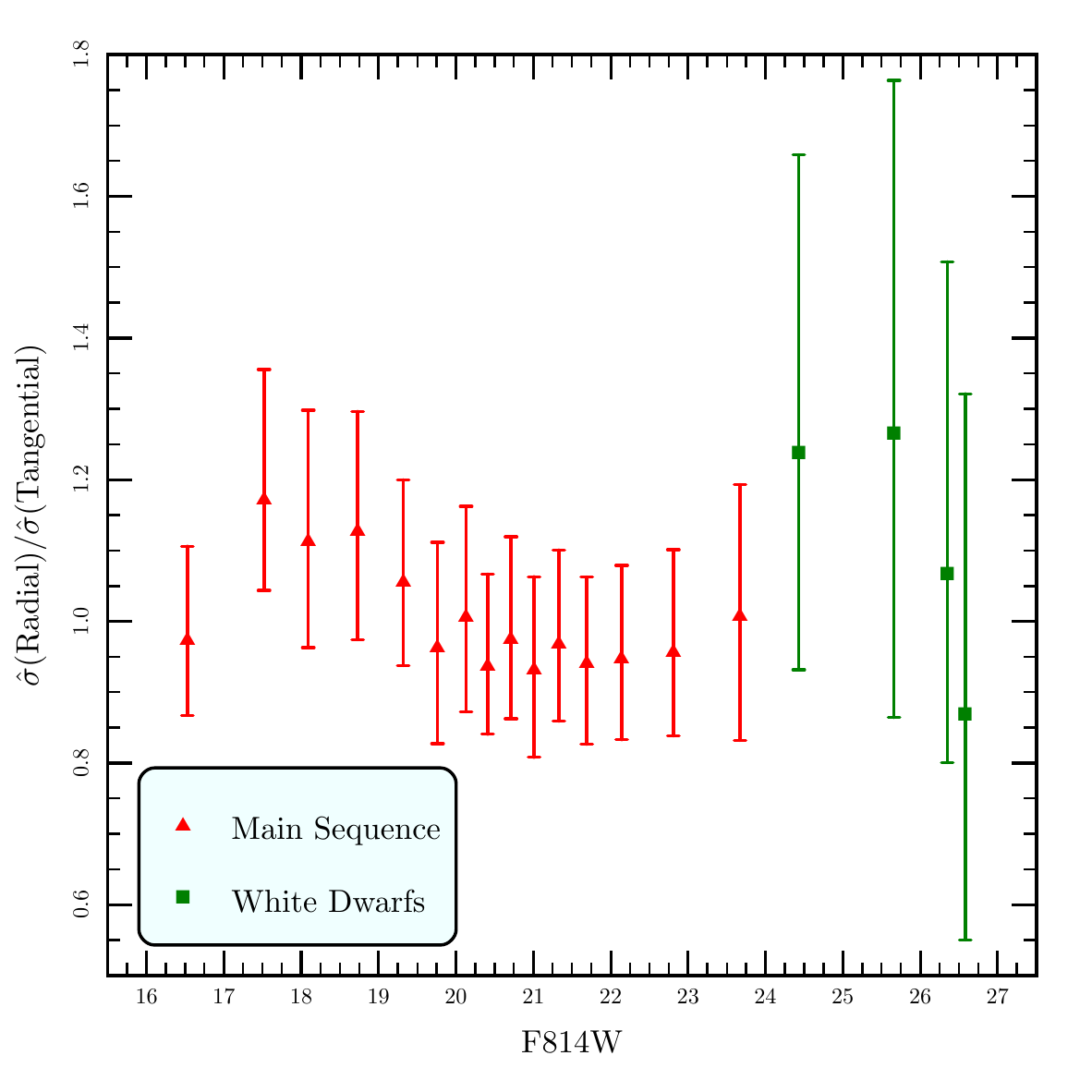}
  \caption{The radial anisotropy of the proper-motion ellipse as a function
    of apparent magnitude for main-sequence and white-dwarf stars.
    Only stars with total proper-motion error less than 0.4 mas/yr are
    included.}
  \label{fig:qnratmio}
\end{figure}

\begin{figure}
  \includegraphics[width=\columnwidth]{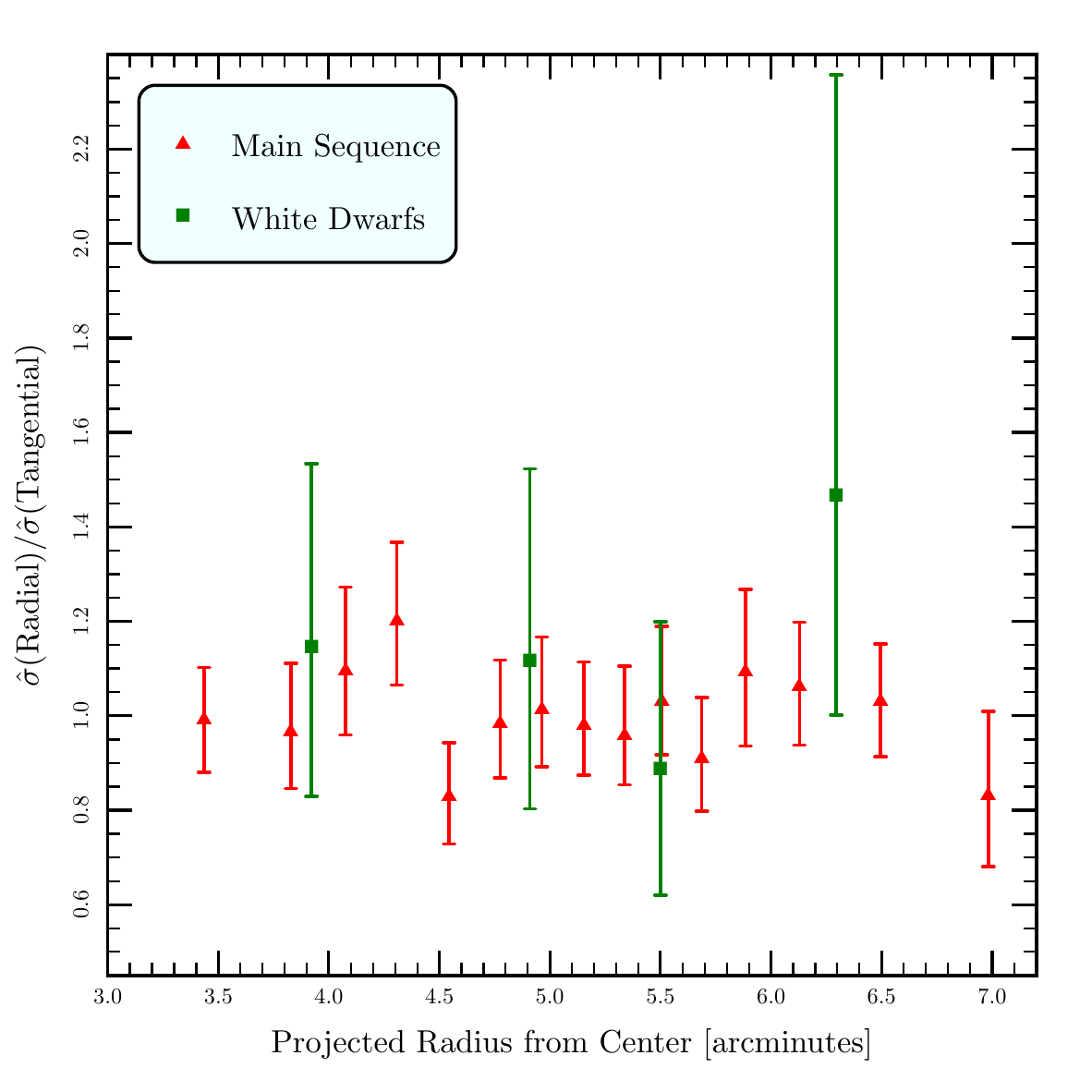}
  \caption{The radial anisotropy of the proper-motion ellipse as a function
    of projected radius from the center for main-sequence and white-dwarf stars.
    Only stars with total proper-motion error less than 0.4 mas/yr are
    included.  }
  \label{fig:qnratrad}
\end{figure}

\begin{figure}
  \includegraphics[width=\columnwidth]{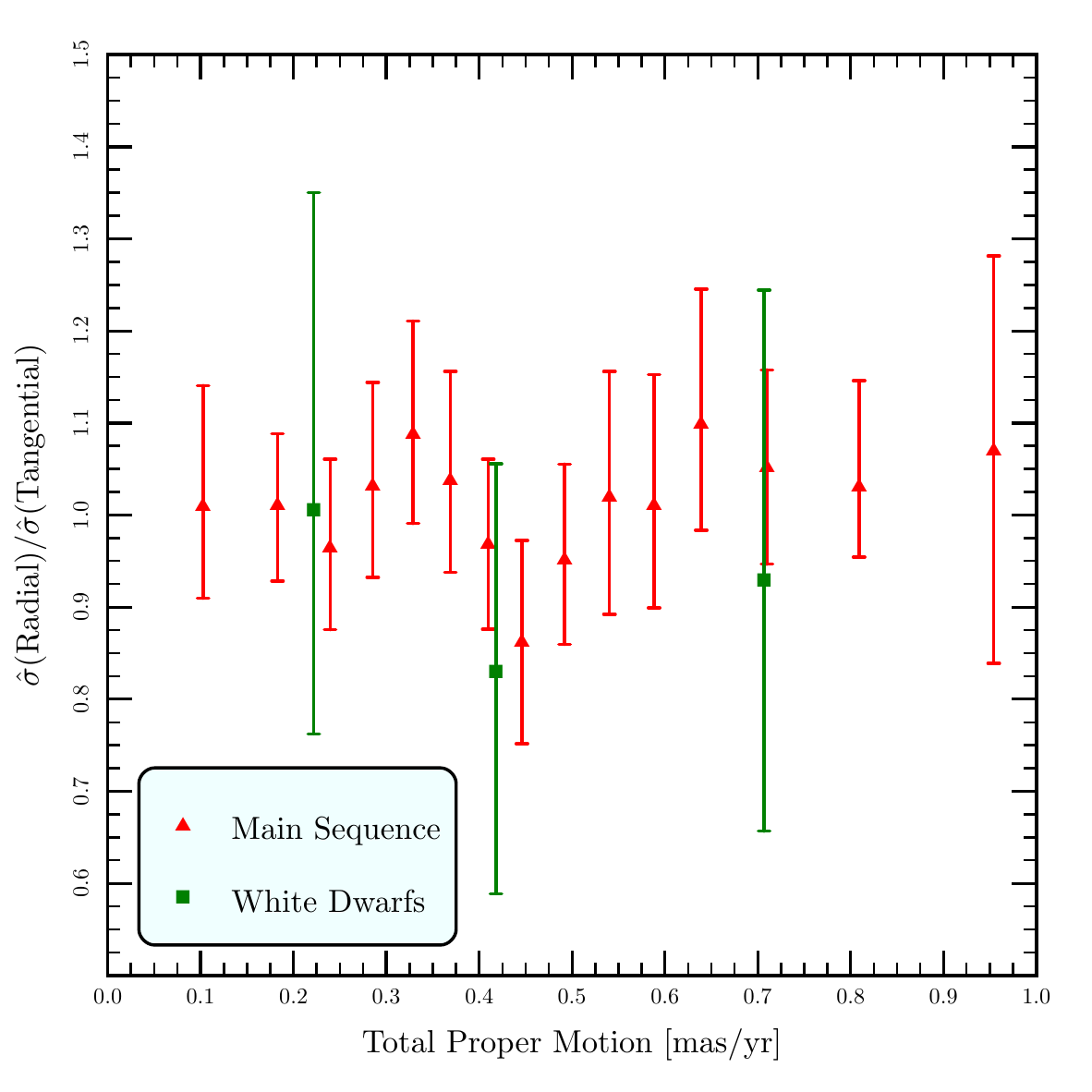}
  \caption{The radial anisotropy of the proper-motion ellipse as a function
    of total proper motion for main-sequence and
    white-dwarf stars.  Only stars with total proper-motion error less
    than 0.4 mas/yr are included.}
  \label{fig:qnratpm}
\end{figure}

We can also compare the velocity dispersion along the radial direction
($\hat \sigma_R$) to the tangential direction ($\hat \sigma_T$) for
the main-sequence stars with the most precisely measured proper
motions ($19.5 < \mathrm{F814W} < 24.5$) and find that ${\hat
  \sigma}(\text{radial})/{\hat \sigma}(\text{tangential}) = 0.96\pm
0.04$, i.e. just consistent with isotropy at 90\% confidence.  To
examine this in greater detail we sort the main-sequence stars and
white-dwarf stars by apparent magnitude into subsamples of 200 stars
for the main sequence and 50 stars for the white dwarfs (the faintest
subsample may contain fewer).  The proper motions are measured in the
radial and tangential directions (relative to the center of the
cluster) and the ratio of the dispersion in the radial direction to
that in the tangential direction is determined and depicted in
Fig.~\ref{fig:qnratmio}.  We can divide the sample by the projected
radius (Fig.~\ref{fig:qnratrad}) or total proper motion
(Fig.~\ref{fig:qnratpm}).  As for the distribution of all the proper
motions for stars with $19.5 < \mathrm{F814W} < 24.5$, we see no
strong evidence for anisotropy in any of these subsamples. We also
relaxed the assumption that the proper-motion ellipse is aligned with
the radial or tangential direction and even in this more general case
found no evidence for anisotropy in either the entire sample or the
subsamples.  The aforementioned results describe the proper-motion
ellipse relative to the center of the cluster.  We can also examine
the proper-motion ellipse relative to the field boundaries.  For the
total proper-motion sample and the various subsamples, we do not find
any evidence for anisotropy of the proper motions relative to the
boundary of the field.

\subsection{Proper-Motion Dispersion}
\label{sec:prop-moti-disp}
To examine the proper-motion dispersion, we sort the main-sequence and
white-dwarf samples by apparent magnitude into subsamples of 200 and
50 stars, as before.  We calculate the value of ${\hat \sigma}$ for each
subsample in the radial (${\hat \sigma}_R$) and tangential directions
(${\hat \sigma}_T$) and define the
one-dimensional value of 
\begin{equation}
{\hat \sigma} = \frac{\sqrt{{\hat \sigma}_R^2+{\hat \sigma}_T^2}}{2}.
\label{eq:43}
\end{equation}
Because the proper motions are nearly isotropic, the value in either
direction is well approximated by the one dimensional value of ${\hat
  \sigma}$ as defined above.  Furthermore, for a spherically symmetric
stellar system even if the velocity distribution is not isotropic, by
extending the results of \citet{1989ApJ...3339..195L} we find that
\begin{equation}
\sigma_R^2 + \sigma_T^2 = 2 \sigma_z^2
\end{equation}
where $\sigma_z$ is the velocity dispersion along the line of sight;
therefore, Eq.~\ref{eq:43} gives the natural value to compare with
radial-velocity measurements, such as MM91.
The root-mean-squared error in the proper motion is scaled and subtracted
from each estimate of the dispersion (and its confidence interval) in
quadrature (see \S~\ref{sec:proper-motions} for more details).
\begin{figure}
  \includegraphics[width=\columnwidth]{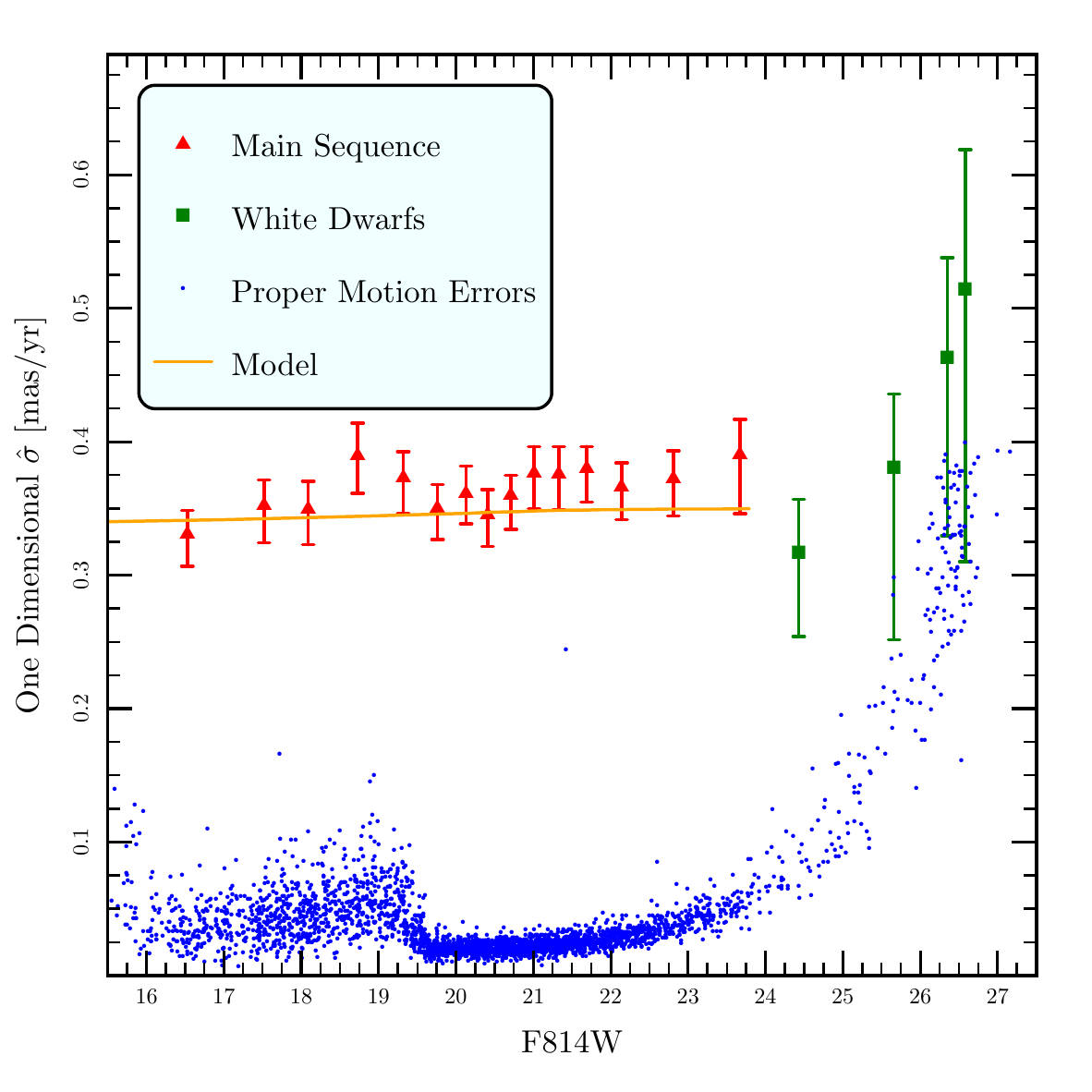}
  \caption{The value of ${\hat \sigma}$ as a function of apparent magnitude.}
  \label{fig:qnmio}
\end{figure}
Fig.~\ref{fig:qnmio} shows that the proper-motion dispersion is nearly
constant, possibly increasing slowly with fainter magnitudes at the
bright end and nearly constant for $\mathrm{F814W}>19$ along the main
sequence.  Such a weak trend is expected for a sample beyond the
half-light radius of the cluster because the velocity distribution is
cut off by escaping stars as shown by the model curve.  The trend in
the data is a bit stronger than the model.  The observed proper
motions of the brightest fifty white dwarfs are smaller by nearly two
standard errors than those of the faintest eighty main-sequence stars
--- $\hat\sigma_{WD} - \hat\sigma_{MS} = -67~\mu\mathrm{as/yr}$ with a
standard error of $36~\mu\mathrm{as/yr}$.  The distribution of the
differences in the dispersion over the bootstrapped resamplings is
approximately normal and the probablility that $\hat\sigma_{WD} >
\hat\sigma_{MS}$ is about three percent.  The main-sequence stars with
masses similar to the white dwarfs have $\mathrm{F814W} \approx 18.8$.
If we take the two hundred main-sequence stars with a median magnitude
of 18.8 and calculate $\hat\sigma_{WD}-\hat\sigma_{MS}$, we obtain
$-90~\mu\mathrm{as/yr}$ with a one-percent chance that
$\hat\sigma_{WD}>\hat\sigma_{MS}$.  If we take the brightest two
hundred main-sequence stars, we obtain a closer agreement with
$\hat\sigma_{WD}-\hat\sigma_{MS} = -26~\mu\mathrm{as/yr}$ and a 17\%
chance that $\hat\sigma_{WD}>\hat\sigma_{MS}$.  If we take the
brightest fifty main-sequence stars, the difference decreases to
$-14~\mu\mathrm{as/yr}$, and the probability increases to 38\%.
Looking along the main sequence to brighter magnitudes, it is apparent
that the best agreement is with the brightest stars in the
proper-motion sample; these stars most resemble the progenitors of the
white dwarfs.

\begin{figure}
  \includegraphics[width=\columnwidth]{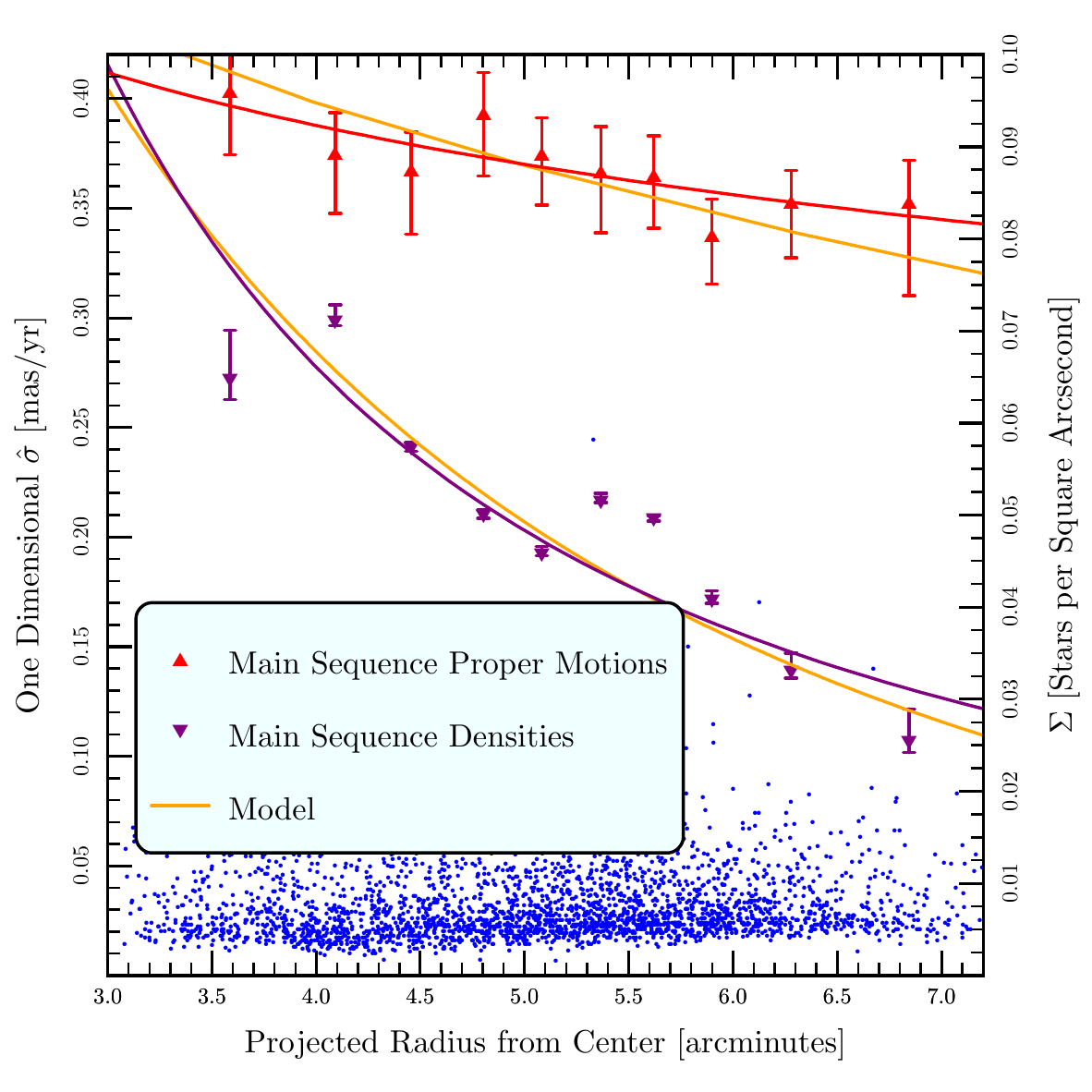}
  \caption{The value of ${\hat \sigma}$ and stellar density as a function
    of projected radius.  The red and purple curves trace the
    best-fitting power-law relations in radius to the data of the same
  color. The upper green curve traces the value of the proper-motion
  dispersion as a function of radius according to the model.  The
  lower curve depicts the value of the projected stellar density
  according to the model.}
  \label{fig:qn_npm_r}
\end{figure}

With a sample of about 2000 main-sequence stars with $19.5 <
\mathrm{F814W} < 24.5$, we sort the stars in projected radial distance
from the center of the cluster and determine the dispersion 
as a function of radius as shown in Fig.~\ref{fig:qn_npm_r}. In this
case each subsample consists of 200 stars. Again the confidence
intervals are determined by resampling, and the root-mean-square error in
proper motion is subtracted from ${\hat \sigma}$ as described in
\S~\ref{sec:proper-motions}.  Both the density and proper-motion
dispersion decrease with increasing radius. The curves give the
best-fit power-law functions of radius for the value of ${\hat \sigma}$
and column density.  We have
\newcommand{\powerlawsone}{
{\hat \sigma} &\approx&
 \left ( 0.37 \pm 0.06 \right ) \left( \frac{R}{5'} \right )^{\!\! -0.21 \pm 0.09} \!\!\!\!\! \mathrm{mas~yr,}^{\!-1}
\nonumber \\
\Sigma &\approx&
 \left ( 0.048 \pm 0.030 \right ) \left( \frac{R}{5'} \right )^{\!\! -1.40 \pm 0.38} \!\!\!\!\! \frac{\mathrm{stars}}{\mathrm{arcsecond}^2}.
}
\newcommand{\disppowerlaw}{$\sigma_\mu \approx 0.36 (R/5')^{-0.20}$}
\begin{eqnarray}
\powerlawsone
\label{eq:27}
\end{eqnarray}
We can also use the proper motions and positions themselves without
binning in radius to determine the best fitting power-law relations
for the column density and velocity dispersion, yielding
\begin{eqnarray}
\sigma_\mu &\approx& 0.36 \left (
  \frac{R}{5'} \right )^{\!\!-0.18}\!\!\!\!\! \mathrm{mas~yr,}^{\!-1}
\nonumber \\
\Sigma &\approx& 0.053
   \left (   \frac{R}{5'} \right )^{\!\!-1.32}\!\!\!\!\! \mathrm{stars~
     arcsecond.}^{\!-2}
\label{eq:28}
\end{eqnarray}
in good agreement with the fitted parameters for ${\hat \sigma}$ and the column density in Eq.~(\ref{eq:27}) and the velocity dispersion fitted from
Fig.~\ref{fig:disp_radius} (\disppowerlaw) 

For clarity we have not quoted errors on these parameters because they
are intermediate results in the determination of the mass of NGC~6397,
which appears in \S~\ref{sec:mass-ngc-6397} with error bars.
Although the power-law fit to the value of ${\hat \sigma}$ characterizes
the data well, the fit for the projected stellar density is poor.  The
observed star counts vary up and down with radius where any
reasonable model would predict a monotonic decrease in radius.  This
behaviour appears to be larger than expected given the errors on the
densities, so it probably reflects either inhomogeneities in the
cluster or underestimated errors in our analysis.  However, the models
developed in \S~\ref{sec:modelling-ngc-6397} rely only weakly on the
projected stellar densities across the ACS field because they also
account for the surface-brightness profile of the entire cluster from
\citet{1995AJ....109..218T}.

The stellar density and velocity dispersion can provide an estimate of
the relaxation time for a group of stars,
\citep{1971ApJ...164..399S}
\begin{equation}
\tau_r = \frac{0.065 v_\mathrm{RMS}^3}{3 G^2 m_*^2 n \ln (0.4 M/m_*)}
\end{equation}
where $v_\mathrm{RMS}$ is root-mean-square velocity of the stars,
$n$ is the number density, $M$ is the
total mass of the cluster and $m_*$ is the mass of a typical star
(taken to be 0.35~M$_\odot$).  Using the results of Eq.~\ref{eq:28}
with our kinematic distance (Table~\ref{tab:data} and
\S~\ref{sec:kinem-dist-ngc})
yields
\begin{equation}
\tau_r \approx 0.6~d_2^6~\mathrm{Gyr} 
\label{eq:39}
\end{equation}
at a radius of five arcminutes
where $d_2$ is the distance to NGC~6397 divided by 2~kpc and we have
deprojected the velocity and projected stellar densities using an
inverse Abel transform \citep{1989ApJ...3339..195L}.  If we use the
velocities measured by MM91 at five arcminutes (the approximate radius
of our field) with the standard candle distance of 2.53~kpc
\citep{2003A&A...408..529G} yields the larger value of
\begin{equation}
\tau_r \approx 0.8~d_{2.53}^3~\mathrm{Gyr}.
\end{equation}
We will revisit these timescales in the following section.
%
%
%
%
%
%

\subsection{Proper-Motion Distribution}
\label{sec:prop-moti-distr}

\begin{figure*}
  \includegraphics[width=\columnwidth]{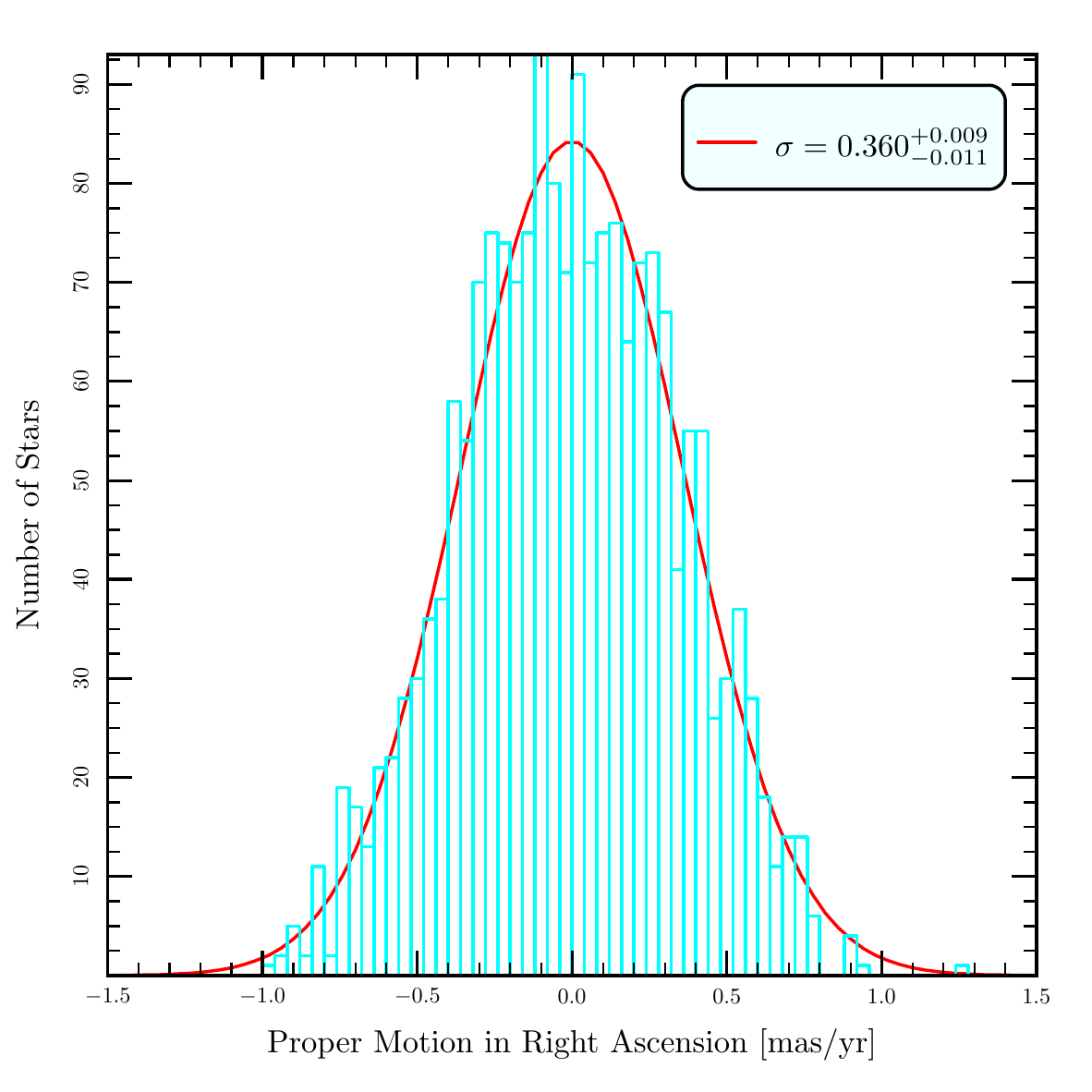}
  \includegraphics[width=\columnwidth]{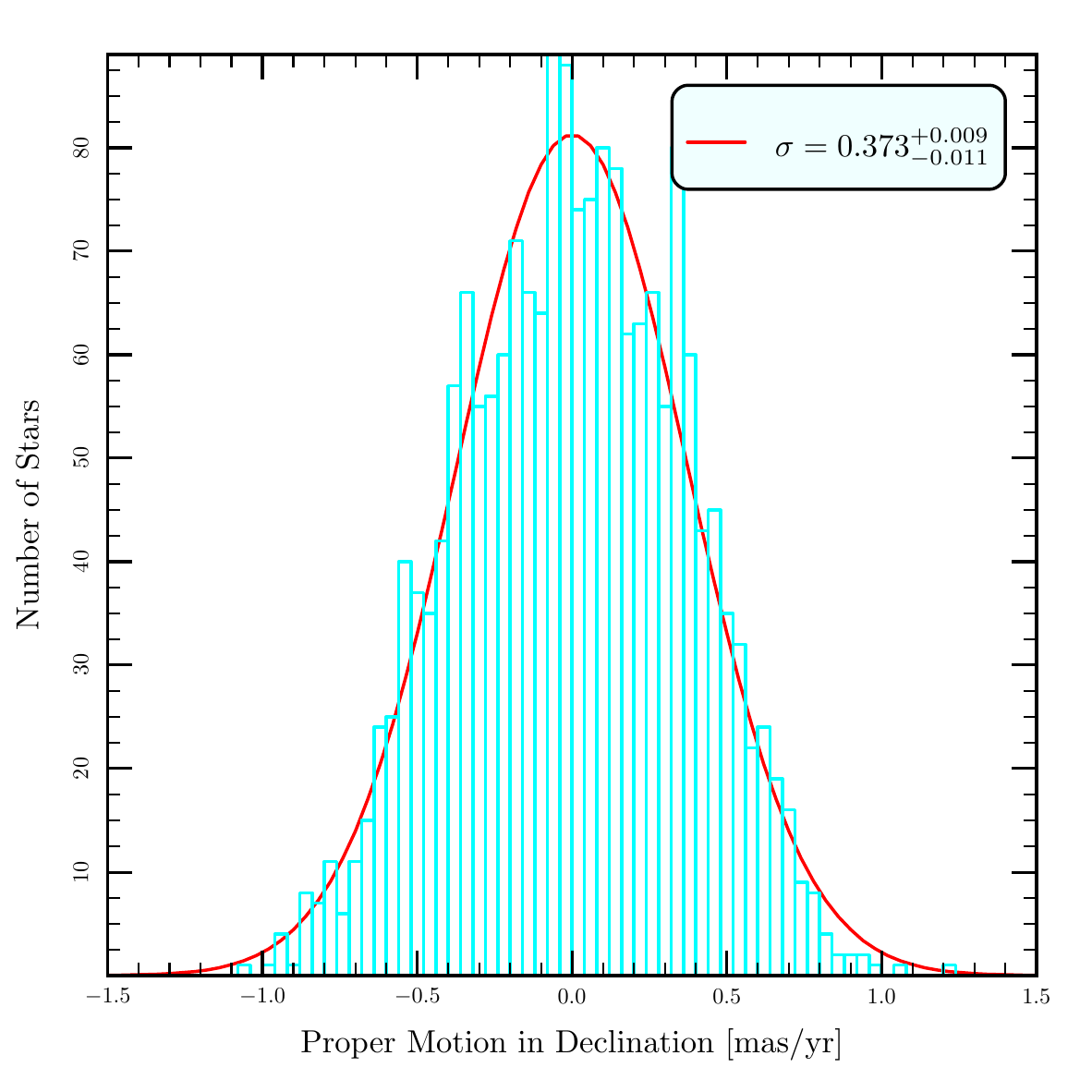}

  \includegraphics[width=\columnwidth]{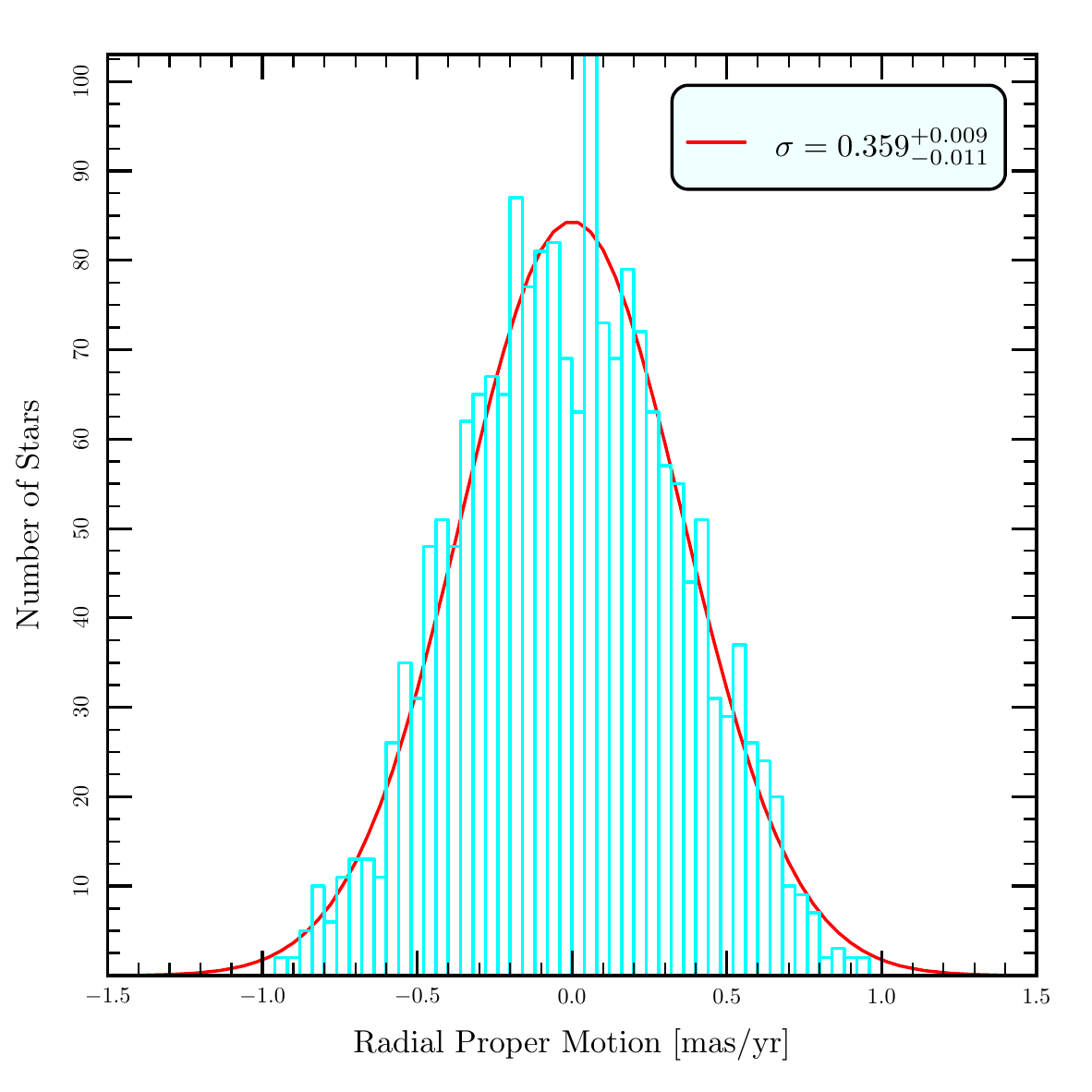}
  \includegraphics[width=\columnwidth]{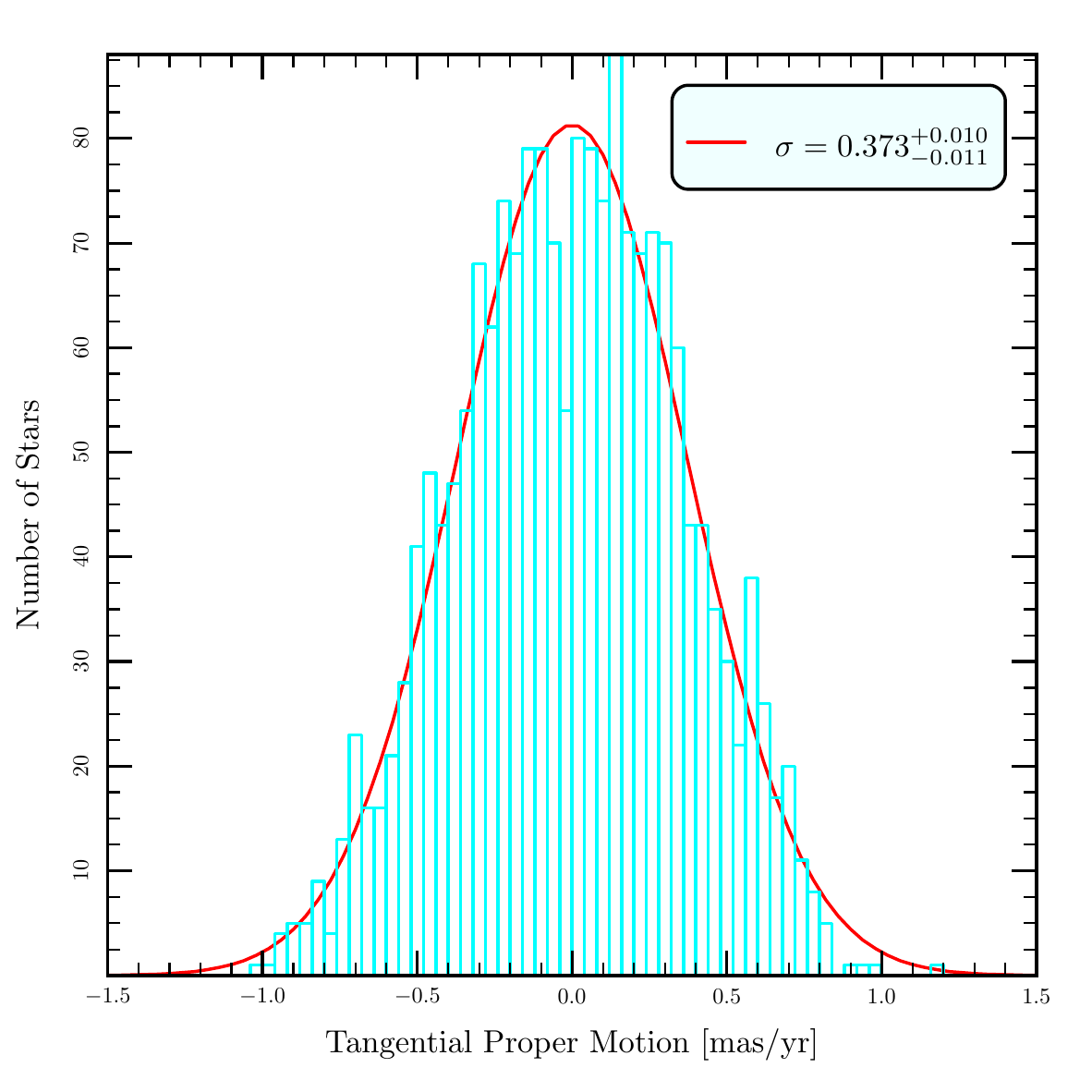}
\caption{Differential distributions of the observed proper motions in
  various directions and best-fitting Gaussians.  The values of
  $\sigma$ are given in milliarcseconds per year.}
\label{fig:msdist_dir}
\end{figure*}

Our first focus will be the sample of about 2000 main-sequence stars
with $19.5 < \mathrm{F814W} < 24.5$ as in the previous section.  As
MAM06 did, we look at the differential distribution of proper motions
in various directions as well as the total proper motion. 
These directional distributions are presented in Fig.~\ref{fig:msdist_dir}.
Each of the distributions has been fit with a Gaussian (Maxwellian)
distribution of proper motions in one dimension; the value of the
standard deviation of the distribution is taken to be the value of
$\hat \sigma$ for the sample of proper motions.  One can conclude from
an Anderson-Darling test (\citeyear{Ande52}) that the proper motions in the
radial and right ascension directions are not sampled from a Gaussian
distribution at the 95\% confidence level.  The conclusion that the
proper motions in the tangential and declination directions are not
drawn from a Gaussian exceeds 99\% confidence.  The values of $\sigma$
for the various directions do not differ significantly from each other
as one would expect from the results of \S~\ref{sec:prop-moti-isotr}.

We can directly compare the various distributions with each other and
look for significant differences using a KS test.  Because the KS test
is most sensitive to changes in the median values (here expected to be
zero for all of the one-dimensional distribution), we compare the
distribution of the absolute values of the proper motions.  In
particular the distribution of the proper motion toward or away from
the cluster center (radial) or perpendicular to this direction
(tangential) do not differ significantly as the KS test yields a
$p-$value of 57\%.  On the other hand, the hypothesis that the proper
motions in right ascension and declination are drawn from the same
distribution can be rejected with nearly 99\% confidence according to
a KS test, indicating that there may be some residual systematics in
the proper-motion measurements; however, the values of the dispersions
of the best-fitting Gaussian agree well within the ninety percent
confidence regions, so the differences are more subtle than the width
of the distributions.

The theoretical model that we developed in \S\S\ref{sec:modelling}
and~\ref{sec:modelling-ngc-6397} exhibits an isotropic velocity
distribution, so it makes most sense to compare the model with the
distribution of the total proper motion.  We shall examine two
theoretical models.  First, the model yields an estimate of the
proper-motion distribution over the entire field as depicted in
Fig.~\ref{fig:cum_pmtot}.  This proper-motion distribution results
from convolving Eq.~\ref{eq:16} in the appendix with the window
function in spherical radius in Fig.~\ref{fig:windowfunk}.  This
proper model distribution yields a KS probability of two percent for
the null hypothesis that the stars from the proper-motion sample are
drawn from it.
The
results of Fig.~\ref{fig:qn_npm_r} indicate a second possible
treatment.  The figure depicts the model results for the stellar
column density and the dispersion (${\hat \sigma}$). Although the fit
to the proper motion appears good, the model does not track the
observed column density well.  In fact the observed column density
does not actually decrease monotonically with radius, indicating that
possibly the cluster is a bit clumpy.  
However, we can also convolve
the distribution in Eq.~\ref{eq:17} with the number of
stars as a function of radius in Fig.~\ref{fig:qn_npm_r}.  This
results in the ``Scaled Model'' in Fig.~\ref{fig:cum_pmtot} that
yields a KS probability of thirty-three percent for the null hypothesis that
the observed proper motions of the nearly two thousand main-sequence
stars in the proper-motion sample are drawn from the theoretical
model.
\begin{figure}
 \includegraphics[width=\columnwidth]{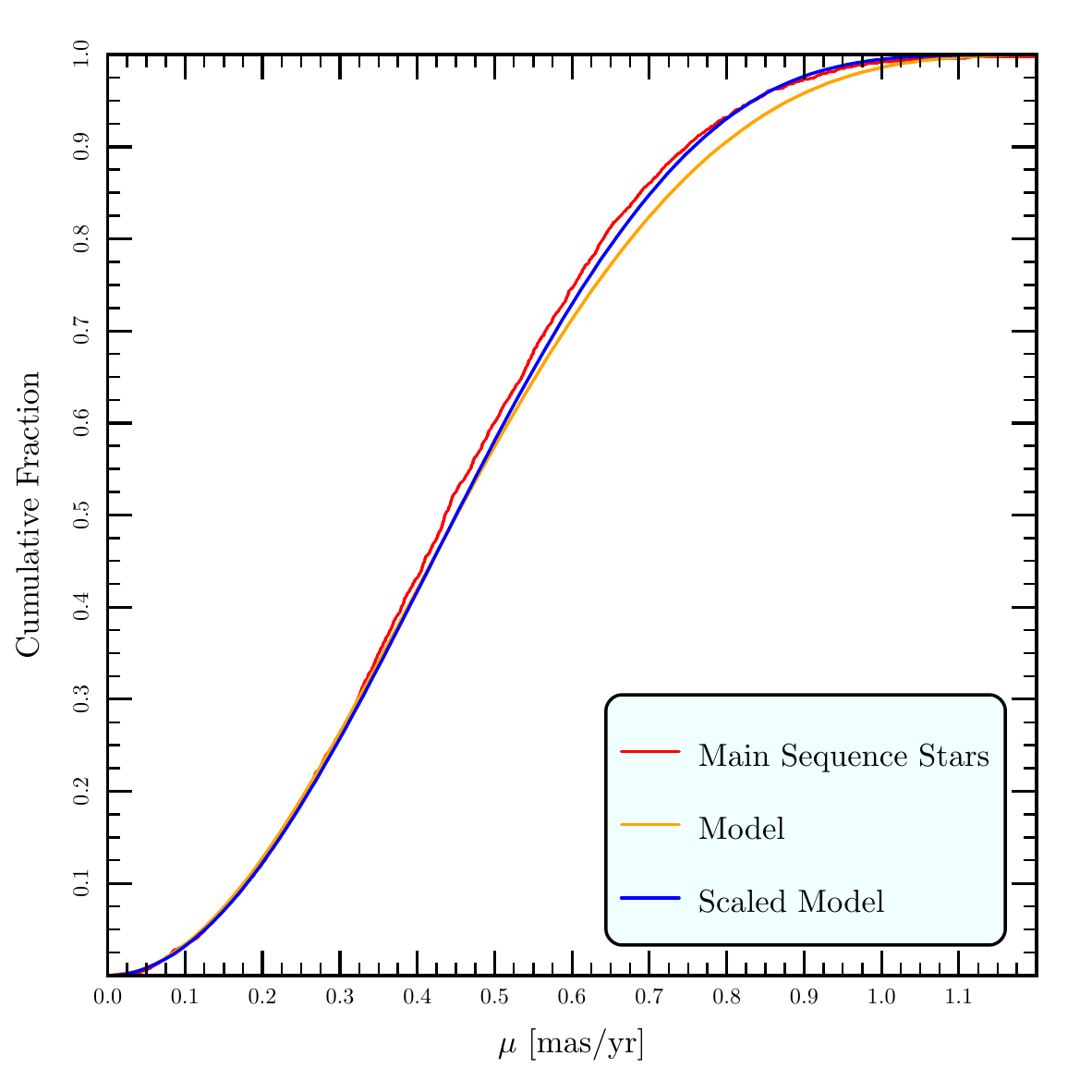}
 \caption{Cumulative distribution of total proper motion for the
   main-sequence stars, the basic model and a model scaled to yield
   the observed column density in Fig.~\ref{fig:qn_npm_r}. }
\label{fig:cum_pmtot}
\end{figure}

\begin{figure}
  \includegraphics[width=\columnwidth]{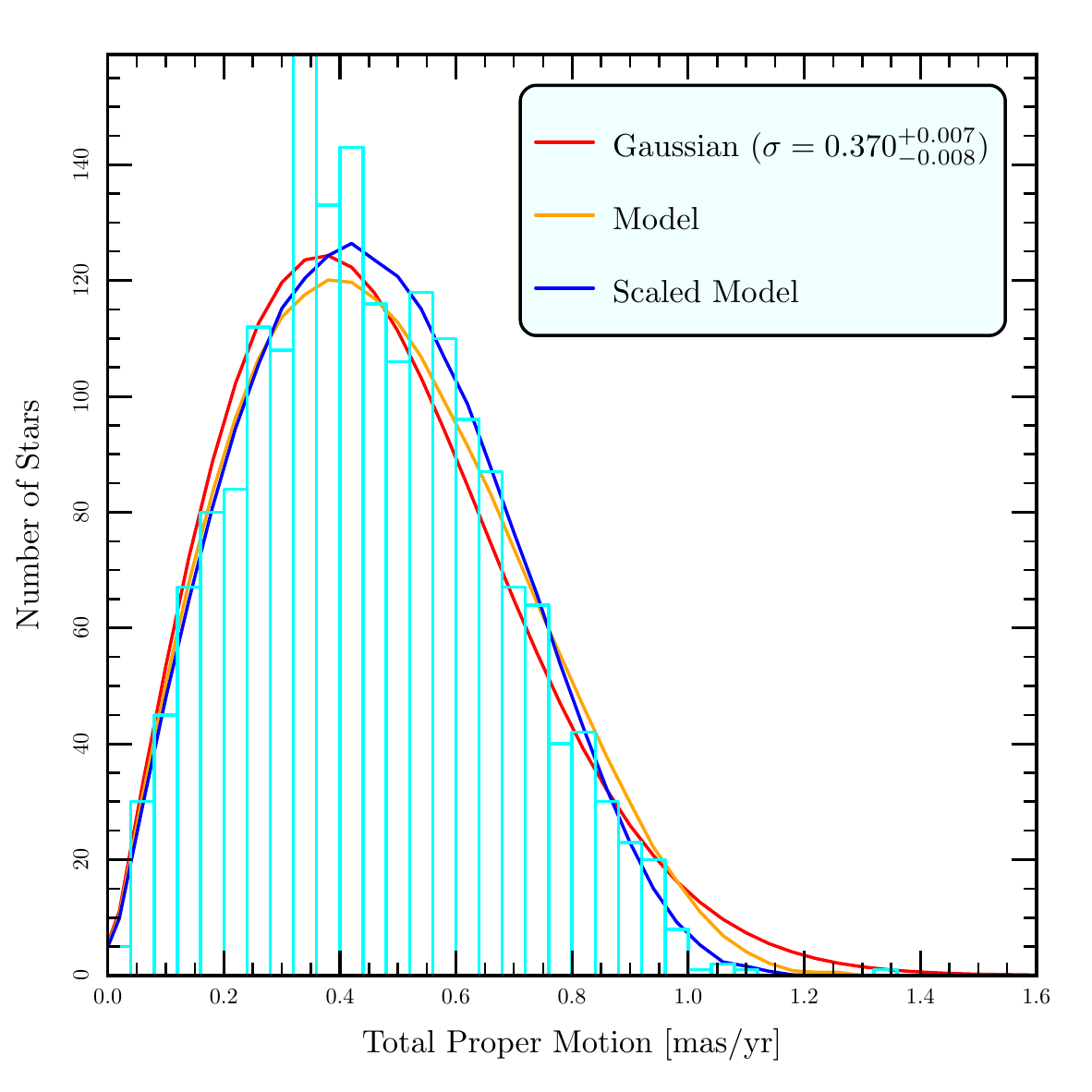}
\caption{Differential distributions of total proper motion for the
   main-sequence stars, the basic model and a model scaled to yield
   the observed column density in Fig.~\ref{fig:qn_npm_r}, and the
   best-fitting Gaussian.
  The value of $\sigma$ is given in milliarcseconds per year.}
\label{fig:msdist_tot}
\end{figure}

An examination of the differential distribution of the total proper
motion yields additional details (Fig.~\ref{fig:msdist_tot}).
The width is similar
to the one-dimensional results in Fig.~\ref{fig:msdist_dir}.  The
distribution of stars cuts off much more quickly than the best-fitting
Rayleigh distribution (the distribution of speeds if the velocity in
each of two dimensions follows a Gaussian).  The latter gives 49 stars
with total proper motions exceeding one milliarcsecond per year; there
were only six observed in the sample.  If we assumed that the best
fitting Rayleigh distribution was the underlying true distribution the
chance of finding six when we expect 49 is about $10^{-14}$, assuming
Poisson statistics.  The cutoff at about one milliarcsecond per year
results because the cluster has a finite escape proper motion at this
radius of roughly one milliarcsecond per year. The results of
\S~\ref{sec:modelling-ngc-6397} produce a distribution of total proper
motions using Eq.~(\ref{eq:16}) convolved with the window function
shown as the ``Model'' curve in Fig.~\ref{fig:msdist_tot}) or convolved with
the observed number of stars as a function of radius from
Fig.~\ref{fig:qn_npm_r} shown as the ``Scaled Model'' curve. 

\begin{figure}
  \includegraphics[width=\columnwidth]{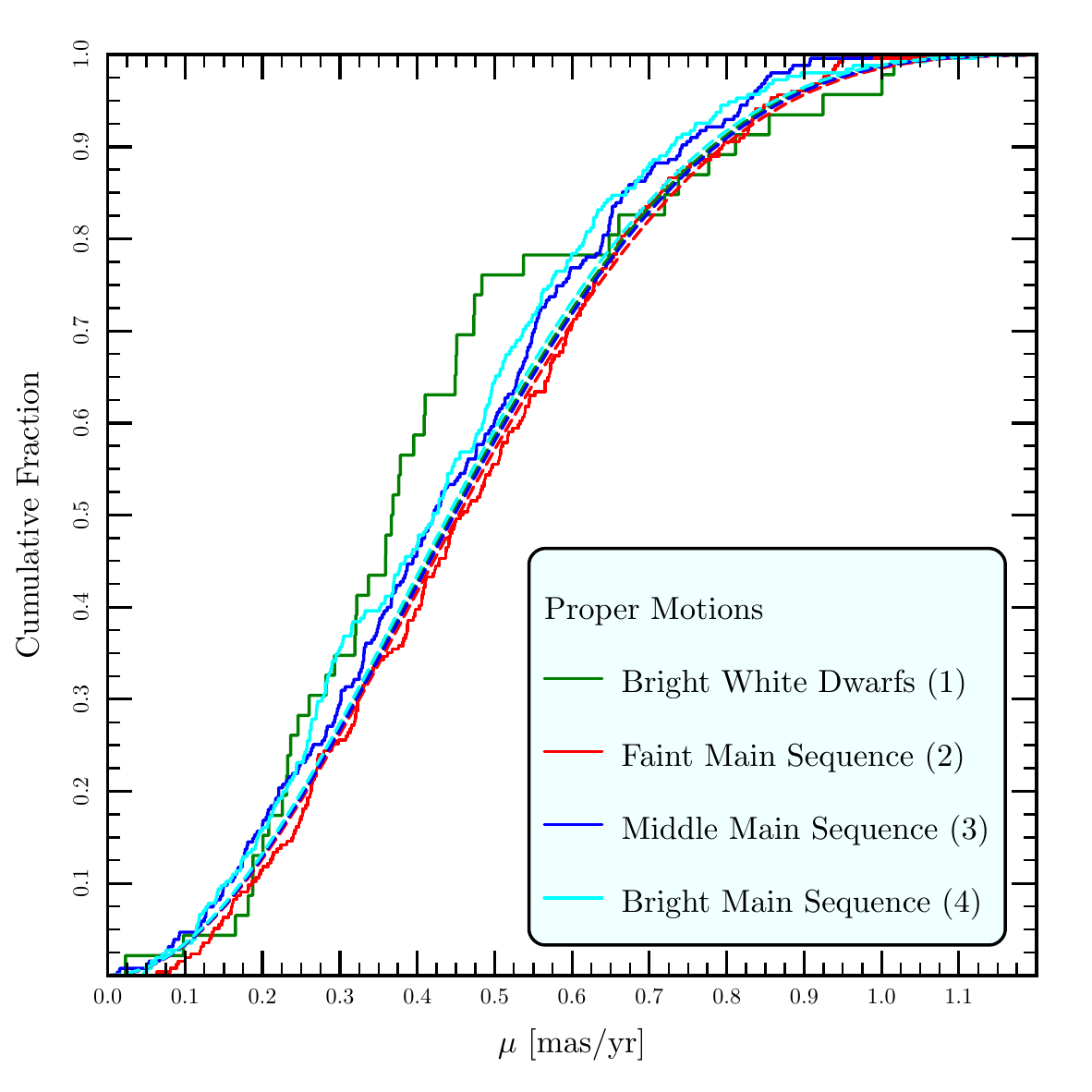}
  \caption{Cumulative distributions of total proper motion for the
    subsamples.  The observed distributions are given by the solid
    curves, and the model distributions using the masses in
    Table~\ref{tab:rmserr} are traced by the dashed curves}.
  \label{fig:cum_dist}
\end{figure}

To contrast the distribution of white dwarfs and main sequence stars,
we examine the subsamples delineated in Fig.~\ref{fig:pmcmd} and
Table~\ref{tab:rmserr}. Fig.~\ref{fig:cum_dist} depicts the cumulative
proper-motion distributions for the four subsamples, and
Table~\ref{tab:prob} gives the KS and Wilcoxon probabilities for each
pair.  Lower probabilities indicate that the null hypothesis that the
two samples are drawn from the same underlying distribution is unlikely.
The stars
in the bright WD sample and faint MS sample span the same magnitude
range, but a comparision of their proper-motion distributions 
yields $p-$value of 0.02 by the KS test.  The main sequence stars with
$\mathrm{F814W} \approx 19$ (middle MS) are likely to have similar
masses to those of the brightest white dwarfs. These stars yield
a larger $p-$value of about 0.12.
\diffblock{
\begin{table}
\begin{center}
\caption{KS and Wilcoxon probabilities}
\label{tab:prob}
\begin{tabular}{lrrrrrl}
\hline \hline
                         & BWD      & BWDN       & FMS      & MMS     & BMS    & \\
\hline               
\multicolumn{1}{l|}{BWD}  &         &           & {2.4}    & {12}    & {22}   & \multirow{4}{*}{$\Bigg\}$ KS} \\
\multicolumn{1}{l|}{BWDN} &         &           & 2.0      & 11      & 22     &  \\
\multicolumn{1}{l|}{FMS} & {7.7}   & 5.6       &          & {22}    & {7.3}  & \\
\multicolumn{1}{l|}{MMS} & {41}    & 26        & {8.3}    &         & {53}   & \\
\multicolumn{1}{l|}{BMS} & {57}    & 44        & {1.8}    & {52}    &        & \\
& \multicolumn{4}{c}{$\underbrace{\hspace{1.33in}}$} &  & \\
& \multicolumn{4}{c}{Wilcoxon} &  & 
\end{tabular}
\end{center}

Kolmogorov-Smirnov (above diagonal) and Wilcoxon (below
  diagonal) probabilities in percent from comparing the proper-motion
  distribution of one of each pair of
samples with the other. The BWDN values compare the BWD sample with the
other samples with added Gaussian errors such that the root-mean-square error in
other samples equals that of the bright-white-dwarf sample.
\end{table}}

The bright main-sequence stars yield the largest $p-$value in a
comparison with the bright white dwarfs of 0.22.  The null hypothesis
that the bright white dwarfs and the bright main-sequence stars are
drawn from the same distribution is the
most difficult to reject by these tests.
If we suppose that these two groups of stars
indeed follow the same underlying proper-motion distribution, we would
conclude that the bright WD stars have not yet relaxed to reflect their
smaller masses.  The observed luminosities of these stars provide
constraints on their ages.  The median magnitude of the
bright-white-dwarf sample is 24.27 in $I-$band with the
transformations of \citet{2005PASP..117.1049S} from F814W and F606W to
$I$.  Using the white-dwarf distance of 2.53~kpc
\citep{2007ApJ...671..380H}, a white-dwarf mass of 0.53~M$_\odot$
\citep{2004A&A...420..515M,2009ApJ...705..408K} and the cooling models
of \citet{1995PASP..107.1047B} yields a median age of the bright WD
sample of 0.5~Gyr.  The shorter kinematic distance (see
\S~\ref{sec:kinem-dist-ngc}) yields a median age of 0.9~Gyr.  We will
simply take the mean of these two determinations yielding a median age
of about 0.7~Gyr.


If the white dwarfs with a median age about 0.7~Gyr have a similar
proper-motion distribution to their progenitors even though their
masses are substantially less, these stars have not yet relaxed to
energy equipartition, and this fact yields a direct estimate of the
relaxation time of the cluster in the ACS field.  This conclusion
agrees with dynamical estimate of the relaxation time for the white
dwarfs from Eq.~\ref{eq:39} of 1~Gyr at the median projected radius of
the subsample, five arcminutes (Fig.~\ref{fig:cum_rdist}).
Fig.~\ref{fig:qnmio} apparently shows that the older white dwarfs have
a similar velocity dispersion to the main-sequence stars.  The median
F814W magnitudes of the next two fainter white-dwarf points are 25.66
and 26.35, corresponding to white-dwarf ages of about 2.5 and 5.2~Gyr
respectively \citep{1995PASP..107.1047B}, a factor of three to seven
older than the bright WD sample.  This yields a loose upper limit to
the relaxation time in agreement with Eq.~\ref{eq:39}; however, the
errors in proper motions also begin to dominate at these magnitudes,
making it difficult to draw conclusions as we must rely more strongly
on the subtraction of the observational errors. Better constraints on
the proper motions at faint magnitudes from future observations could
allow us to place a more reliable upper limit on the relaxation time.

The proper-motion errors increase at faint magnitudes because the
poisson noise in the pixels used to determine the position increases
as the flux decreases. Brighter than $\mathrm{F814W} = 19.5$
saturation begins to set in for the deep images, so a smaller set of
shorter exposures is used for the astrometry of brighter stars.  The
proper-motion errors increase abruptly here and again decrease with
brighter magnitudes.  This results in a variation of the typical
proper motion error for the various samples as shown in
Table~\ref{tab:rmserr}.  These errors contribute to the observed
dispersions so they have been subtracted from the dispersions in
Fig.~\ref{fig:qnmio} and~\ref{fig:qn_npm_r}.  They also broaden the
cumulative distributions depicted in Fig.~\ref{fig:cum_dist} and bias
the probabilities in Table~\ref{tab:prob}.  From
Table~\ref{tab:rmserr} it is apparent that the typical error in the
white dwarfs is substantially larger than for the other samples and
that the errors in the other samples are nearly equal to each other;
therefore, we artificially increase the proper-motion errors in these
samples by adding normally distributed random numbers to each of the
proper-motion measurements, so that all four distributions have the
same root-mean-square error as the bright white dwarfs.  Although the
faint-main-sequence sample (FMS) and the white-dwarf sample cover the
same range of apparent magnitude, the white dwarfs are typically
fainter within that range than the main-sequence stars (see
Tab.~\ref{tab:rmserr}) and hence exhibit larger proper-motion errors.

We then calculate the KS and Wilcoxon probabilities for the new
subsamples against the white-dwarf distribution.  One thousand new
subsamples are generated in this way, and the average probabilities
are tabulated.  These results are depicted in the column and row
labelled ``WDN'' in Table~\ref{tab:prob}.  By making the distributions
even broader than the white dwarf distribution, the additional error
decreases the $p-$values that result from comparing one sample to
another, verifying that the differences in the distributions do not
result from differences in the errors.

The distributions depicted in Fig.~\ref{fig:cum_dist} merit further
comment.  The faint MS sample appears to have typically larger proper
motions than the more massive middle MS sample, and the brightest MS
stars have slightly smaller values still.  This is in agreement with
the values of ${\hat \sigma}$ in Table~\ref{tab:rmserr} and with the
expectations of equipartition.  However, from
Fig.~\ref{fig:disp_radius},~\ref{fig:qnmio} and \ref{fig:cum_dist} we
can seen that the velocity dispersions within our field expected from
the models vary little with mass (in fact less than is observed), but
observations nearer to the core would be a better probe of the
dynamical state of the various stars in the cluster.

\subsection{The Kinematic Distance of NGC~6397}
\label{sec:kinem-dist-ngc}

Fig.~\ref{fig:disp_radius} depicts the observed proper-motion
measurements as a function of distance from the center of the cluster
along with the measurements of the radial-velocity dispersion
converted to proper motions at a fiducial distance of 2.55~kpc
\citep{2008AJ....135.2141R} and our best-fitting distance of 2~kpc.
\citet{1989ApJ...3339..195L} argue that for a spherical stellar
cluster the proper motions themselves will reveal the hallmark of an
anisotropic velocity ellipsoid.  In \S~\ref{sec:prop-moti-isotr} we
argue that there is no compelling evidence for anisotropy either in
the sample as a whole or in various subsamples; therefore, it is
natural to compare the observed proper-motion dispersions with the radial
velocity dispersions to obtain a kinematic (or geometric) estimate of
the distance to the cluster.   

The best-fitting distance with the smallest statistical error in
Table~\ref{tab:distances} is obtained by estimating the proper-motion
dispersion of all the main-sequence stars in the proper-motion sample,
$0.369 \pm 0.009$~mas/yr.  We then fit the MM91 data linearly with
projected radial distance to estimate the radial-velocity dispersion
at the median distance of the stars in our sample (5.16 arcminutes)
yielding $3.12 \pm 0.11$~km/s.  This yields a distance of $1.78 \pm
0.07$~kpc.  Because it involves the most data, this estimate has the
smallest statistical confidence interval.  However, there are two
possible sources of systematic error that we would like to address.
First MM91 actually measured the radial-velocity dispersion at 5.22
arcminutes near the median radius of our dataset to be $3.3 \pm
0.8$~km/s.  Combining this velocity with the dispersion of the entire
dataset yields a larger distance of $1.9\pm 0.5$~kpc.  Second, the
stars studied by MM91 were among the brightest in the cluster so it is
natural to compare the brightest stars in our sample with their data.
Fig.~\ref{fig:qnmio} shows that the proper-motion dispersion decreases
slightly with increasing flux.  The dispersion ($\hat \sigma$) of the
brightest 200 stars in our sample is $0.33^{+0.01}_{-0.03}$~mas/yr
with a slightly smaller projected median distance of 5 arcminutes.  To
obtain a more robust kinematic estimate of the distance to NGC~6397 we
have used the unpublished radial velocities from MM91 (Meylan, private
communication). We restrict the MM91 dataset to 46 stars with
projected distances from the center of NGC~6397 between 3 and 7.2
arcminutes; the median projected distance of this sample from the
center is slightly smaller at 4.5~arcminutes.  The dispersion of this
bright overlap sample is $3.5^{+1.0}_{-0.6}$~km/s after correcting for
measurement uncertainties, yielding a distance estimate of
$2.2^{+0.5}_{-0.7}$~kpc with a ninety-percent confidence interval.  We
obtain our best distance estimate by fitting the MM91 data in radius
and comparing the dispersion to that of the brightest 200 stars in our
sample ($1.9\pm 0.2$~kpc).  Additionally we correct for the difference
in the median mass of these two samples ($0.84 \mathrm{M}_\odot$ for
the giants and $0.74 \mathrm{M}_\odot$ for the bright proper motion
sample).  This yields a distance estimate of $2.0\pm 0.2$~kpc and
balances biases against statistical errors.  For reference all of
these distances are presented in Table~\ref{tab:distances}.
\begin{table}
\caption{Distance Estimates to NGC~6397}
\label{tab:distances}
\begin{tabular}{cccc}
\hline
\hline
Method & Value [kpc] & References \\
\hline
\multicolumn{3}{c}{Standard-Candle Estimates} \\
\hline
Subdwarf fit in $B$ and $V$ & $2.53 \pm 0.05$ & \protect{\citealt{2003A&A...408..529G}} \\
Subdwarf fit in $b$ and $y$ & $2.58 \pm 0.07$ &  " \\
Subdwarf fit                & $2.67 \pm 0.25$ & \protect{\citealt{1998AJ....116.2929R}} \\
White-dwarf fit             & $2.55 \pm 0.11$ & \protect{\citealt{2007ApJ...671..380H}} \\
\hline
\multicolumn{3}{c}{Kinematic Estimates} \\
\hline
MS sample (radial fit)      & $1.78 \pm 0.07$ & MM91/Here \\
MS sample (at 5.22')  & $1.9 \pm 0.5$ &    " \\
Bright sample (fit)           & $1.9 \pm 0.2$ &  " \\
Bright sample (fit/corr)           & $2.0 \pm 0.2$ &  " \\
Bright overlap sample        & $2.2^{+0.5}_{-0.7}$ & " \\
\hline
\end{tabular}
\end{table}

These successive distance estimates trade larger statistical errors for
reduced systematic errors that arise from the heterogeneity of the
stars and through interpolating the velocity profiles in radius.  The
final estimate of $2.2^{+0.5}_{-0.7}$~kpc compares stars of nearly the same
mass at the same median radius, so it minimizes the systematic errors;
however, it uses only a portion of both samples, so the statistical
errors are necessarily larger.  Increasing the size of the sample of
stars with radial-velocity measurements could dramatically decrease
the uncertainities of this kinematic distance estimate.
 
\subsection{The Mass of NGC~6397}
\label{sec:mass-ngc-6397}

\citet{1989ApJ...3339..195L} outlined several mass estimators for the
open cluster M35 using the observed proper-motions.  Among these the
most straightforward is
\begin{equation}
\label{eq:30}
\langle G M_r \rangle = \frac{16}{\pi} \left \langle R \left (
    \frac{2}{3} v_R^2 + \frac{1}{3} v_T^2 \right ) \right \rangle
\end{equation}
where $v_R$ is the component of the proper-motion toward the center of the
cluster and $v_T$ is the tangential component of the proper-motion.
This estimator yields
\begin{equation}
  \label{eq:31}
  M_r = 8.3 \pm 0.4 \times 10^4 d_{2.53}^3 \mathrm{M}_\odot
\end{equation}
where we have summed over all of the stars along the cluster main
sequence and white-dwarf cooling tracks and excluded those stars with
proper motion errors greater than 0.4~mas/yr or proper-motions greater
than 5~mas/yr.  The confidence interval gives a ninety-percent
confidence region obtained through bootstrapping the sample.

We use the power-law fits to the column density and velocity
dispersion (Eq.~\ref{eq:27}) of the cluster to deproject the
proper-motion dispersion, density and their gradients to obtain an
estimate of the mass enclosed within the outer radius of our field.
We can simply apply Jeans equation
\begin{equation}
v_r^2 - \sigma_r^2 \left [ \frac{d \ln \sigma_r^2}{d \ln r} + \frac{d
    \ln n}{d \ln r} + \beta \right ] = \frac{G M_r}{r}
\label{eq:32}
\end{equation}
to obtain an estimate of the enclosed mass within a given spherical
radius ($M_r$). The symbols $\sigma_r$ and $n$ denote the radial
velocity dispersion and the number density of stars. In this equation
$v_r$ and $\beta$ quantify the rotation of the cluster and the
anisotropy of the velocity distribution; we see evidence for neither
in our data, so we assume these vanish to yield
\begin{equation}
M_r =   8.7 \pm 0.6 \times 10^4 d_{2.53}^3 \left ( \frac{R}{7'} \right
  )^{0.6}
\mathrm{M}_\odot.
\label{eq:33}
\end{equation}
Whether one uses $\sigma$ or $\hat \sigma$ in Eq.~\ref{eq:32} does
not affect the mass estimate (Eq.~\ref{eq:33}) within the errorbars.
We used proper motions and positions themselves without
binning in radius to determine the best fitting power-law relations
(Eq.~\ref{eq:28}) for the column density and velocity dispersion,
yielding a mass estimate of
\begin{equation}
M_r =   8.5 \pm 0.6 \times 10^4 d_{2.53}^3 \left ( \frac{R}{7'} \right
  )^{0.64} 
\mathrm{M}_\odot.
\label{eq:34}
\end{equation}
We can combine these estimates with the light within this projected
radius to yield
\begin{equation}
\label{eq:35} 
L_{V,R} = 4.0 \times 10^4 d_{2.53}^2 \mathrm{L}_{V,\odot}
\end{equation}
and the model for the light distribution outlined in
\S~\ref{sec:modelling-ngc-6397} allows us to estimate the luminosity
within the spherical radius at
\begin{equation}
  \label{eq:36}
L_{V,r} = 3.5 \times 10^4 d_{2.53}^2 \mathrm{L}_{V,\odot}
\end{equation}
Scaling the mass within seven arcminutes (Eq.~\ref{eq:34}) by the
ratio of the total light (Table~\ref{tab:data}) to that within seven
arcminutes (Eq.~\ref{eq:36}) yields an estimate of the total mass of
the cluster of
\begin{equation}
  M = 1.1 \pm 0.1 \times 10^5 d_{2.53}^3 \mathrm{M}_\odot
\label{eq:37}
\end{equation}
in agreement with the model mass (Eq.~\ref{eq:29}) found in
\S~\ref{sec:modelling-ngc-6397}.  This gives a mass-to-light ratio of
$2.4 \pm 0.3 d_2 \mathrm{M}_\odot/\mathrm{L}_{V,\odot}$; 
including
the uncertainty in the distance yields 
$2.4 \pm 0.5 \mathrm{M}_\odot/\mathrm{L}_{V,\odot}$.
This agrees with the previous
result of MM91 of $2.1\pm 0.1$ at an assumed distance of 2.4~kpc.  The
mass-to-light ratio of the totality of the stellar population of
NGC~6397 range from 1.4 to 2.0 according to MM91 and about 1.2
according to \citet{1995ApJS..100..347D}.  Our present mass-to-light
measurement does not require any dark contribution to the mass of
NGC~6397.

\subsection{Stellar Escapers}
\label{sec:stellar-escapers}

Fig.~\ref{fig:disp_radius} also depicts the escape velocity as a
function of distance from the center of the cluster. The escape
velocity (here given as an escape proper-motion) is about three times
the velocity dispersion for the range of radii probed by the
observations.  This is the escape velocity at a given distance (not
projected) from the center of the cluster; therefore, a star with a
proper-motion larger than this at a particular projected radius will
escape.  Exceeding this proper-motion is a sufficient condition for
escape.  Other stars whose actual three-dimensional distances are
further from the cluster center may also escape as can stars with
significant motion along the line of sight. For an isotropic velocity
distribution the three-dimensional velocity is typically only about
twenty percent larger than the proper-motion; therefore, we will use a
conservative criterion and consider stars that exceed the escape
proper-motion as potential escapers.

\input escape.tex

\begin{figure}
\includegraphics[width=\columnwidth,clip,trim=0 0.8in 0 0.8in]{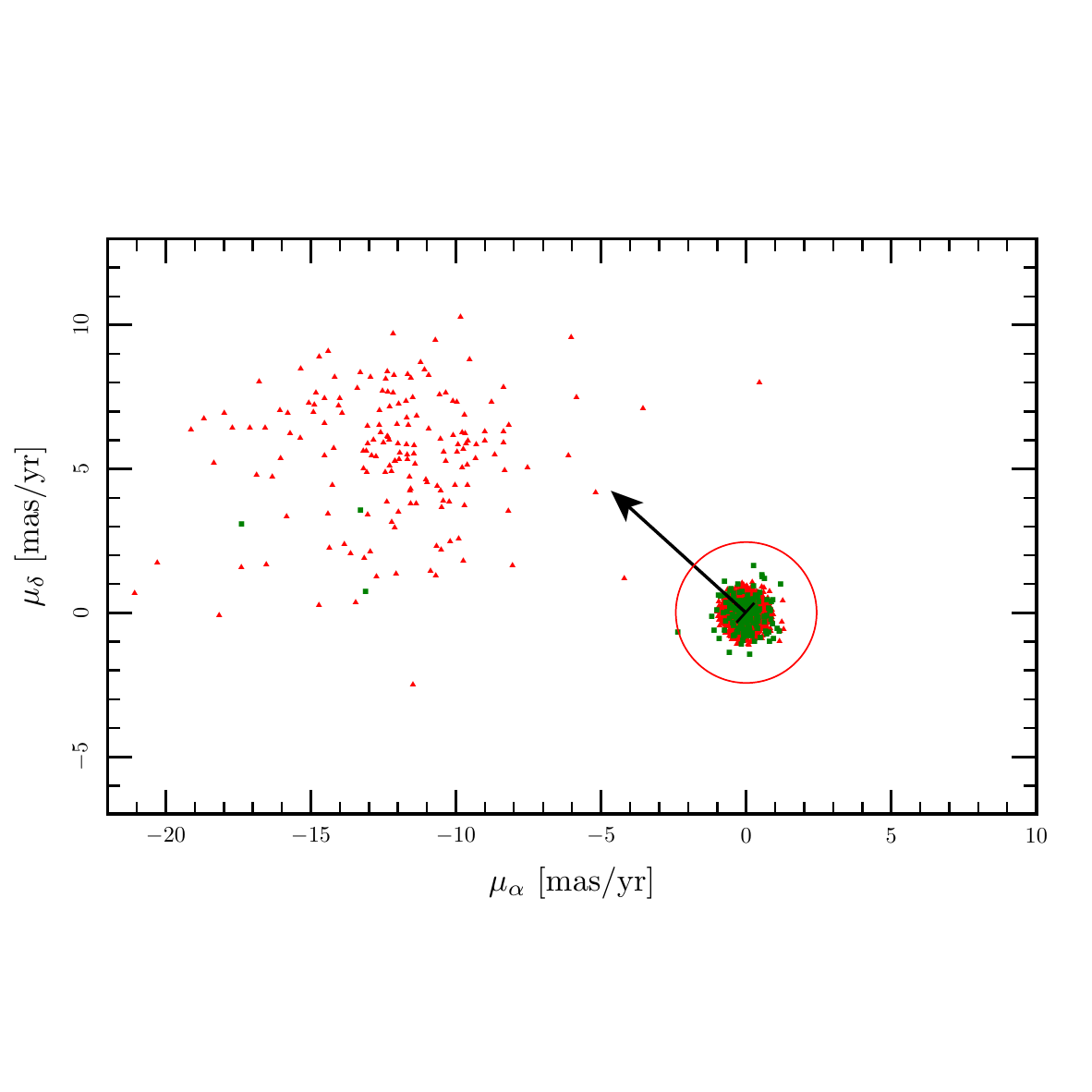}
\caption{Proper motions of the objects identified as stars in the ACS
  field and lying near the main sequence and white dwarf cooling track
  of NGC~6397.  The mean proper-motion of the cluster members has been
  subtracted. The red circle denotes twice the escape proper-motion
  from the inner edge of the ACS field.  
}
\label{fig:pm_mswd}
\end{figure}



Fig.~\ref{fig:pm_mswd} depicts the proper motions of all stars along
the main-sequence and white-dwarf tracks. The tight cloud centered
near zero proper motion are the cluster members whereas the more
diffuse cloud at $\mu_\alpha \approx -12$~mas/yr consists of field
stars that happen to have fluxes similar to cluster main-sequence or
white-dwarf stars.  A comparison with the proper motions of all
stellar sources in Fig.~\ref{fig:allpm} demonstrates that the flux
criteria removes nearly all the field stars while affecting the
potential cluster stars more modestly.  We define the region of
potential escapers as having a total proper motion between once and
twice the escape proper motion for stars at the inner edge of the ACS
field. This upper limit is indicated by the red circle.  Looking at the
density of presumably field stars outside this circle, one expects
fewer than one field star within the narrow range of color and proper
motion.

\begin{figure}
 \includegraphics[width=\columnwidth]{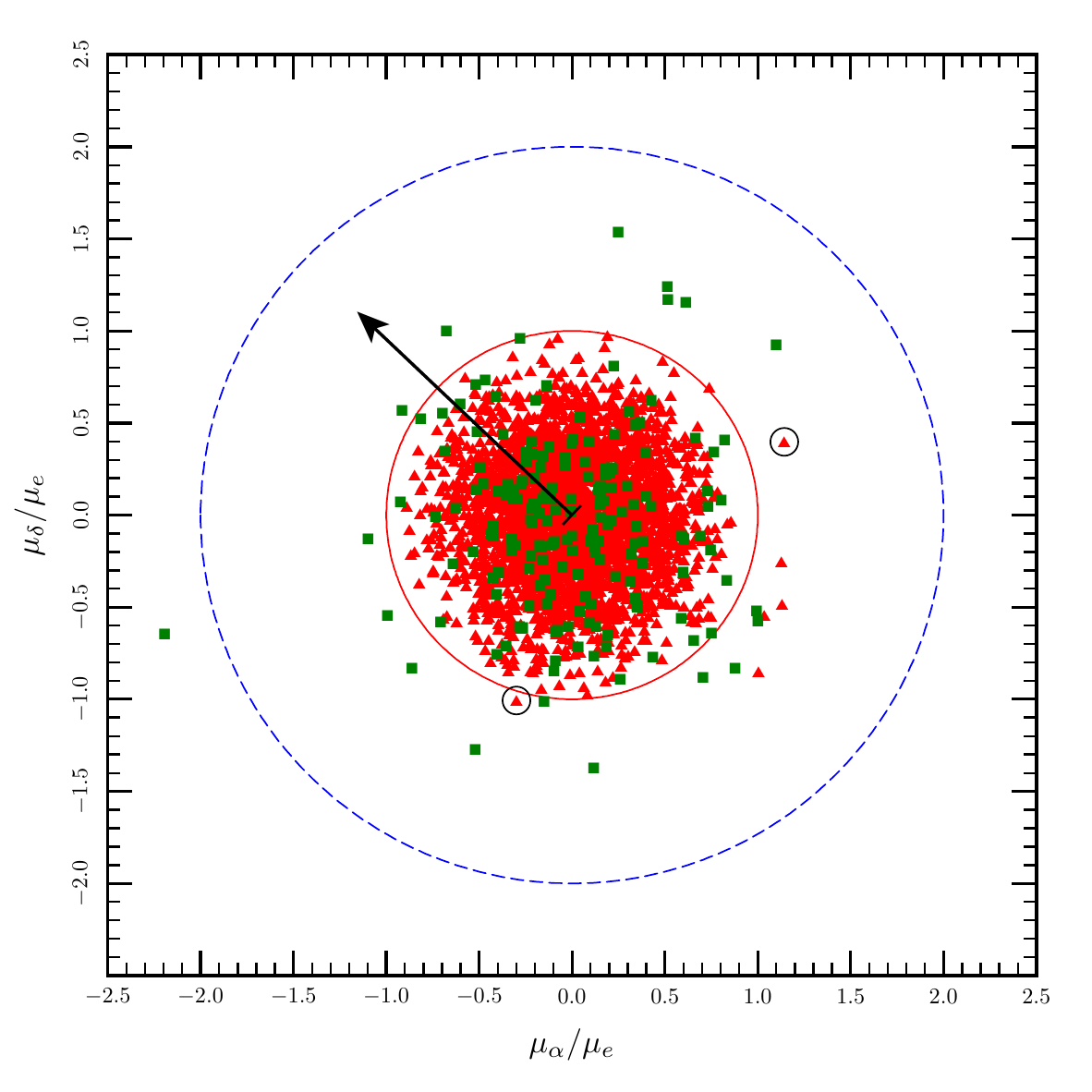}
 \caption{The proper motions of potential stellar escapers.  Only
   objects with proper-motion errors less than 0.4 mas/yr are included
   (the minimum escape proper motion is 1.3 mas/yr).  There is no
   proper-motion cutoff.  The arrow indicates the direction of the
   center of the cluster.  The proper motions of stars along the main
   sequence and white-dwarf track are depicted with red triangles and green
   squares respectively.  The most likely escapers are circled.  The
   escape proper motion is denoted by the solid red circle and twice
   its value by the dashed blue circle.}
  \label{fig:pm_escape}
\end{figure}

\begin{figure}
 \includegraphics[width=\columnwidth]{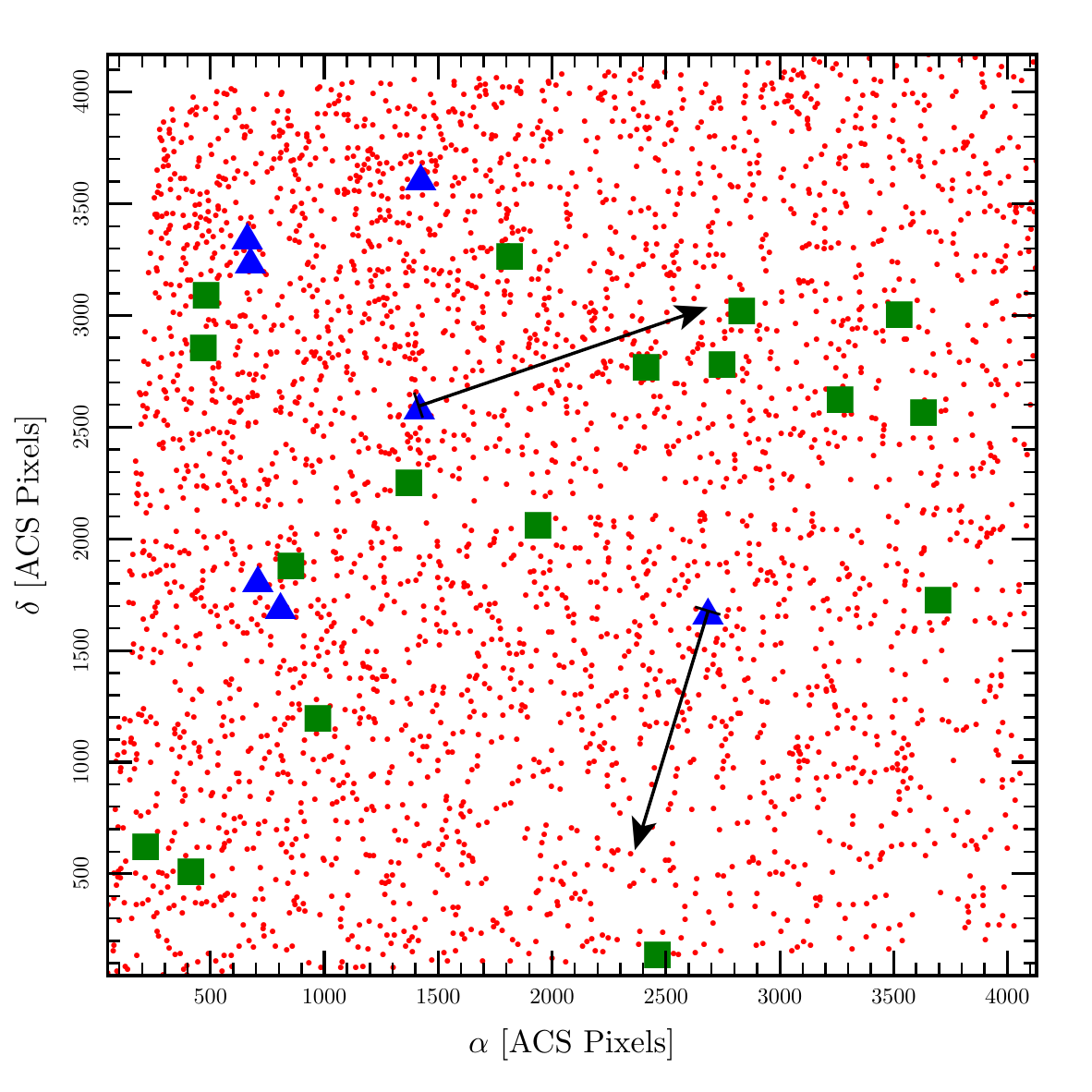}
 \caption{The positions of potential stellar escapers.  The center of
   the cluster is located beyond the upper-left corner of the field. The
   red dots show the positions of all stars in the sample, the
   potential escaping MS stars are blue triangles, and potential
   escaping WD stars are green squares. The motions of two most likely
   escapers over the next fifty thousand years are depicted.}
  \label{fig:xy_escape}
\end{figure}

\begin{figure}
 \includegraphics[width=\columnwidth]{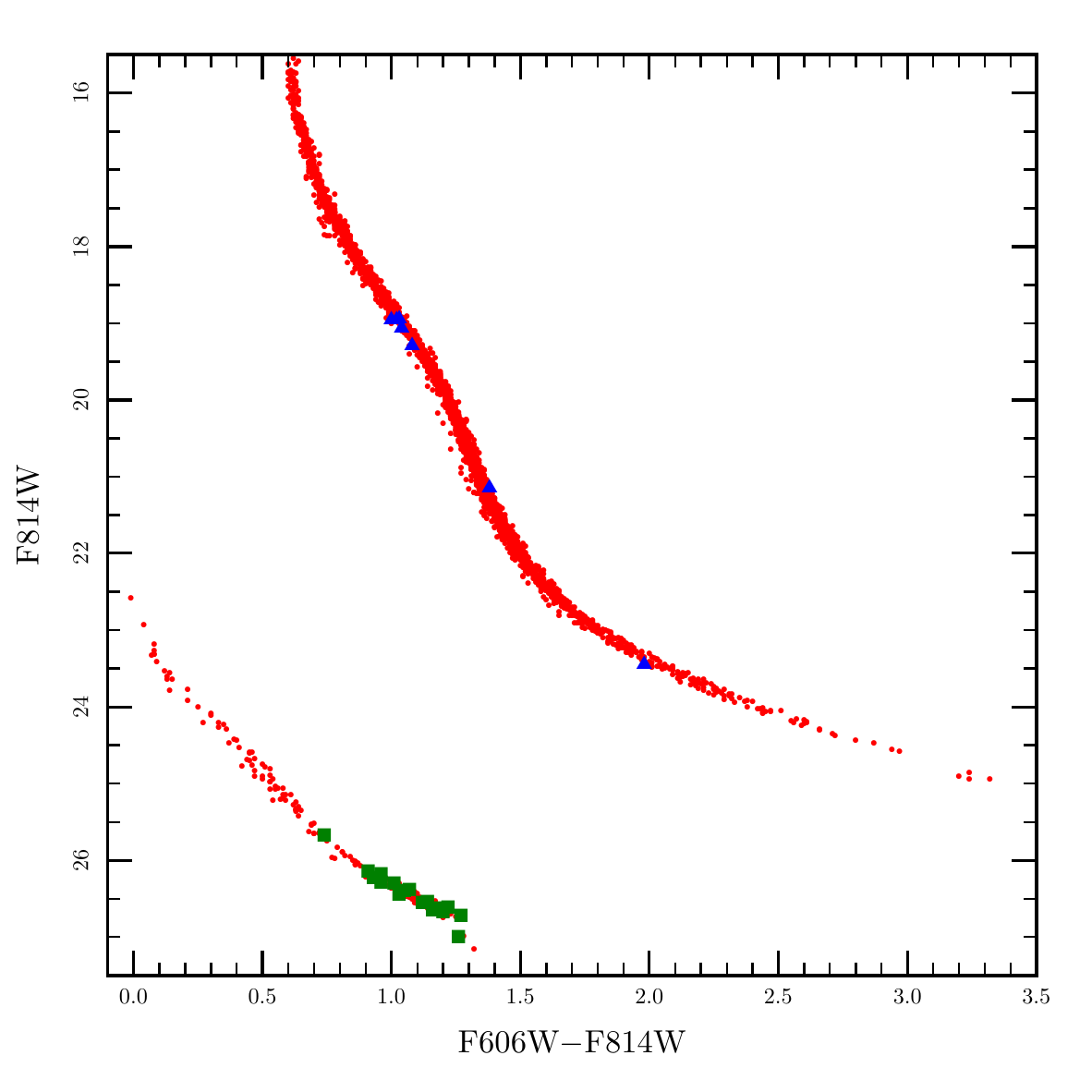}
 \caption{The color and magnitude of
   potential stellar escapers.  The red dots
   show the positions of all stars in the sample, the potential
   escaping MS stars are blue triangles, and WD stars are green
   squares.}
  \label{fig:cmd_escape}
\end{figure}

Table~\ref{tab:escapers} lists those stars that lie on or near the
cluster main-sequence or white-dwarf tracks whose total proper motion
exceeds the escape proper motion at the tangent point for the
projected radius of the star.  We have verified that all of these
potential escapers satisfy the image-quality criteria outlined in
\citet{2008AJ....135.2114A}. Fig.~\ref{fig:pm_escape} depicts the
ratios of the stellar proper motions to the escape proper motion
(Tab~\ref{tab:escapers} lists those stars between the two circles).
Looking at the proper motions themselves in Fig.~\ref{fig:pm_escape},
the potential white-dwarf escapers (green) appear to be a smooth
extension of the rest of the stars; whereas the potential
main-sequence (red) escapers appear to lie beyond the tail of the
main-sequence distribution. 
The
positions of the potential escapers in Fig.~\ref{fig:xy_escape} do not
appear extraordinary; however, their locations on the color-magnitude
diagram in Fig.~\ref{fig:cmd_escape} are.  The potential escaping
white dwarfs are among the faintest stars with measured proper motions
in the sample and therefore are prone to greater error.  This agrees
well with the apparent smoothness of their proper-motion distribution
with the rest of the white-dwarf population.

On the other hand, the location of the potential main-sequence
escapers in Fig.~\ref{fig:cmd_escape} presents a bit more of a puzzle.
All but two of these stars lie between $18.9<\mathrm{F814W}<19.3$
where the astrometry begins to use the shorter exposures in which
these stars are not saturated.  In fact in these short exposures the
proper-motion errors are comparable to those of $\mathrm{F814W}
\approx 24$ in the longer exposures and approach 0.1 mas/yr.  Although
these errors are not as large as those for the potential white-dwarf
escapers, there are many more stars at these magnitudes, so one would
expect a number of outliers.  The final potential escapers lie at
$\mathrm{F814W} \approx 21$ and 23 where the proper-motion errors are
typically less than 0.05 mas/yr.  Statistically it is unlikely that
these objects are field stars for the reasons mentioned earlier.  The
brighter, faster moving star (ID~25395) with a total proper motion of
1.34~mas/yr appears clearly beyond the tail in the distribution shown
in Fig.~\ref{fig:msdist_tot}. From Tab.~\ref{tab:escapers} we see that
its total proper motion exceeds its escape proper motion by sixteen
times its proper-motion error.  On the other hand the fainter, slower
moving star (ID~14687) exceeds its escape proper motion by about the
value of its proper motion error, so its assignment as an escaper is
less certain.  In fact given the number of stars slightly below the
escape threshold it is likely that this star is not an escaper but
rather measurement errors pushed it above the threshold;
therefore, we estimate that there is one escaper out of a sample of
3,245 stars.  Typically a star on an escape trajectory will leave the
cluster within a crossing time, $\tau_c \sim r/v \approx
5^\prime/0.3~\mathrm{mas~yr}^{-1} \approx 10^6$~yr.  This yields an
estimate of the evaporation timescale, $(d\ln N/dt)^{-1} \approx
3$~Gyr, or about three relaxation timescales.  This evaporation
timescale does not indicate that the cluster will vanish over the next
2~Gyr.  On the contrary the escaping stars as apparent from
Fig.~\ref{fig:cmd_escape} are less massive than average so the
mass-loss timescale will be perhaps two or three times longer;
furthermore, this timescale is the time over which the cluster loses
about two-thirds of its stars, not all of them.  This agrees with the
destruction timescale calculated by \citep{1997ApJ...474..223G} of
about 4~Gyr.  This star is moving away from the centre of
the cluster, although it does not travel on a strictly radial trajectory.
This is not surprising as the diffusion of stars in velocity to become
unbound is not dominated by a few strong encounters but by many small
ones.  The final of these small encounters could have occurred
anywhere within the cluster not necessarily near the centre, so the
trajectory of a star as it escapes need not be radial.

\section{Discussion}
\label{sec:discussion}

\subsection{Dynamics}
\label{sec:dynamics}

The proper motion measurements by ACS in NGC~6397 reveal gross
agreement with the features of a mass-segregated lowered isothermal
distribution function for the stars.  On the face of it this
conclusion is not particularly surprising because such a model was
conceived to describe globular clusters such as NGC~6397.  
There is no strong evidence for anisotropy in the proper motion
distribution, neither in the stellar population as a whole nor in
subsamples of stars of particular apparent magnitude or at particular
projected distances from the cluster center.

Three discrepancies with the lowered isothermal distribution are
apparent.  First, the effects of mass segregation appear larger in the
data than predicted in the model under the assumption that
$m_i\sigma_i^2$ is constant. The velocity dispersion in the ACS field
depends modestly on the apparent magnitude (the mass) of the stars, as
expected for the model; however, the mass dependence appears to be
slightly stronger in the data than in our model
(Fig.~\ref{fig:qnmio}).  Furthermore, the radial distribution of stars
in our sample depends more strongly on mass than in the models
(Fig.~\ref{fig:cum_rdist}); these differences are stronger than we
expect from incompleteness (Fig.~\ref{fig:radmio}).
Second, the proper motion distribution of young white dwarfs
(younger than 0.7 Gyr) appear to not satisfy equipartition with the
other stars of the cluster.  Their proper motions are probably smaller
than those of stars of similar masses, and their proper motions are
most similar to the motions of the most massive stars observed in the
cluster, {\em i.e.} stars similar to their progenitors.  The young
white dwarfs do not appear to have relaxed, yielding the {\em first
  direct measurement of the relaxation time in a globular cluster} of
about or greater than 0.5~Gyr.  This value agrees with the theoretical
expectations from the number of stars in NGC~6397 near the half-light
radius and the velocity dispersion of the cluster.  Thirs, there are
some stars with sufficient energy to escape the cluster whereas the
lowered isothermal distribution function contains no escaping stars.
The abundance of these stars along with the observed crossing time
yields an estimate of the evaporation time of the cluster of about
3~Gyr.  Because the typical escaping stars have masses smaller than
average, the mass-loss timescale is longer perhaps by a factor of a
few.  This constraint agrees with theoretical expectations
\citep{1997ApJ...474..223G} of a destruction timescale of about 4~Gyr.

\subsection{Properties}
\label{sec:properties}

Because the ACS field lies beyond the half-light radius of the
cluster, the measurements yield an estimate for the total mass of the
globular cluster with only modest extrapolation.  The resulting mass
is 
$1.1 \pm 0.1 \times 10^5 d_{2.53}^3 \mathrm{M}_\odot$
in agreement with
the result of MM91.  Proper motions, unlike radial velocity
measurements, probe the isotropy of the velocity distribution, so they
give a more robust mass estimate \citep{1989ApJ...3339..195L}.
Because the proper motions show no evidence for anisotropy, the
line-of-sight velocity measurements of MM91 who assumed isotropy gave
a similar mass estimate and mass-to-light estimate.  This
mass-to-light ratio (of about 1.5 in solar units) does not indicate a
need for dark matter in the cluster to explain the observed dynamics.

The major difference with previous results is the kinematic estimate
of the distance to NGC~6397.  These estimates range from $1.78\pm
0.07$~kpc to $2.2^{+0.5}_{-0.7}$~kpc as summarized in
Table~\ref{tab:distances}.  The larger estimates mitigate against
various known biases at the expense of increased statistical
uncertainties.  Our best estimate with the current data of $2.0\pm
0.2$~kpc with ninety-percent confidence was obtained by fitting a
linear model to the radial velocity dispersion measured by MM91
and comparing this with the dispersion of the proper motions of the
brightest stars in our sample and by correcting for the difference in
mass between the bright-star sample and the giants.  Although our
longest distance (and most uncertain) estimate does agree with the
previous standard-candle estimates, the most precise of which is $2.53
\pm 0.05$~kpc within their individual ninety-percent confidence
intervals, there is some tension between our best estimate ($2.0\pm
0.2$~kpc) and the standard-candle estimates.  A short kinematic
distance to NGC~6397 is similar to the short distance found by MAM06
for 47~Tuc that was also about twenty percent smaller than the
standard candle estimates.  This kinematic distance estimate is still
plagued by potential systematic errors because the stars with proper
motions are not the stars with radial velocities; therefore, some
extrapolation over position in the cluster and stellar mass is
required to get the distance estimate.  Such a short distance, if
indeed correct, could pose a challenge for stellar evolution and white
dwarf cooling models; however, it does indicate a need to obtain more
precise and model-independent measurements of distances to globular
clusters.

\subsection{Future Directions}
\label{sec:future-directions}

In light of the data presented, further modelling of the cluster is
warranted.  In particular a larger $N-$body model than that recently
performed by \citet{2008AJ....135.2129H} to account for both the
observed star count and the observed proper motions could dramatically
increase our understanding of this well-studied cluster.  In
particular the improved model of the cluster should include the
effects of disk shocking to consider the effects of the observed orbit
\citep{2007ApJ...657L..93K} of the cluster.  As mentioned in the
introduction, the number of the stars in the outskirts of the
\citet{2008AJ....135.2129H} model were insufficient for a detailed
comparison with the data.  A model resulting in a final cluster with
more than 150,000 stars would be sufficient for a comparison in this
field.  An observational study closer to the core of NGC~6397 would be
easier to compare with numerical models because the higher stellar
densities both in the models and the data may allow a detailed
comparison with current models.  Combination of the outer and the
inner observations for this cluster or 47 Tuc would be important.  Our
attempts to obtain a kinematic estimate of the distance to NGC~6397
have been thwarted by potential systematic errors in particular, a
lack of overlap between the stars with proper motions (presented here)
and those with radial velocities (MM91).  This obstacle also affected
the work of MAM06 for 47~Tuc.  A natural way forward would be to
obtain radial velocity measurements for the brightest stars in our
proper motion sample with F814W $\sim 16$.  This would yield the first
significant sample of stars with known velocities in all three
directions for a globular cluster and would prove a powerful tool in
probing the cluster as well as obtaining a distance estimate.

Given the relative vicinity of NGC~6397 to Earth (and 47~Tuc as well
for that matter) and the high-precision astrometry of ACS, a natural
next step would be to measure the distances to these clusters with
parallax. Both astrometric epochs for this work were obtained at the
same time of year, removing the effects of parallax to focus on the
proper motion of the cluster and its stars.
\citet{2007ApJ...657L..93K} found a sufficient number of background
galaxies to fix the astrometry to a precision of about 0.2~mas.  Even
with ACS observations at a different time of year, this level of
precision is insufficient to improve on the current precision of the
measurement of the distance to NGC~6397. Two potential ways forward
would be to find another way to fix the astrometry or to obtain proper
motions and radial velocities for the same stars.  In the more distant
future GAIA should be able to measure the distance to NGC~6397 and
other nearby clusters.  Distances to these nearby metal-rich and
metal-poor clusters would have dramatic implications beyond the study
of globular clusters themselves from detailed stellar models to the
extragalactic distance ladder, extending to the edge of the observable
Universe.



\begin{acknowledgements}
We would like to thank Georges Meylan for providing unpublished radial
velocity data for NGC~6397.
The research discussed is based on NASA/ESA Hubble
Space Telescope observations obtained at the Space Telescope Science
Institute, which is operated by the Association of Universities for
Research in Astronomy Inc. under NASA contract NAS5-26555. These
observations are associated with proposals GO-10424 (PI: Richer) and
GO-11633 (PI: Rich). This work was supported by NASA/HST grants
GO-10424 and GO-11633, the Natural Sciences and Engineering Research
Council of Canada, the Canadian Foundation for Innovation, the British
Columbia Knowledge Development Fund.  It has made used of the NASA ADS
and arXiv.org.
\end{acknowledgements}


\newpage
\appendix
\newcounter{Aequ}
\newenvironment{Aequation}
  {\stepcounter{Aequ}%
    \addtocounter{equation}{-1}%
    \renewcommand\theequation{A\arabic{Aequ}}\equation}
  {\endequation}

The observations presented here probe beyond the half-light radius of
NGC~6397.  In particular a useful approximation is the limit where the
value of $v_e \ll \sigma$ as is appropriate in this region.  In this
limit we have
\begin{Aequation}
J_n \approx \frac{1}{\sigma^2} \int_0^{v_e} 
    \frac{v_e^2-v^2}{2} v^{n+2} dv  \approx  \frac{1}{\sigma^2}\frac{v_e^{n+5}}{(n+3)(n+5)}
\label{eq:6}
\end{Aequation}
so ratios such as $\langle v^2 \rangle = J_2/J_0$ no longer depend on
$\sigma$.  This particular limit applies near the edge of the cluster
regardless of the underlying distribution function as long as the
distribution function is approximately linear in energy near the
escape threshold.  Furthermore, the linear approximation of the
distribution function underestimates the number of stars with small
velocities; therefore, it will overestimate the value of higher
moments of the distribution function relative to lower ones.
In particular 
\begin{Aequation}
\langle v^2 \rangle = \frac{3}{7} v_e^2 ~\mathrm{or}~ 
 \langle v_{1D}^2 \rangle = \frac{v_e^2}{7}
\label{eq:7}\end{Aequation}
yielding an estimate of the escape velocity in terms of the local
one-dimensional velocity dispersion.  Let us define the
line-of-sight integrated quantities
\begin{Aequation}
K_n(R) \equiv \int_{-\infty}^\infty J_n d z = 2 \int_0^{\pi/2}\!\! J_n(R \sec\theta)
\sec^2\theta d \theta
\approx 2 \int_0^{\pi/2}
  \frac{1}{\sigma^2}\frac{v_e^{n+5}}{(n+3)(n+5)} \sec^2\theta d\theta.
\label{eq:9}
\end{Aequation}
Let us assume that the bulk of the matter lies within the radius $R$,
so
\begin{Aequation}
\Psi = \frac{v_e^2}{2} \approx  \frac{GM}{r} - \frac{GM}{r_t} = \frac{GM}{R} \left ( \cos\theta
  - \cos\theta_t \right )
\label{eq:10}\end{Aequation}
where $\cos\theta_t=R/r_t$, so in this
limit
\begin{Aequation}
K_n(R) \approx \frac{2}{\sigma^2} \frac{\left ( 2GM/R \right
  )^{(n+5)/2}}{(n+3)(n+5)} \int_0^{\theta_t} \frac{\left ( \cos\theta - \cos\theta_t \right )^{(n+5)/2}}{\cos^2\theta}
 d\theta.
\label{eq:11}
\end{Aequation}
If $R\ll r_t$ we get
\begin{Aequation}
K_n(R) = \frac{\pi^{1/2}}{\sigma^2} \frac{v_e(R)^{n+5}}{(n+3)(n+5)}
\frac{\Gamma\left (\frac{n+3}{4} \right)}{\Gamma\left (\frac{n+5}{4}\right)}
\label{eq:12}\end{Aequation}
so
\begin{Aequation}
v_e(R)
\approx \frac{\Gamma^2\left(\frac{3}{4}\right)}{\pi} \sqrt{ 42 \langle v_{1D}^2 \rangle_{\mathrm{LOS}} }
\approx 3.1 \sqrt{ \langle v_{1D}^2 \rangle_{\mathrm{LOS}} }. \label{eq:13}
\end{Aequation}
As the ratio of the projected radius to the tidal radius increases,
$v_e(R)$ approaches $\sqrt{8} \sqrt{ \langle v_{1D}^2
  \rangle_{\mathrm{LOS}} }$.  This not only provides a useful
empirical estimator of the local escape velocity to find stars that
may be escaping but also demonstrates that the observed velocity
dispersion integrated along the line of sight for the outer regions of
the cluster is independent of the value of $\sigma$; therefore, it is
also independent of the mass of the stars even if equipartition holds
and $\sigma \propto m^{-1/2}$.  Further examination of Eq.~\ref{eq:6}
and~\ref{eq:12} shows that the density of stars in the outskirts of
the cluster is proportional to $\sigma^{-2}$ but the radial dependence
is independent of the value of $\sigma$; therefore, within a field
where $v_e\ll \sigma$ we expect that the signatures of mass
segregation will be weaker than regions where $v_e \gtrsim \sigma$.

This paper looks at the distribution of the magnitude of
proper motions that results from integrating the phase-space density
over the allowed range of velocities along the line of sight,
\begin{Aequation}
n_2(v|r) = \frac{\rho_1}{2 \pi \sigma^2} ~\mathrm{erf} \left (
    \sqrt{\frac{v^2_e -v^2}{2\sigma^2}} \right )
  e^{(v_e^2-v^2)/(2\sigma^2)} - \frac{2\rho_1 \sqrt{v^2_e - v^2}}{\left(2\pi \sigma^2\right)^{3/2}}
\label{eq:15}
\end{Aequation}
where $\rho_1$ is a normalizing factor.
Of particular interest is the cumulative distribution of the magnitude 
of the proper motion of the stars. The number density of stars with
proper motions greater than a given value $v$ is
\begin{Aequation}
n(>v|r) = \rho_1 \Biggr [ \exp \left ( \frac{v^2_e-v^2}{2\sigma^2} \right ) \mathrm{erf} \left (
  \sqrt{\frac{v^2_e-v^2}{2\sigma^2}} \right ) - \sqrt{\frac{2}{\pi} \frac{v_e^2-v^2}{\sigma^2} } \left (  1 + 
\frac{v_e^2-v^2}{3 \sigma^2}  \right ) \Biggr ].
\label{eq:16}
\end{Aequation}
One should note that one can get Eq.~\ref{eq:16} simply by replacing
$v_e$ in Eq.~\ref{eq:3} (of the main text) with $\sqrt{v_e^2-v^2}$;
therefore, in the limit as the minimum velocity $v$ vanishes,
Eq.~\ref{eq:16} yields Eq.~\ref{eq:3} as expected.  The observable is
of course the velocity distribution of the sample along the line of
sight
\begin{Aequation}
N(>v|R) = \int_R^{r_t} n(>v|r) \frac{r dr}{\sqrt{r^2-R^2}},
\label{eq:17}\end{Aequation}
which must be calculated numerically.  Alternatively we can integrate
$n(>v|r)$ against the spherical window function defined in
\S~\ref{sec:acs-field} to obtain the proper-motion distribution for
the entire field.

\end{document}

%% file: escape.tex
\begin{table*}
\begin{center}
\caption{Cluster stars with proper-motions that exceed the local escape proper motion}
\label{tab:escapers}
\begin{tabular}{rrrrrrrrrc}
\hline \hline 
& & & 
\multicolumn{1}{c}{$\mu_\mathrm{tot},\mu_e$} & 
\multicolumn{1}{c}{$\alpha_{2000}$} & 
\multicolumn{1}{c}{$\delta_{2000}$} & 
\multicolumn{1}{c}{$\mu_\alpha$} & 
\multicolumn{1}{c}{$\mu_\delta$} &
\multicolumn{1}{c}{$\mu_R$} &
\\ 
\multicolumn{1}{c}{ID} & 
\multicolumn{1}{c}{F814W} & 
\multicolumn{1}{c}{F606W-F814W} &
\multicolumn{1}{c}{[mas yr$^{-1}$]} & 
\multicolumn{1}{c}{[arcsec]} & 
\multicolumn{1}{c}{[arcsec]} & 
\multicolumn{1}{c}{[mas yr$^{-1}$]} & 
\multicolumn{1}{c}{[mas yr$^{-1}$]} & 
\multicolumn{1}{c}{[mas yr$^{-1}$]} & 
\multicolumn{1}{c}{Comment} 
\\ 
\hline

 14960 & 18.92~~ & 1.03~~~~~~~ &  1.27, 1.10~ &  40.38 &  84.95 &  1.24$\pm$0.08 & -0.28$\pm$0.08 & $ 0.77\pm0.08$ & MS, short \\
 35641 & 18.93~~ & 1.00~~~~~~~ &  1.52, 1.15~ &  33.10 & 167.75 &  1.16$\pm$0.05 & -0.98$\pm$0.05 & $ 1.00\pm0.05$ & MS, short \\
 39398 & 18.93~~ & 1.02~~~~~~~ &  1.32, 1.13~ &  71.20 & 180.95 &  1.17$\pm$0.08 & -0.62$\pm$0.08 & $ 1.00\pm0.08$ & MS, short \\
 34104 & 19.04~~ & 1.04~~~~~~~ &  1.41, 1.15~ &  33.82 & 162.30 &  1.30$\pm$0.05 & -0.56$\pm$0.05 & $ 0.95\pm0.05$ & MS, short \\
 16272 & 19.27~~ & 1.08~~~~~~~ &  1.12, 1.10~ &  35.40 &  91.00 &  0.81$\pm$0.03 &  0.76$\pm$0.03 & $-0.11\pm0.03$ & MS, short \\
 25395 & 21.12~~ & 1.38~~~~~~~ &  1.34, 1.11~ &  70.85 & 129.75 &  1.26$\pm$0.01 &  0.44$\pm$0.01 & $ 0.48\pm0.01$ & MS \\
 14687 & 23.42~~ & 1.98~~~~~~~ &  1.11, 1.06~ & 134.25 &  83.70 & -0.32$\pm$0.05 & -1.06$\pm$0.05 & $ 0.41\pm0.05$ & MS \\
 19308 & 25.67~~ & 0.74~~~~~~~ &  1.10, 1.08~ &  96.95 & 102.95 & -0.16$\pm$0.21 & -1.09$\pm$0.21 & $ 0.56\pm0.21$ & WD \\
 21043 & 26.14~~ & 0.91~~~~~~~ &  1.32, 1.10~ &  68.60 & 112.45 & -0.74$\pm$0.35 &  1.09$\pm$0.35 & $-0.98\pm0.35$ & WD \\
 27645 & 26.18~~ & 0.96~~~~~~~ &  1.19, 1.07~ & 137.35 & 139.00 & -1.18$\pm$0.27 & -0.14$\pm$0.27 & $-0.77\pm0.27$ & WD \\
 34340 & 26.23~~ & 0.93~~~~~~~ &  1.26, 1.11~ &  90.70 & 163.20 & -1.10$\pm$0.33 & -0.61$\pm$0.33 & $-0.46\pm0.33$ & WD \\
 9666 & 26.29~~ & 0.96~~~~~~~ &  1.21, 1.08~ &  48.60 &  59.70 &  1.07$\pm$0.32 & -0.56$\pm$0.32 & $ 0.89\pm0.32$ & WD \\
 15213 & 26.30~~ & 1.01~~~~~~~ &  1.12, 1.04~ & 184.80 &  86.15 & -0.95$\pm$0.36 &  0.59$\pm$0.36 & $-1.00\pm0.36$ & WD \\
 4333 & 26.39~~ & 1.07~~~~~~~ &  1.29, 1.08~ &  10.77 &  30.96 & -0.93$\pm$0.33 & -0.90$\pm$0.33 & $ 0.30\pm0.33$ & WD \\
 28585 & 26.40~~ & 1.05~~~~~~~ &  1.32, 1.14~ &  23.43 & 142.55 &  1.14$\pm$0.36 & -0.66$\pm$0.36 & $ 0.95\pm0.36$ & WD \\
 25008 & 26.44~~ & 1.03~~~~~~~ &  1.45, 1.05~ & 181.60 & 128.10 &  0.12$\pm$0.26 & -1.44$\pm$0.26 & $ 0.57\pm0.26$ & WD \\
 3445 & 26.54~~ & 1.14~~~~~~~ &  1.37, 1.07~ &  20.72 &  25.39 &  0.55$\pm$0.31 &  1.25$\pm$0.31 & $-0.62\pm0.31$ & WD \\
 30992 & 26.55~~ & 1.12~~~~~~~ &  1.30, 1.07~ & 141.70 & 151.00 &  0.94$\pm$0.28 & -0.90$\pm$0.28 & $ 0.97\pm0.28$ & WD \\
 433 & 26.61~~ & 1.22~~~~~~~ &  1.35, 1.03~ & 123.15 &   6.78 &  0.63$\pm$0.37 &  1.19$\pm$0.37 & $-0.38\pm0.37$ & WD \\
 27465 & 26.62~~ & 1.18~~~~~~~ &  1.55, 1.08~ & 120.70 & 138.20 &  1.19$\pm$0.35 &  1.00$\pm$0.35 & $ 0.27\pm0.35$ & WD \\
 31935 & 26.65~~ & 1.16~~~~~~~ &  1.30, 1.15~ &  24.08 & 154.50 &  0.81$\pm$0.28 & -1.02$\pm$0.28 & $ 0.99\pm0.28$ & WD \\
 16969 & 26.67~~ & 1.20~~~~~~~ &  1.52, 1.10~ &  42.69 &  93.90 & -0.57$\pm$0.34 & -1.40$\pm$0.34 & $ 0.49\pm0.34$ & WD \\
 30725 & 26.72~~ & 1.27~~~~~~~ &  1.42, 1.06~ & 176.30 & 150.00 &  0.54$\pm$0.30 &  1.31$\pm$0.30 & $-0.09\pm0.30$ & WD \\
 25724 & 27.00~~ & 1.26~~~~~~~ &  1.65, 1.06~ & 163.30 & 131.10 &  0.26$\pm$0.39 &  1.62$\pm$0.39 & $-0.38\pm0.39$ & WD \\
\hline
\end{tabular}
\end{center}
\smallskip
The quantities $\alpha_{2000}$ and $\delta_{2000}$ give the offset in
arcseconds from the origin of the coordiates of the field at
$\alpha=17^\mathrm{h}40^\mathrm{m}56^\mathrm{s}\!\!.72$ and $\delta=-53^\circ45'36''\!\!.8$.
\end{table*}